\documentclass[twocolumn, tighten]{aastex701}

\usepackage[utf8]{inputenc}
\usepackage{natbib,
graphics,
graphicx,
amsmath,
amssymb,
threeparttable,
xcolor}
\usepackage{times}
\usepackage{newtx}
\usepackage{hyperref}

\setcitestyle{semicolon,notesep={}}

\defcitealias{straatman16}{S16}
\defcitealias{eisenstein26}{E26}
\defcitealias{momcheva16}{M16}
\defcitealias{cowie18}{C18}

\newcommand{\ci}{[C\,{\sc i}]\,}

\newcommand{\fsm}{\ensuremath{f_{{\rm 850\, \mu m}}}}
\newcommand{\falma}{\ensuremath{f_{{\rm 870\, \mu m}}}}
\shortauthors{McKay et al.}

\begin{document}

\title{A Deep ALMA Survey of the Redshift Distribution of Dusty Star-forming Galaxies}

\correspondingauthor{Stephen J. McKay}
\email{sjmckay3@wisc.edu}

\author[0000-0003-4248-6128]{S.~J.~McKay}
\affiliation{Department of Physics, University of Wisconsin--Madison, 1150 University Avenue,
Madison, WI 53706, USA}
\email{sjmckay3@wisc.edu}

\author[0000-0002-3306-1606]{A.~J.~Barger}
\affiliation{Department of Astronomy, University of Wisconsin--Madison, 475 N. Charter Street, Madison, WI 53706, USA}
\affiliation{Department of Physics and Astronomy, University of Hawaii, 2505 Correa Road, Honolulu, HI 96822, USA}
\affiliation{Institute for Astronomy, University of Hawaii, 2680 Woodlawn Drive, Honolulu, HI 96822, USA}
\email{barger@astro.wisc.edu}

\author[0000-0002-6319-1575]{L.~L.~Cowie}
\affiliation{Institute for Astronomy, University of Hawaii, 2680 Woodlawn Drive, Honolulu, HI 96822, USA}
\email{cowie@ifa.hawaii.edu}

\author[0000-0002-8686-8737]{F.~E.~Bauer}
\affiliation{Instituto de Alta Investigaci\'on, Universidad de Tarapac\'a, Casilla 7D, Arica, 1010000, Chile}
\email{franz.e.bauer@gmail.com}

\begin{abstract}
We present an Atacama Large Millimeter/submillimeter Array (ALMA) spectroscopic follow-up survey of an 870~$\mu$m-selected sample of dusty star-forming galaxies (DSFGs) in the GOODS-S field. We use these linescans to identify or confirm spectroscopic redshifts (spec-zs) for 26 sources. Including spec-zs from the literature, there are now secure or tentative spec-zs for 54 out of 75 DSFGs (72\%). At $\falma>2.5$~mJy, the sample is 97\% spectroscopically complete, allowing us to model the  full DSFG redshift distribution down to nearly the confusion limit for a 15-m telescope at 850~$\mu$m. This is the highest completeness for an unbiased sample at this flux limit to date. We find that nearly all of the DSFGs in our sample that were targeted with JWST/NIRSpec were spectroscopically identified, without much dependence on near-infrared or submillimeter flux or redshift. However, only 29\% of our sample have JWST spectroscopic coverage. We use the spec-zs to evaluate various photometric redshift (photo-z) estimates, finding that all methods exhibit an outlier fraction of at least $>20$\%. Nearly all of the photo-z methods tend to overshoot the redshifts, leading to overestimates of the number of DSFGs at high redshift ($z>4$). Our results suggest that $\lesssim6$\% of $\falma \geq 2.25$~mJy DSFGs lie at $z>4$ and $\lesssim1$\% lie at $z>5$, reflecting a steep decline in the abundance of massive dusty galaxies in the first 1.5~Gyr.
\end{abstract}

\section{Introduction}

Submillimeter-bright ($\falma \gtrsim 1$--2~mJy) dusty star-forming galaxies (DSFGs) represent an extreme phase of dust-obscured galaxy evolution that was most prevalent roughly 10 billion years ago ($z\sim2.5$; see reviews by \citealt{casey14, hodge20}). 
While their extreme far-infrared (FIR) luminosities are comparable to those of local ultraluminous infrared galaxies ($L_{\rm IR} \gtrsim 10^{12}\,{\rm L}_\odot$), their space densities are $\sim$1000 times higher and they contribute $\gtrsim$30\% of the total cosmic star formation rate density (SFRD) above $z>1$ \citep[e.g.,][]{cowie17}, making them a crucial component of the overall framework of galaxy evolution. 

Our current understanding of DSFGs suggests that they represent the progenitors of the most massive elliptical galaxies observed in the local Universe, after quickly building up their stellar mass and passing through a compact quiescent phase. This rapid evolution could be triggered by major mergers \citep[e.g.,][]{lilly99, swinbank06, toft14, simpson14, ikarashi15}, although there is evidence that a fraction of the population may evolve through more secular processes \citep[e.g.,][]{valentino20, birkin21, araya-araya26} and large JWST imaging studies have generally found that less than half of DSFGs appear to be involved in interactions, with most exhibiting disk-like morphologies \citep[e.g.,][]{gillman23, gillman24, ren25, mckay25, bodansky26, smail26}.
Even as recent state-of-the-art simulations have successfully matched some of the broad characteristics of the population \citep[][]{lacey16, mcalpine19, hayward21, lovell21, kumar25}, reproducing the flux and redshift distributions and correlations for DSFGs continue to be a challenge for models \citep[e.g.,][]{chen22}. 
Testing theories of DSFG evolution further requires properly accounting for their volume density, clustering, and contribution to the cosmic SFRD; accurately measuring properties such as their stellar and gas masses; and understanding their environments and descendants. 

Crucially, all of these endeavors rely on accurate redshifts.
Although (sub)millimeter observations are effective at detecting DSFGs out to $z\gtrsim8$, thanks to the very negative $K$-correction at these wavelengths, measuring spectroscopic redshifts (spec-zs) has proved difficult due to the challenge of successfully detecting rest-frame optical emission lines and/or FIR molecular transitions for faint and dusty sources \citep[e.g.,][]{barger99b, casey17, danielson17, birkin21, chen22}. 

For most galaxies, we are forced to rely on photometric redshifts (photo-zs) from either optical/near-infrared (NIR) or full panchromatic spectral energy distribution (SED) modeling \citep[e.g.,][]{brammer08, dacunha15, battisti+19, boquien19}, or from various FIR color- or template-based methods \citep[e.g.,][]{hughes97, barger00, aretxaga03, casey13, ivison16, casey20, barger22}. 
While state-of-the-art photo-z codes achieve excellent redshift accuracy for the broader galaxy population, for very dusty sources the results are more likely to be uncertain, or to fail catastrophically \citep[e.g.,][]{zavala23, jin24}. Highly complete spectroscopic coverage is essential in order to test and refine our current photo-z methods for DSFGs.

Early DSFG studies were somewhat biased by the necessity of using probabilistic optical counterparts or by requiring radio counterpart identifications \citep[e.g.,][]{barger99, chapman05, aretxaga07, chapin09, yun12} and therefore reported varying results for the overall DSFG redshift distribution.
As unbiased studies have been made possible by interferometers such as the Atacama Large Millimeter/submillimeter Array (ALMA), they have converged on a median redshift of around $z\sim2.6$, increasing slightly with increasing flux and selection wavelength. Most show an asymmetric redshift distribution spanning roughly $1\lesssim z \lesssim 4$, with a small subset at higher redshifts \citep[][]{dacunha15, cowie18, simpson20, dudzeviciute20}.

Currently, however, the only flux-complete and nearly or fully spectroscopically complete ALMA samples sit at the very bright end ($\falma>25$~mJy; {the 81 galaxies} in the South Pole Telescope sample, which are mostly strongly lensed; \citealt{reuter20}) and the very faint end ({32 secure ASPECS sources}, with only two having $\falma>2.25$~mJy; \citealt{aravena20}). 
By contrast, for ALMA follow-up studies of James Clerk Maxwell Telescope (JCMT)/SCUBA-2 850~$\mu$m surveys, with single-dish flux limits of $\fsm \sim $3--10~mJy, the total spec-z fraction is typically $\sim$5--40\% \citep[][]{danielson17, dudzeviciute20, simpson20, mckinney25}, even in very well-studied fields, with full spectroscopic completeness only at the very brightest ends of the samples \citep[$\gtrsim$10--12~mJy; e.g.,][]{birkin21, chen22}.

With longer wavelength coverage, better sensitivity, and no atmospheric transmission concerns, JWST appears to be a powerful new tool for obtaining spec-zs for DSFGs.
Four years into JWST's operation, there are now thousands of published JWST/NIRSpec spectra across key legacy fields, including for very dusty and high-redshift objects \citep[e.g.,][]{cooper25, degraaff25, barrufet25}. 
However, to date, it is unclear how successful JWST has been at confirming redshifts for the overall DSFG population and what the prospects are for completing the spectroscopic census, particularly at the highest redshifts and the faint end of the luminosity function.

In this paper, we present an ALMA spectroscopic survey of the large and deep sample of DSFGs in the GOODS-S field \citep{giavalisco04} presented by \citet[][; hereafter, \citetalias{cowie18}]{cowie18}. These sources benefit from many archival UV/optical/NIR spectroscopic campaigns, including medium-to-deep JWST/NIRSpec surveys, offering an opportunity to test how well JWST and ALMA can jointly map out the DSFG population. 
The excellent multiwavelength archival data also provide extremely well-sampled SEDs from the UV to the radio, including wide and medium JWST/NIRCam bands and multiple ALMA bands for most sources. These enable us to compare the performance of various photo-z methods against the spec-zs for a sample with ``ideal'' photometric coverage.

This paper is organized as follows: In Section~\ref{sec:data}, we describe our sample, the ancillary data, and our analysis of the ALMA data. 
In Section~\ref{sec:speczs}, we report the spec-zs obtained from the ALMA linescans and discuss our spectroscopic completeness. 
In Section~\ref{sec:photozs}, we describe the various photo-z methods and evaluate their performance. 
In Section~\ref{sec:discussion}, we discuss the redshift distribution of the sample, the high-redshift population, and the sources that are not yet spectroscopically confirmed.
Finally, in Section~\ref{sec:summary}, we summarize our results.

We assume a flat $\Lambda$CDM cosmology with $\Omega_{m}= 0.3$, $\Omega_{\Lambda} = 0.7$, and $H_0 = 70.0$~km~s$^{-1}$~Mpc$^{-1}$. 
All magnitudes are given in the AB system \citep{oke83}.

\section{Data and Observations}
\label{sec:data}

\subsection{ALMA sample and multiwavelength photometry}
\label{sec:sample}

The sample we analyze consists of 75 DSFGs in the GOODS-S field, originally identified from ALMA 870~$\mu$m continuum follow-up of JCMT/SCUBA-2 850~$\mu$m sources from the SUPER GOODS program (hereafter, SG; \citetalias{cowie18}). {
These ALMA observations targeted all of the $>4\sigma$ SCUBA-2 sources in the central 100~arcmin$^2$, corresponding to an 850~$\mu$m flux limit of 2.25~mJy. 
Within the ALMA half power radius, the 870~$\mu$m data cover 8.5~arcmin$^2$ of the SCUBA-2 area (88 pointings) and yield 75 sources in total, with 30 above 2.75~mJy, 33 above 2.5~mJy,
and 44 above 2.25~mJy. Hereafter, we refer to these sources as SG-ALMA-1 to SG-ALMA-75, ordered as in Table~4 of \citetalias{cowie18}.}

Since some SCUBA-2 sources were resolved into multiple contributions, the resulting ALMA sample has 870~$\mu$m fluxes ranging from 0.84~mJy to 8.93~mJy. {We expect it to be nearly flux-complete at $\geq$2.25~mJy (\citetalias{cowie18}). In order to test the completeness, we used the ALMA fluxes convolved with the SCUBA-2 850~$\mu$m point-spread function (PSF) to compute
a ``true'' ALMA-based image. We then subtracted this from the SCUBA-2 image and searched the residual cleaned image for additional 850~$\mu$m sources. We found no additional sources above 2.75~mJy, 1 above 2.5~mJy and 3 above 2.25~mJy. 
It is likely these are
flux-boosted fainter sources, but accepting them
at face value we have 97\% completeness above 2.5~mJy and 94\% above 2.25~mJy.}

GOODS-S contains some of the richest multiwavelength coverage on the sky, meaning that our sources have excellent photometric data from UV to millimeter wavelengths.
In this work, we adopt the $K_s$-band magnitudes presented in \citetalias{cowie18} {(their Table~5)}, which were measured from deep CFHT/WIRCAM imaging \citep{wang10}. 
We also use the JWST/NIRCam and MIRI photometry from JADES Data Release 5 (DR5: \citealp[][]{robertson26, eisenstein26, alberts26}), which incorporates NIRCam imaging from FRESCO \citep{oesch23} and JEMS \citep{williams23}, as well as MIRI mid-infrared (MIR) imaging from SMILES \citep[][]{rieke24, alberts24}. The JADES catalogs include the HST photometry from the Hubble Legacy Field mosaics \citep{illingworth16, whitaker19}, measured consistently with the JWST photometry \citep{eisenstein26}. 
Where these data are unavailable, we use the fluxes from the CANDELS catalog \citep[][]{guo13}. {We present color cutouts of the entire sample in Appendix~A. We take the cutouts from JWST/NIRCam imaging, where available, and HST/ACS and WFC3 imaging otherwise.} All of the DSFG counterparts are clearly identified in the JWST and/or HST images.

In the MIR and FIR, we use Spitzer/MIPS {24~$\mu$m} and Herschel/PACS {100~$\mu$m and 160~$\mu$m} fluxes from \citet{elbaz11}, if available, and Herschel/SPIRE {250~$\mu$m and 350~$\mu$m} fluxes from \citet{oliver12}, which were based on Spitzer/MIPS prior positions. We also use SCUBA-2 450~$\mu$m fluxes from the SG program \citep{barger22, mckay23}. {In addition to the ALMA data from our programs, we include 1.1~mm fluxes from the GOODS-ALMA survey \citep{gomez-guijarro22}, where available. With these, plus our ALMA data at 870~$\mu$m, 1.2~mm, 2~mm, and 3~mm, our sample is covered by up to 11 bands at $\lambda \geq 24\,\mu$m. Fifty-five sources (out of 75) are detected in nine or more of these bands. For more detail on the available FIR data, we refer the reader to \citet{mckay23} and \citet{cowie23}.}

\subsection{Literature redshifts}
\label{sec:lit_redshifts}

Spec-zs for our sample were previously reported in \citetalias{cowie18} and \citet{mckay23}. We searched the literature for newly reported spec-zs, which primarily come from recently published JWST/NIRSpec catalogs (\citealp{bunker24, deugenio25, barrufet25}), taken either directly from the corresponding paper or from the Dawn JWST Archive\footnote{https://dawn-cph.github.io/dja} (DJA; \citealp[][]{heintz25}) NIRSpec catalog (version 4). The reductions for the DJA catalogs use {\tt msaexp} \citep[][]{brammer23} and are described in \citet{degraaff25} and \citet{heintz25}. It is worth noting that the redshift for SG-ALMA-68, which was originally reported to be at $z=5.579$ based on a single line in the JWST/NIRCam F444W/GrismR data from the FRESCO survey \citep[][]{oesch23, xiao24}, has since been confirmed instead to be at $z_{\rm spec} = 3.2461$ via NIRSpec integral field unit (IFU) observations \citep[][]{xiao26}.

We use photo-zs from several different surveys, with the aim of comparing the accuracy of different methods and selecting the best estimate on a case-by-case basis. We use the ZFOURGE catalog (\citealp[][]{straatman16}; hereafter, \citetalias{straatman16}), which used the {\tt EAZY} code to estimate redshifts based on the available UV to FIR data (ground-based optical to NIR data, plus HST and Spitzer/IRAC data). 
{All but three of the DSFGs have ZFOURGE photo-zs.}
We also use photo-zs from the JADES DR5 catalog (\citealp[][]{eisenstein26}; hereafter, \citetalias{eisenstein26}), which used {\tt EAZY} to fit just the HST and JWST photometry.
{Only SG-ALMA-46 does not have an E26 photo-z.}
Some sources in the sample also have grism redshifts (grism-zs) measured using the HST/WFC3 G141 and G800L grisms as part of the 3D-HST program (\citealp[][]{brammer12}; \citealp[][]{momcheva16}; hereafter, \citetalias{momcheva16}).

\begin{deluxetable}{cccccc}
\tablewidth{\textwidth}
\tabletypesize{\scriptsize}
\tablecaption{\label{tab:obs_summary}
Summary of ALMA Observations}

\tablehead{
\colhead{Program}&\colhead{Band}&\colhead{No. sources}&\colhead{Avg. rms range, 3.9~MHz channel}&\colhead{Avg. beam size}&\colhead{Freq. range$^*$}\\
&&&\colhead{(mJy)}&\colhead{($''$)}&\colhead{(GHz)}}

\startdata
2021.1.00024.S & 3 & 39 & 1.58--1.90 & 0.79 & 84.11--111.21 \\
 & 4 & 41 & 1.68--2.41 & 0.95 & 139.51--162.87 \\
 & 6 & 44 & 3.62--6.00 & 0.94 & 228.26--251.49 \\
2024.1.01213.S & 3 & 24 & 0.87--1.04 & 1.15 & 85.07--111.50 \\
 & 4 & 19 & 1.19--1.63 & 0.77 & 136.57--162.87 \\
\enddata
\tablecomments{*\,Eleven objects in program 2021.1.00024.S and nine objects in program 2024.1.01213.S were targeted in just one tuning based on a known spec-z, meaning that they do not have the full frequency coverage. 
}
\end{deluxetable}

\subsection{ALMA linescans}

We carried out two ALMA programs to perform millimeter spectral scans of sources drawn from the parent 870~$\mu$m sample. The primary goal of these programs was to obtain spec-zs from CO and/or \ci emission lines and measure molecular gas properties for a flux-complete sample of DSFGs.

In the initial program (project code \#2021.1.00024.S), we targeted {the 51 brightest ALMA sources without existing CO detections down to a flux limit of $\falma\sim2$~mJy. }
The sources were observed in four tunings in band 3 spanning $\sim$84--111~GHz, three tunings in band 4 spanning $\sim$140--163~GHz, and two tunings in band 6 spanning $\sim$228--252~GHz. These tunings were chosen to cover 2+ CO and/or \ci emission for the majority of redshifts at $z\gtrsim1$. {However, for efficiency, we included 11 objects with spec-zs known at the time of observation only in the specific tunings in which we expected to detect CO emission. In addition, our ALMA pointings covered six additional fainter (1--2.25 mJy)~sources} from the \citetalias{cowie18} sample, for a total of 57 galaxies. We estimated the expected line fluxes (and required sensitivities) assuming empirical $L_{\rm IR}$--$L'_{\rm CO}$ conversions \citep{daddi15,liu15}, a FWHM of 300\,km\,s$^{-1}$, and the CO line brightness temperature ratios from \citet{bothwell13}. For efficiency, we divided the sample into three subsamples (bright, medium, and faint) based on the observed $\falma$ and $K_s$-band fluxes, with the rms sensitivity requirements adjusted for each subsample. We calibrated the data with {\tt CASA} version 6.2.1.7 using the default ALMA pipeline parameters. The continuum detections obtained from the ALMA data were initially discussed in \citet{mckay23} and \citet{cowie23}; in this paper, we provide a more comprehensive analysis of the line detections in the dataset. 

Later, in a follow-up program (project code \#2024.1.01213.S), we obtained $\sim2.5\times$ deeper scans in ALMA Bands 3 and 4 {for the 24 sources with $\falma > 2.5$~mJy that did not have multiple lines detected in the initial survey. }
The tunings were set up in a similar manner to the previous program, with four tunings in each band, spanning $\sim$85--111~GHz in band 3 and $\sim$137--163~GHz in band 4 (thus covering 2+ lines for nearly all redshifts from $2<z<5$, where most of the unconfirmed sources were likely to fall). The spacing between the tunings was reduced slightly compared to the previous program in order to remove any gaps between spectral windows. In this program, we divided the sample into two subsets based on the expected line flux to maximize efficiency. {We also included nine sources with known optical/NIR spec-zs in the appropriate tunings, in order to detect one or more CO lines.}
We calibrated the data from this program with {\tt CASA} version 6.6.1.17 and the default parameters. 

For both programs, we produced cleaned spectral cubes and continuum images for each spectral window (without combining overlapping tunings) using the {\tt tclean} routine, adopting multithreshold automasking at the standard thresholds, Briggs weighting with {\tt robust}~=~0.5, and the standard clipping at the edges of spectral windows. The cleaned spectral cubes have a frequency resolution of 3.9~MHz. We measured continuum fluxes in the cleaned continuum images at the known source positions using 1$\farcs$6 apertures, with errors measured by placing random apertures in other regions of the image with the source masked out. 

In Table~\ref{tab:obs_summary}, we list the number of pointings, average rms per channel across the frequency range, average beam size, and frequency coverage in each ALMA band for each program.

\subsection{Line detection procedure}

To identify emission lines in the cleaned ALMA spectral cubes, we use a matched-filtering algorithm. Our method is similar to that of the {\tt LineSeeker} code \citep{gonzalez-lopez19}, but designed for the format of our dataset, i.e., multiple spectral cubes for targets with known positions. 
In brief, we first convolve the data cubes along just the spectral dimension with a series of 24 1-D Gaussian kernels ranging in FWHM from 200~km\,s$^{-1}$ to 1500~km\,s$^{-1}$. We then search for peaks in the smoothed cube at the coordinate of the targets from the 870~$\mu$m continuum imaging \citepalias{cowie18}. 
The rms noise in the peak channel is determined by sampling random positions within 15$''$ of the 870~$\mu$m position, after masking out the central 2$\farcs$0 region. We compare the peak signal-to-noise ratio (S/N) values obtained with each kernel and select the kernel producing the highest S/N for our final line measurement. 
If the maximum S/N exceeds our initial detection threshold of 3.5, we then identify the brightest pixel in the 2D peak channel map within 1$\farcs$0 of the original coordinate and extract the final spectrum at this position, binning to a spectral resolution of 80\,km\,s$^{-1}$. We measure the central frequency and FWHM of the line from a Gaussian fit to the spectrum.

Assuming Gaussian statistics and the minimum smoothing kernel of 200\,km\,s$^{-1}$, the number of expected false positives at a threshold of $4.5\sigma$ is $\lesssim$1 in all three ALMA bands (larger smoothing kernels/wider binning would reduce the number of expected false positives). 
{However, as a more complete check, we perform our identical smoothing and extraction procedure with all kernels at 2000 random positions in each data cube}, masking out the known source(s).  We then do the same for the inverted cubes (multiplied by $-1.0$).
We find that at 4.5$\sigma$, the average probability of a false detection is {$\lesssim$2.3$\%$ in both the original and the inverted cubes, and above 5$\sigma$ it is $\lesssim$0.3\%}. These are largely independent of the observed frequency or the radius used for the rms noise measurement. Based on this evidence, for simplicity, we adopt 4.5$\sigma$ as our significance threshold for secure emission lines across all bands, {since we expect nearly all of these detections to be real.}

\begin{figure*}
    \centering
    \includegraphics[width=\linewidth]{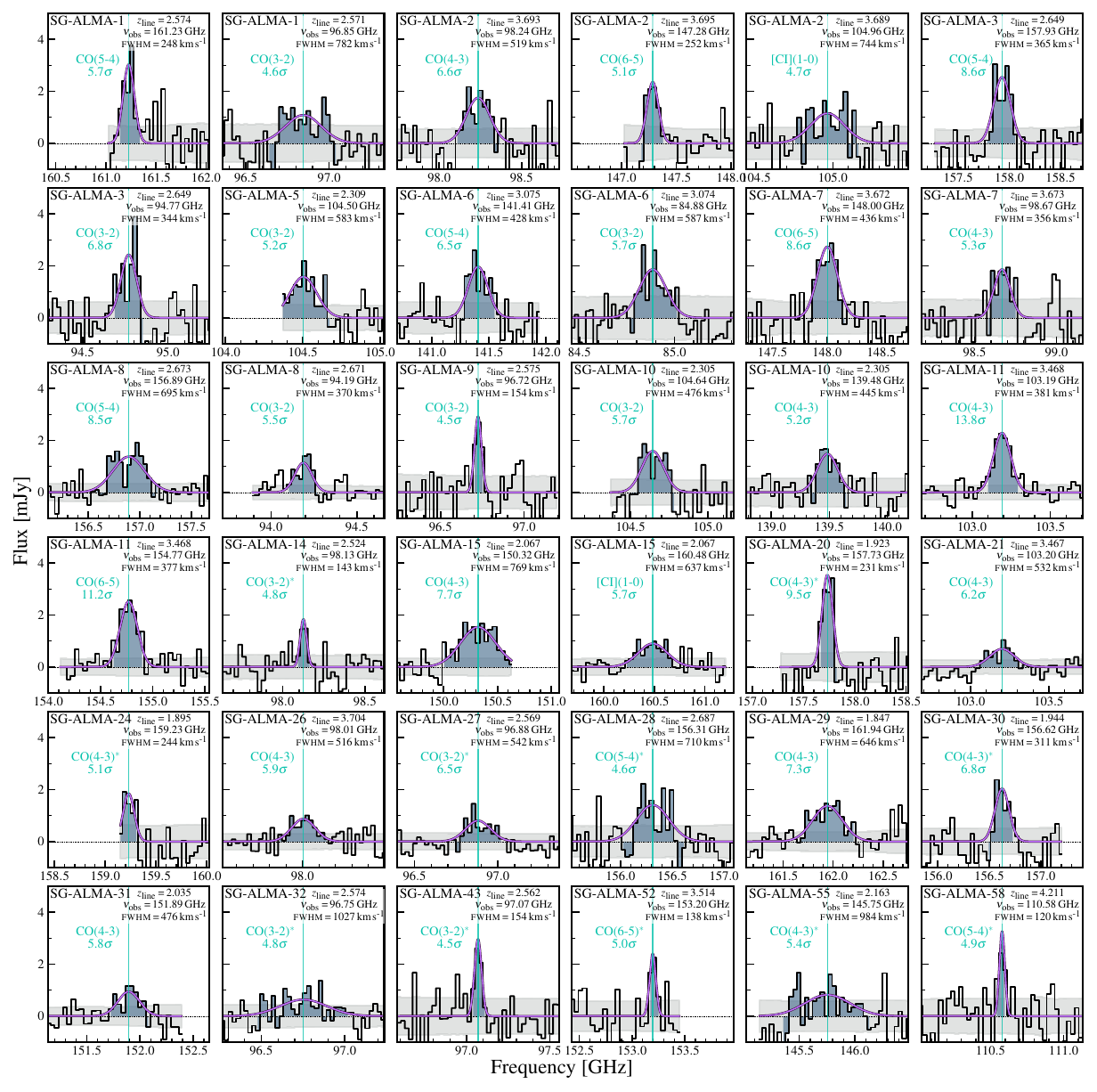}
    \caption{All confirmed $>4.5\sigma$ line detections (source names given in upper left corner of each panel). We show the spectra (black line) and error per channel (gray shaded region) binned to a resolution of 80~km\,s$^{-1}$. The Gaussian fit to the line is shown (purple curve) with channels within $\pm2\sigma$ shaded blue for clarity. We mark the adopted frequency of the line (cyan line) and label the transition and S/N value. Line transitions marked with * are tentative, determined using photo-z(s) and the ALMA line. In the figure legend, we label the redshift, $z_{\rm line}$, determined from the line frequency; the observed line frequency, $\nu_{\rm obs}$; and the FWHM of the line (obtained from the Gaussian fit). }
    \label{fig:secure_lines}
\end{figure*}

\section{Spectroscopic Redshifts}
\label{sec:speczs}

\subsection{Redshift identification}
\label{sec:redshift_ids}

In all, we find $\geq4.5\sigma$ peaks in 27 DSFGs. 
We use these detections to identify redshifts for the sample in several quality tiers, taking into account the preexisting photometric and spectroscopic data.
For nine sources, we determine a secure redshift using two or more $>4.5\sigma$ lines (four of these redshifts were reported in \citealt{cowie23} from the same dataset, and three have other literature spec-zs). For six additional sources, we confirm a previous spec-z from the literature with at least one $>4.5\sigma$ millimeter line (several have additional fainter lines as well). For SG-ALMA-9, we revise the redshift to $z_{\rm spec}=2.575$ rather than $z_{\rm spec}=2.322$, as originally reported by \citet[][; the incorrect redshift was based on a second tentative line in the ALMA data identified using a different detection method, which has since been confirmed to be spurious]{cowie23}. Our updated redshift is in agreement with a previous VLT/FORS2 redshift from \citet{vanzella08}.

{Next, for 11 sources with a single $>4.5\sigma$ ALMA line but no confirmed spec-z, we use the grism-zs from \citetalias{momcheva16} and the photo-zs from \citetalias{straatman16}, \citetalias{eisenstein26}, and our own {\tt MAGPHYS} fits (Section~\ref{sec:lit_redshifts}) to assign the most probable solution out of the possible line identifications, taking into account solutions ruled out by non-detections in the ALMA data. We consider these spec-zs tentative, but we weight them higher if multiple photo-z estimates are consistent with a single solution within the uncertainties. To test these spec-z assignments, we perform the same procedure on the sources with secure CO lines and reliable redshifts (15 sources), taking just the brightest CO detection and the various photo-z estimates. We find that we would have correctly identified the true spec-z in 12 out of 15 cases, suggesting that $\sim$80\% of our tentative spec-zs are correct.}
Several of these tentative redshifts fall between $1.8 < z < 2.0$, where we expect to detect only one line given our spectral window setups.

{In total, we have identified redshifts based on or in agreement with $>4.5\sigma$ line(s) for 26 sources. We note that in four sources, we find additional spurious lines at $\gtrsim4.5\sigma$, i.e., they do not correspond to any known FIR emission line, given the known spec-z that was verified by another method or from brighter lines. This number is higher than expected based on our false positive estimates}; however, we note that the spurious lines typically appear as narrow spikes near the edges of spectral windows. They may be due to poor bandwidth calibration at the edges, or to incomplete continuum subtraction. 

{For SG-ALMA-13 and SG-ALMA-17, where we detect no $>4.5\sigma$ lines, we were able to detect a lower significance line using a previously confirmed spec-z. Additionally, SG-ALMA-4 has two different published spec-zs: $z_{\rm spec}=2.252$ (VLT/ISAAC H$\alpha$ detection; \citealp{casey11}) and $z_{\rm spec}=1.965$ (VLT/KMOS H$\alpha$ detection; \citealp{birkin24}). Our ALMA scans did not detect CO(3--2) at 106.33~GHz or CO(4--3) at 141.77~GHz, which we would expect to see if the source was at $z=2.252$. However, there is a single $4.4\sigma$ peak at 155.55~GHz (just below our secure detection threshold), corresponding to CO(4--3) at $z=1.965$, in line with the \citet{birkin24} redshift. This suggests the $z=2.252$ H$\alpha$ line may be spurious, and we adopt $z_{\rm spec}=1.965$ for this source.}

In Figure~\ref{fig:secure_lines}, we show the secure $>4.5\sigma$ emission lines obtained by our ALMA programs. The line redshift, observed frequency, FWHM, line identifications, and S/N are listed in each panel. In Table~\ref{tab:line_detections}, we list the redshifts either identified or confirmed by these lines, along with a flag denoting how secure the redshift identification is (4 = secure, two millimeter lines; 3\,=\,secure, confirmed previous spec-z; 2\,=\,tentative, agreement with multiple photo-zs; 1\,=\,tentative, agreement with single photo-z). We also give the line identifications, observed frequencies, S/N, and line redshifts in the table. For completeness, we separately list additional $>4.0\sigma$ lines that are confirmed via an independent spec-z.

\begin{deluxetable}{ccccccc}
\tabletypesize{\scriptsize}
\tablecaption{\label{tab:line_detections}
ALMA Emission Lines and Redshifts}

\tablehead{
\colhead{Source}&\colhead{$z_{\rm spec}$}&\colhead{Flag}&\colhead{Line}&\colhead{$\nu_{\rm obs}$}&\colhead{S/N}&\colhead{$z_{\rm line}$}\\
&&&&\colhead{(GHz)}&&}
\colnumbers

\startdata
\multicolumn{6}{c}{Secure emission lines ($>4.5\sigma$)}\\
SG-ALMA-1&2.574&4&CO(5--4)&161.228&5.69&2.574\\
&&&CO(3--2)&96.846&4.56&2.571\\
SG-ALMA-2&3.693&4&CO(4--3)&98.236&6.58&3.693\\
&&&CO(6--5)&147.284&5.07&3.695\\
&&&[C \sc{i}](1--0)&104.964&4.75&3.689\\
SG-ALMA-3&2.649&4&CO(5--4)&157.929&8.64&2.649\\
&&&CO(3--2)&94.770&6.76&2.649\\
SG-ALMA-5&2.309&3&CO(3--2)&104.503&5.21&2.309\\
SG-ALMA-6&3.075&4&CO(5--4)&141.406&6.45&3.075\\
&&&CO(3--2)&84.885&5.72&3.074\\
SG-ALMA-7&3.672&4&CO(6--5)&148.003&8.62&3.672\\
&&&CO(4--3)&98.667&5.28&3.673\\
SG-ALMA-8&2.673&4&CO(5--4)&156.893&8.48&2.673\\
&&&CO(3--2)&94.189&5.48&2.671\\
SG-ALMA-9&2.575&3&CO(3--2)&96.724&4.54&2.575\\
SG-ALMA-10&2.304&4&CO(3--2)&104.643&5.70&2.305\\
&&&CO(4--3)&139.483&5.19&2.305\\
SG-ALMA-11&3.468&4&CO(4--3)&103.191&13.82&3.468\\
&&&CO(6--5)&154.770&11.22&3.468\\
SG-ALMA-14&2.524&2&CO(3--2)&98.126&4.84&2.524\\
SG-ALMA-15&2.067&4&CO(4--3)&150.319&7.71&2.067\\
&&&[C \sc{i}](1--0)&160.484&5.71&2.067\\
SG-ALMA-20&1.923&2&CO(4--3)&157.733&9.48&1.923\\
SG-ALMA-21&3.467&3&CO(4--3)&103.202&6.17&3.467\\
SG-ALMA-24&1.895&2&CO(4--3)&159.231&5.06&1.895\\
SG-ALMA-26&3.704&3&CO(4--3)&98.005&5.92&3.704\\
SG-ALMA-27&2.569&2&CO(3--2)&96.881&6.54&2.569\\
SG-ALMA-28&2.688&1&CO(5--4)&156.311&4.58&2.687\\
SG-ALMA-29&1.847&3&CO(4--3)&161.943&7.26&1.847\\
SG-ALMA-30&1.944&2&CO(4--3)&156.624&6.82&1.944\\
SG-ALMA-31&2.035&3&CO(4--3)&151.894&5.84&2.035\\
SG-ALMA-32&2.574&2&CO(3--2)&96.752&4.79&2.574\\
SG-ALMA-43&2.562&2&CO(3--2)&97.069&4.53&2.562\\
SG-ALMA-52&3.514&1&CO(6--5)&153.195&5.02&3.514\\
SG-ALMA-55&2.163&1&CO(4--3)&145.754&5.41&2.163\\
SG-ALMA-58&4.211&2&CO(5--4)&110.579&4.94&4.211\\
&&&&&\\
\multicolumn{6}{c}{Additional emission lines ($>4.0\sigma$)}\\
SG-ALMA-2&3.693&4&H$_2$O(2$_{11}$--2$_{02}$)&160.243&4.15&3.693\\
SG-ALMA-4&1.965&3&CO(4--3)&155.525&4.40&1.964\\
SG-ALMA-5&2.309&3&CO(4--3)&139.360&4.14&2.308\\
SG-ALMA-10&2.304&4&[C \sc{i}](1--0)&149.004&4.08&2.303\\
SG-ALMA-13&4.426&3&CO(5--4)&106.215&4.42&4.425\\
SG-ALMA-17&3.581&3&CO(4--3)&100.635&4.40&3.581\\
\enddata
\tablecomments{Columns: (1) Source name, (2) Adopted spec-z, (3) Flag denoting $z_{\rm spec}$ quality (4 = secure, two millimeter lines; 3\,=\,secure, confirmed previous spec-z; 2\,=\,tentative, agreement with multiple photo-zs; 1\,=\,tentative, agreement with single photo-z), (4) Name of emission line, (5) Observed frequency, (6) Signal-to-noise ratio, (7) Redshift of the individual line.\\
}
\end{deluxetable}

\subsection{Spectroscopic completeness}
\label{sec:specz_completeness}

\begin{figure}
    \centering
    \includegraphics[width=\linewidth]{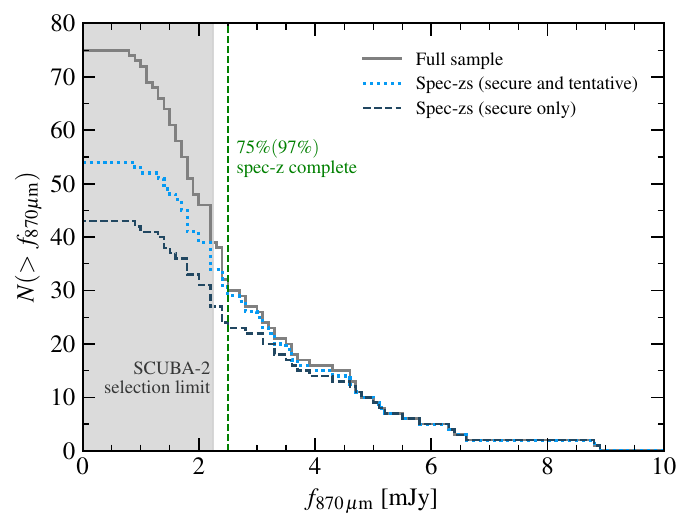}
    \caption{Cumulative number of ALMA sources above a given $\falma$ versus $\falma$. We show the full sample (gray solid line), sources with secure or tentative spec-zs (light blue dotted line), and sources with secure spec-zs (dark blue dashed line). Above $\falma=2.5$~mJy (green dashed line), 75\% (97\%) of the sources have secure (tentative) spec-zs. {Our ALMA survey is nearly flux-complete above $\falma=2.25$~mJy, the 4$\sigma$ limit of the original SCUBA-2 850~$\mu$m survey; we shade the region below this limit gray to illustrate where our sample begins to suffer from lower completeness.} }
    \label{fig:cdf_f870}
\end{figure}

We combine the ALMA spec-zs with the additional literature spec-zs described in Section~\ref{sec:lit_redshifts}.
Assuming the tentative spec-z identifications are correct, {54 out of the 75 DSFGs (72\%)} are now spectroscopically confirmed. This fraction is 57\% counting only secure spec-zs. We discuss the sources that are not confirmed in Section~\ref{sec:missing_sources}.
For now, we note that the brighter ALMA sources have much higher spectroscopic completeness. In Figure~\ref{fig:cdf_f870}, we illustrate this by plotting the cumulative number of sources above a given $\falma$ versus $\falma$. We show the entire sample (gray solid line), secure or tentative spec-zs (light blue dotted line), and just secure spec-zs (dark blue dashed line). Above $\falma=2.5$~mJy, the spectroscopic completeness reaches 75\% (97\%) for secure (tentative) spec-zs; and above $\falma=3$~mJy, 81\% have secure spec-zs. 

{We emphasize that due to the depth of the SCUBA-2 850~$\mu$m selection, the ALMA 870~$\mu$m survey is nearly flux-complete down to $\falma=2.25$~mJy.} 
In other words, we are close to achieving a complete spectroscopic redshift distribution for DSFGs down to the SCUBA-2 850~$\mu$m single-dish confusion limit; i.e., the ``classical'' SMG population. This is a remarkable success given that past spec-z fractions were typically in the range $\sim$5--40\% for DSFG samples \citep[][]{danielson17, dudzeviciute20, birkin21, chen22, mckinney25}, and the only flux- and spectroscopically complete DSFG surveys to date are of bright lensed sources \citep[][]{reuter20} or very faint sources \citep[][]{aravena20}, which are both highly subject to cosmic variance. 

As an aside, we observe that a number of our spec-zs align with known overdensities in the GOODS-S. 
DSFGs are known to form in protocluster environments, both from observations \citep[e.g.,][]{chapman01, chapman09, tamura09, casey16, nicandro_rosenthal25} and from simulations \citep[][]{narayanan15, kumar26}. However, lacking spectroscopic redshifts, it has been difficult to ascertain whether most DSFGs are associated with dense environments. {Given our spec-zs, at least 29 galaxies, or 54\% of the total spec-zs}, appear to be associated with known overdensities at $z\approx1.61$, $z\approx2.30$, $z\approx2.55$, $z\approx2.69$, $z\approx3.47$, and $z\approx3.70$ \citep[][]{castellano07, guaita20, shah24}. {Moreover, many of them are clearly located close to ($\lesssim$5~cMpc) a high-density peak of the respective structure, when such a peak is reported in the literature.} Our total fraction associated with overdensities is at the high end of the (wide) range ($\sim30$--60\%) given by other observational studies \citep[e.g.,][]{smolcic17, alvarez_crespo21} {and suggests that the majority of DSFGs might be found in such environments. As a caveat, however, we note that confirming protocluster membership is complex and several of the reported structures are multi-peaked or diffuse, so we cannot be certain that all the sources are in fact associated. On the other hand, the above list of overdensities is incomplete, so there may be additional protocluster members in our sample.}

\section{Photometric Redshifts}
\label{sec:photozs}

We next estimate the best redshifts for the remainder of our sample using several photo-z fitting codes, taking advantage of the high fraction of spec-zs in our sample to test the performance of each method. We first describe the various methods we use to estimate photo-zs, and then we discuss their performance against the spec-zs.

\subsection{Photo-z methods}

The photo-z estimates that we compare are as follows:

 \textbf{EAZY}: As described in Section~\ref{sec:lit_redshifts}, most of our sample have {\tt EAZY} redshifts published in the \citetalias{straatman16} and/or \citetalias{eisenstein26} catalogs. {\tt EAZY} constructs a redshift grid using a set of pre-defined templates and then performs a simple $\chi^2$-minimization routine to determine the best-fitting redshift. 
 
It is important to note that the different sets of {\tt EAZY} photo-zs considered here used different template sets and photometric bands (see Section~\ref{sec:lit_redshifts}). The complementary datasets used by these studies make it possible to see how well deep optical/NIR legacy surveys constrain redshifts for the full submillimeter-selected DSFG population.

\textbf{FIR redshifts}: FIR color-based or dust emission template-based photo-z methods are particularly attractive for submillimeter samples, because they do not require ancillary optical/NIR counterpart matching and can therefore be easily applied to wide single-dish surveys.

Here we test two FIR photo-z methods: The first is the {\tt MMPZ} code (\citealt{casey20}), which takes into account the shape of the dust SED and the total luminosity, using an empirical relation between the rest-frame peak wavelength, $\lambda_{\rm peak}$, and integrated FIR luminosity, $L_{\rm IR}$, to attempt to break the redshift-temperature degeneracy.
The second is the {\tt mbb} package \citep{mckay_mbb}, a simple Bayesian modified-blackbody (MBB) fitting code that can fit the redshift along with the other dust parameters. The package supports a variety of dust emission models and accompanying parameter sets, although here we use it mainly as a single-template fitter.

Fitting is performed across the excellent photometric coverage from 24~$\mu$m to 3~mm {(Section~\ref{sec:sample})}, adding a 10\% relative flux calibration error in quadrature to the error for each band.
{\tt MMPZ} does not require any user-specified parameters aside from the input photometry. Since the uncertainties on the Herschel/SPIRE flux deblending has been known to bias the fits, we increase the errors (which already include confusion noise) by a factor of 50\%, finding slightly better accuracy compared to the spec-zs with this factor included.
For {\tt mbb}, we use an optically thin MBB with fixed $T=35$~K and $\beta=1.8$, and we include a MIR power law joined where the blackbody slope is equal to the power-law slope, $\alpha=2.0$ \citep[e.g.,][]{drew22}.
We find that the results with uniform priors on $z$ are comparable to those with more informative priors (e.g., Gaussian); therefore, for simplicity, we adopt the uniform $z$ priors, and we adopt log-uniform priors on $L_{\rm IR}$.

\textbf{MAGPHYS}: Lastly, we perform fits to the full optical to millimeter photometry using the {\tt MAGPHYS} code with the {\tt photo-z} extension \citep{dacunha08, dacunha15, battisti+19}. This code is unique among energy-balance SED codes in that it is specifically designed to include the FIR to radio data along with the optical/NIR data in estimating $z_{\rm phot}$, and it was designed with a particular focus on estimating DSFG properties. {\tt MAGPHYS} does not require user-specified parameters; instead, it marginalizes over a large library of stellar and dust emission models using a Bayesian framework, enforcing energy balance between the dust attenuation and emission. It has been used to derive photo-zs for most of the other large, unbiased ALMA 870~$\mu$m surveys of DSFGs \citep[e.g.,][]{dacunha15, dudzeviciute20, simpson20}.

For the {\tt MAGPHYS} fits, we use the available FIR to millimeter data (including the increased error on the
Herschel/SPIRE bands),
as well as the HST and JWST fluxes from JADES DR5, where available.
For the one source not covered by the JWST mosaics, we use the CANDELS photometry \citep{guo13} instead. We also take the Spitzer/IRAC 3.6--8.0~$\mu$m photometry from CANDELS.

In Table~\ref{tab:redshifts}, we give the best-fit photo-zs and errors that we obtained with each method, along with the spec-zs from our ALMA data and the literature surveys (Section~\ref{sec:lit_redshifts}). For convenience, we also list the ALMA 870~$\mu$m fluxes and FIR luminosities, which we integrated from an additional {\tt mbb} fit performed with the redshift fixed to the best available estimate.

\begin{figure*}
    \centering
    \includegraphics[width=0.98\linewidth]{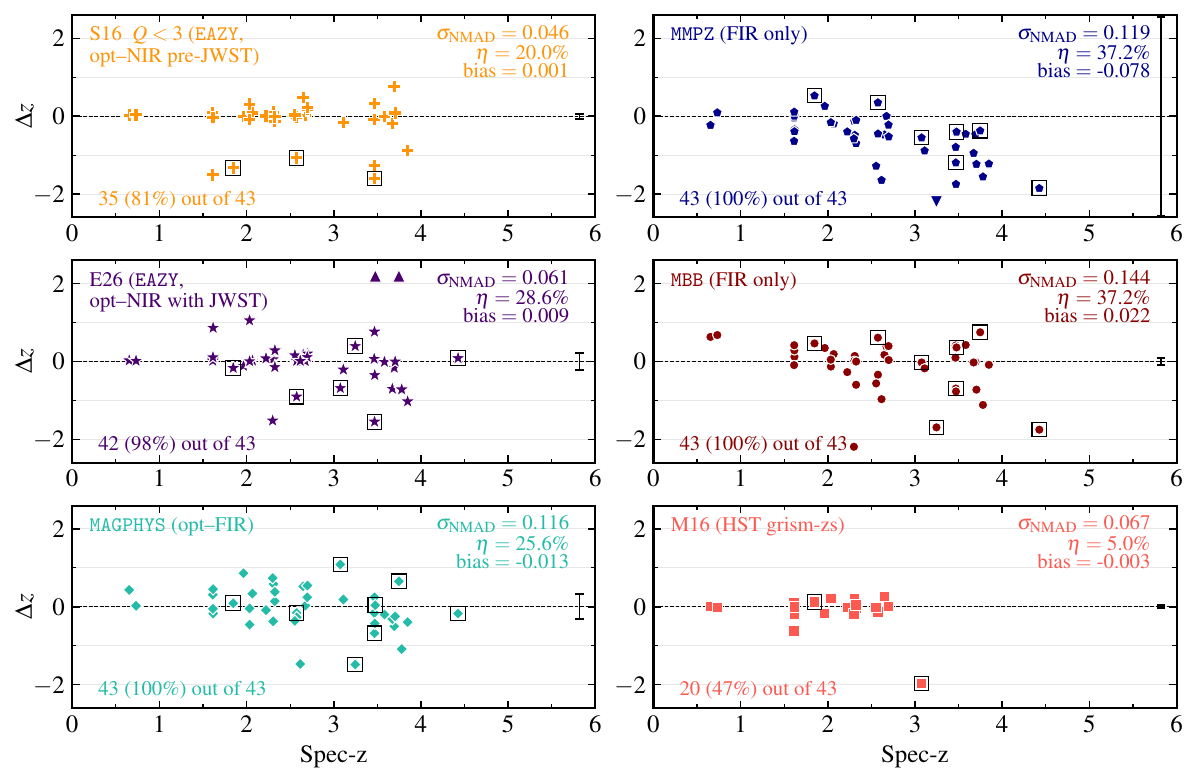}
    \caption{Difference between photo-zs/grism-zs and spec-zs ($\Delta z \equiv z_{\rm phot} - z_{\rm spec}$) vs. $z_{\rm spec}$ for sources with both to quantify the effectiveness of various photo-z estimation methods. Each panel indicates the {photo-z method, data, and the number of sources covered by that method (out of the 43 with secure spec-zs)}; as well as the normalized median absolute deviation, $\sigma_{\rm NMAD}$; the outlier fraction, $\eta$; and the bias for that method and subsample. Tentative spec-zs that we inferred using photo-zs are not included. Objects which lie above or below the y-axis ranges are denoted by upward- or downward-facing triangles, respectively.  Sources identified as either blends or $K_s$-dark in \citetalias{cowie18} are denoted by black open squares. The median error on the photo-zs is shown on the right-hand side of each panel, centered at zero (black error bars). 
    }
    \label{fig:photz_comparison}
\end{figure*}

\subsection{Comparison with spec-zs}
\label{sec:photz_comp}

Here we compare the results from each of the above photo-z codes to the available spec-zs. The latter do not include tentative spec-zs that we inferred using photo-zs, to avoid biasing the results. For completeness, we include the \citetalias{momcheva16} HST grism-zs as one of the comparison samples.

In Figure~\ref{fig:photz_comparison}, we plot $\Delta z \equiv z_{\rm phot} - z_{\rm spec}$ vs. $z_{\rm spec}$ for each of the photo-z methods, labeling the code and the wavelength range used. We mark with black open squares the sources previously identified in \citetalias{cowie18} as either blended with foreground objects or $K_s$-dark, since these are objects where we would expect optical/NIR-based photo-zs to perform poorly. We plot the median reported error on $z_{\rm phot}$ at the far right in each panel (black error bars).
    
To quantify the performance of each method, we measure the precision (scatter) by computing the normalized median absolute deviation (NMAD), defined as:
    
    \begin{equation}\label{eq:nmad}
        \sigma_{\rm NMAD} \equiv 1.483 \times \textnormal{median} \left(\left|\frac{\Delta z - {\rm median}(\Delta z)}{1 + z_{\rm spec}}\right|\right).
    \end{equation}
    
\noindent We also compute the outlier fraction, $\eta$, which we define as the fraction of sources with $|\Delta z|/(1 + z_{\rm spec}) > 0.15$, and the bias, which we define as the median $\Delta z/(1+z_{\rm spec})$ for sources that are not classified as outliers. We report $\sigma_{\rm NMAD}$, $\eta$, and the bias computed for each subsample in the respective panels of Figure~\ref{fig:photz_comparison} for reference.

We can draw several broad conclusions from Figure~\ref{fig:photz_comparison}. First, all of the photo-z methods have outlier fractions of at least $\eta > 20\%$ and scatter of $\sigma_{\rm NMAD} \gtrsim 0.05$. {Comparing these to the overall results for the JADES catalog, which had an outlier fraction of $\eta = 5$\% and a $\sigma_{\rm NMAD} = 0.024$ \citep[][]{rieke23}, we see---unsurprisingly---that these dusty galaxies present a unique challenge to photo-z codes, even with the best available data.}

Furthermore, we see that both sets of {\tt EAZY} photo-zs perform better overall than either {\tt MAGPHYS} or the FIR methods, although they still have significant scatter and outlier fractions. They do not perform well for the blends or $K_s$-dark sources, generally underpredicting their redshifts (for blends) or overpredicting them (for $K_s$-dark sources). 

{The {\tt MAGPHYS} photo-zs have roughly twice as much scatter ($\sigma_{\rm NMAD} = 0.116$) as the {\tt EAZY} photo-zs, despite including the JWST and FIR to millimeter data.  These results are fairly consistent with past results for DSFG samples, with similar statistics to those obtained by \citet{dudzeviciute20} for the subset of 44 AS2UDS DSFGs with spec-zs.
As a test, since many photo-z estimates are performed without any FIR data, we also check the effect of running {\tt MAGPHYS} without any of the photometry at $\lambda_{\rm obs} \geq 24\,\mu$m. In this case, we find that the outlier fraction increases from 26\% to $>$40\%.} In other words, the inclusion of the FIR data reduces the overall outlier fraction.

The FIR photo-zs with {\tt MMPZ} and {\tt mbb} perform worse ($\sigma_{\rm NMAD}\approx0.12$--0.14 and $\eta \approx 37$\%) than the optical/NIR photo-zs. This is not very surprising given the intrinsic variation in the dust SED and the smaller number of bands to fit compared to the optical/NIR. The overall scatter in the {\tt MMPZ} results is slightly lower, but they suffer from a strong bias ($-0.08$) that is clearly correlated with $z$, as the code tends to favor solutions close to $z\sim 2.0$--2.5. By contrast, the {\tt mbb} fits are less biased but have the highest scatter of any of the methods tested. While the error bars from {\tt MMPZ} are extremely wide, reflecting the large range of dust models employed, the errors on the {\tt mbb} fits are significantly underestimated, since we fit a single isothermal template.

Finally, the \citetalias{momcheva16} grism-zs {have a scatter fairly comparable to that of the best photo-zs. They achieve a low bias ($-0.003$) and outlier fraction (5\%).} Nevertheless, they are not entirely free from outliers and fail to access high redshifts ($z>3$) due to selection and sensitivity issues. 

{
For a more careful assessment of the photo-z methods in order to select the best redshifts for DSFGs lacking spec-zs, we compare the performance of the codes on several preselected subsamples. For the 35 spectroscopically confirmed objects whose photo-zs were assigned quality flag $Q<3$ (high quality) in the \citetalias{straatman16} ZFOURGE catalog, we find that the \citetalias{straatman16} {\tt EAZY} photo-zs achieve the lowest scatter and bias, followed by the \citetalias{eisenstein26} {\tt EAZY} photo-zs and then our {\tt MAGPHYS} results. Perhaps surprisingly, for the 42 sources with both spec-zs and \citetalias{eisenstein26} {\tt EAZY} photo-zs {\it regardless} of ZFOURGE quality flag, the ordering is the same; i.e., even with poor quality photo-zs included\footnote{For all six of the sources with spec-zs and \citetalias{straatman16} $Q>3$ photo-zs, the photo-zs are classified as outliers, including two optically dark sources with $z_{\rm phot} > 7$.}, the \citetalias{straatman16} results without JWST data outperform the \citetalias{eisenstein26} results based on HST and JWST alone. 
However, for the subset of sources with $K_s > 23$~mag, it is instead the {\tt MAGPHYS} photo-zs which perform best, followed by \citetalias{eisenstein26} and then \citetalias{straatman16}. This is likely because the combination of both the JWST data and the FIR data are helping to constrain the redshift solution for sources that are more optically faint. }

{To summarize, for objects with $K_s < 23$ and which have $Q\leq3$ in the \citetalias{straatman16} catalog, we find that the \citetalias{straatman16} {\tt EAZY} redshifts are the most accurate. For the remainder, {\tt MAGPHYS} appears to perform best. We reiterate, however,} that no photo-z method tested here is able to reduce the outlier fraction for DSFGs below 20\%, even with high-quality JWST data and FIR data included. This emphasizes the necessity of deep spectroscopy for future large DSFG surveys.

\section{Discussion}
\label{sec:discussion}

\subsection{Redshift distribution}

As we have discussed, our sample's redshift distribution is largely spectroscopically complete, especially at higher fluxes. Since all of the DSFGs have robust counterparts and our survey is nearly flux-complete at $\falma \geq 2.25$~mJy, we are able to map the redshift distribution of DSFGs down to nearly the confusion limit for a single-dish 15-m telescope at 850~$\mu$m \citep[e.g.,][]{cowie17}.

We construct the redshift distribution of our sample using photo-zs for the 21 DSFGs that do not have spec-zs. {For these, we adopt as the best available redshift the ZFOURGE photo-z from \citetalias{straatman16}, if it has quality flag $Q<3$ and if the source has $K_s \leq 23$~mag (11 sources). For sources without a high-quality ZFOURGE photo-z or with $K_s > 23$~mag, we instead adopt our {\tt MAGPHYS} photo-z (10 sources). This ordering is based on the relative performance of the methods on the subsets tested in Section~\ref{sec:photz_comp}.}

\begin{figure}[t]
    \centering
    \includegraphics[width=\linewidth]{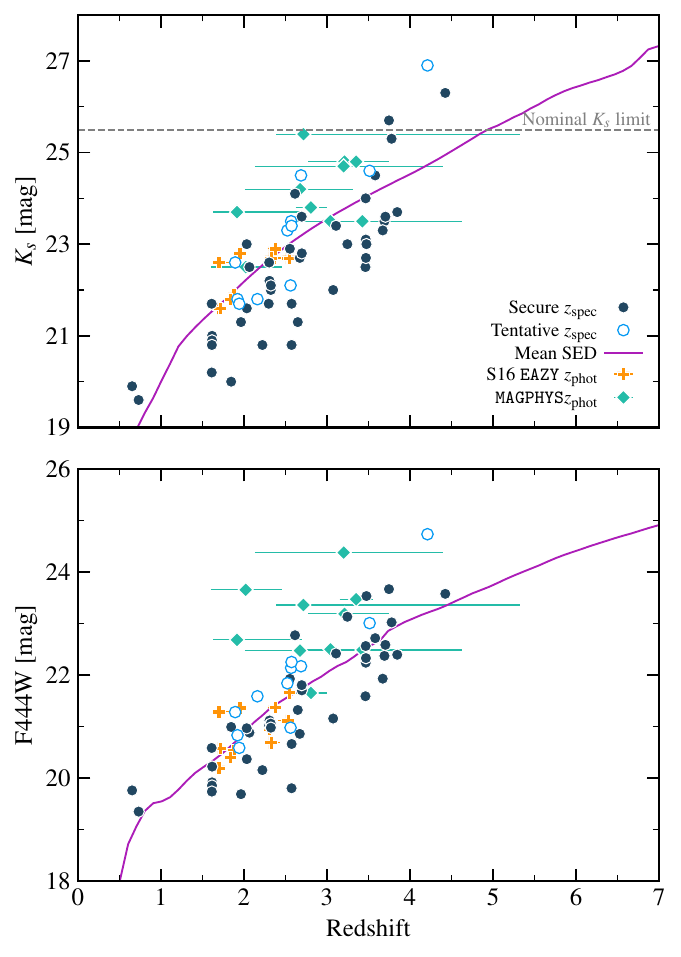}
    \caption{$K_s$ (top) and F444W (bottom) magnitudes vs. best redshift. We show secure spec-zs (dark blue points), tentative spec-zs (open light blue circles), and photo-zs from \citetalias{straatman16} (orange crosses) and our {\tt MAGPHYS} fits (teal diamonds). The errors on the photo-zs are the 68\% credible interval. We plot the nominal $K_s$-band flux limit in the top panel (gray dotted line). We also plot the expected track for a source with fixed luminosity (purple line), using our mean {\tt MAGPHYS} SED normalized to our sample's median F444W magnitude and median redshift. }
    \label{fig:Ks_f444_z}
\end{figure}

In Figure~\ref{fig:Ks_f444_z}, we plot the $K_s$ (top) and F444W (bottom) magnitudes versus the best redshift estimates.
This figure confirms that there is a correlation between the NIR brightness and redshift, as shown previously for DSFGs (e.g., \citealt{danielson17}; \citetalias{cowie18}; \citealt{dudzeviciute20, mckay25}). The relation is reasonably tight, with a $\pm$1~mag spread at a given redshift for F444W and a slightly larger dispersion for $K_s$, considering the spec-zs alone.

For comparison, we plot the expected track for our sample's mean {\tt MAGPHYS} SED with fixed luminosity normalized to our median F444W magnitude and redshift. This track agrees fairly well with the observed fluxes from $1\lesssim z \lesssim 4$, suggesting that the evolution of the NIR flux with redshift can largely be accounted for by $K$-correction and distance effects, with the scatter reflecting the intrinsic spread in stellar mass and dust content for DSFGs. {However, there is a slight skew between the expected track and the observed spec-zs for both $K_s$ and F444W, with low-redshift sources brighter and high-redshift sources fainter than expected, on average. Without a larger spec-z sample, we cannot draw strong conclusions, but it appears that there is some evolution in the NIR SED that is not captured in the {\tt MAGPHYS} assumptions, possibly driven by a change in the average stellar mass and/or dust content with redshift.}

\begin{figure}[t!]
    \centering
    \includegraphics[width=\linewidth]{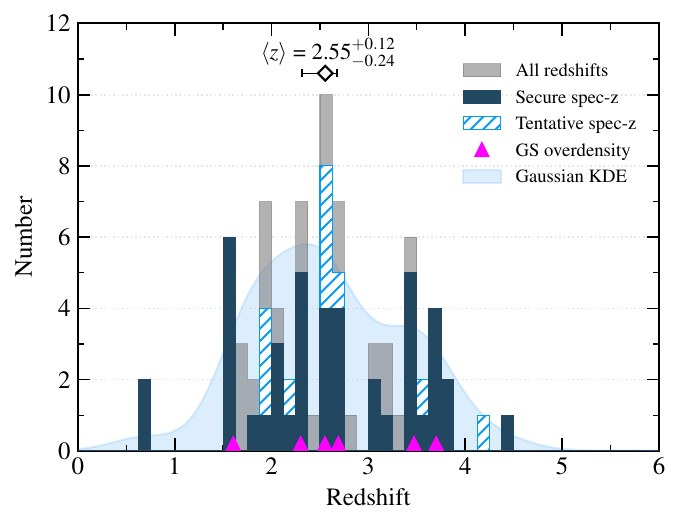}
    \caption{Redshift distribution of our DSFG sample, including secure (dark blue) and tentative (hatched light blue) spec-zs 
    and photo-zs (light gray). We plot the median redshift and bootstrapped error for the full sample above the histogram (open diamond).  {We also plot the Gaussian kernel density estimate (KDE) of the distribution (light blue shaded curve).  We mark several known overdensities in the GOODS-S \citep[purple triangles;][]{castellano07, guaita20, shah24}, which align with spikes in our redshift distribution. 
    }
    }
    \label{fig:zhist}
\end{figure}

In Figure~\ref{fig:zhist}, we show the combined distribution of redshifts for our DSFG sample. 
The median redshift is $\langle z \rangle = 2.55^{+0.12}_{-0.24}$ (open diamond), with the errors obtained by bootstrapping the sample 1000 times. The 16th--84th percentile range of the sample is $1.85 <z<3.47$. The median remains largely unchanged ($\langle z \rangle = 2.57^{+0.12}_{-0.27}$) if we only consider the subset where our survey is nearly flux-complete ($\falma\geq2.25$~mJy). It also does not change substantially if we only consider sources with spec-zs.

As in other studies \citep[e.g.,][]{yun12, simpson14}, we model the redshifts as a log-normal distribution,
    \begin{equation}
        P(z) = \frac{1}{(1+z) \sigma \sqrt{2 \pi}} e^{-[\ln(1+z) - \ln(1+z_{\mu})]^2 /2\sigma^2},
    \end{equation}
\noindent where $\ln(1+z_\mu)$ and $\sigma$ represent the mean and standard deviation of the normally distributed variable $\ln (1 + z)$. This distribution captures the asymmetric shape of the DSFG redshift distribution, with a steep fall-off below $z\lesssim1$ and a tail out to $z\gtrsim5$.

We fit the model to the redshift distribution of just the $\falma\geq2.25$~mJy subset, using a Markov chain Monte Carlo (MCMC) algorithm to sample from the log-normal distribution 15000 times, assuming log-uniform priors on $z_\mu$ and uniform priors on $\sigma$. By limiting to this subset, we avoid major biases due to incompleteness, at the cost of a smaller sample size (44 galaxies instead of 75). We note that using a normal distribution instead of the log-normal one also gives a good fit to our data. Using leave-one-out cross-validation, we confirm that the log-normal is slightly preferred, but the difference is not statistically significant.

\begin{figure}[t!]
    \centering
    \includegraphics[width=\linewidth]{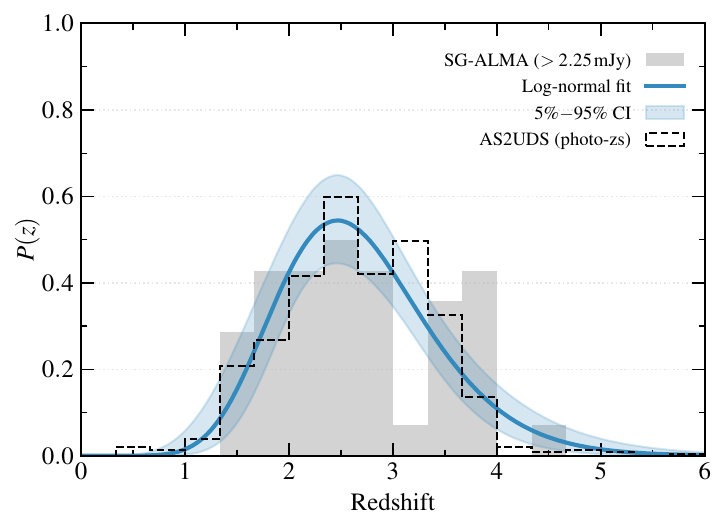}
    \caption{Redshift distribution (scaled to unit probability) of the $\falma \geq 2.25$~mJy subset (gray histogram) and best-fit log-normal distribution with 5\%--95\% credible interval (blue curve and shaded region). For comparison, we show the distribution of redshifts (mainly {\tt MAGPHYS} photo-zs) for the 707 AS2UDS DSFGs \citep[dashed histogram;][]{dudzeviciute20}. }
    \label{fig:lognormal}
\end{figure}

In Figure~\ref{fig:lognormal}, we show the median and 5\%--95\% posterior range of the log-normal fit. The median (and 68\% credible interval) of the individual parameter posteriors are $z_\mu=2.61^{+0.12}_{-0.11}$ and $\sigma=0.20^{+0.03}_{-0.02}$. This is in good agreement with previous results for AzTEC 1.1~mm galaxies ($z_\mu=2.6$; \citealt{yun12}) and for ALESS ($z_\mu \sim 2.5$; \citealt{simpson14}). 

For comparison, we plot the redshift distribution for the 707 AS2UDS DSFGs \citep[][]{dudzeviciute20}. {Since we cover a smaller sky area than AS2UDS, our sample is somewhat more prone to cosmic variance; however, it is deeper and much more spectroscopically complete (the published spec-z fraction for AS2UDS is $\sim$6\%; \citealt{dudzeviciute20}). It is therefore useful to compare the two distributions.}

We see that our best-fit log-normal is in good agreement with the distribution of the {\tt MAGPHYS} photo-zs for AS2UDS.      
This agreement is encouraging, since it implies that large photo-z-based studies may not suffer from substantial biases due to incorrect photo-zs---despite the fact that we find individual photo-z estimates result in catastrophic outliers in $>$20\% of DSFGs regardless of method.
This also provides valuable confirmation for simulations that attempt to match the DSFG redshift distribution, since these typically benchmark their models against the photo-z distributions for ALMA samples {(e.g., \citealt{elliott25} and \citealt{kumar25} test predictions from GALFORM and FLAMINGO, respectively, against the AS2UDS redshift distribution).}
{For smaller samples or less accurate photo-zs, however, the bias due to photo-zs may be significant. }

\subsection{Variation with submillimeter flux}

Past submillimeter studies have reported a mild evolution in average redshift with limiting 850~$\mu$m flux, with brighter sources lying at higher redshifts. Although initial trends for uniform ALMA samples (e.g., \citealt{simpson14}; \citetalias{cowie18}) were not statistically significant, recent studies have used either larger photo-z-based samples across a wider dynamic range in $\falma$ \citep[e.g.,][]{stach19, simpson20} or spectroscopically confirmed subsets of the parent samples \citep[e.g.,][]{danielson17, birkin21, chen22} to demonstrate a significant trend, with $z$--$\falma$ slopes typically between $0.06$--$0.09$ \citep[e.g.,][]{stach19, chen22}.

\begin{figure*}[t!]
    \centering
    \includegraphics[width=0.75\linewidth]{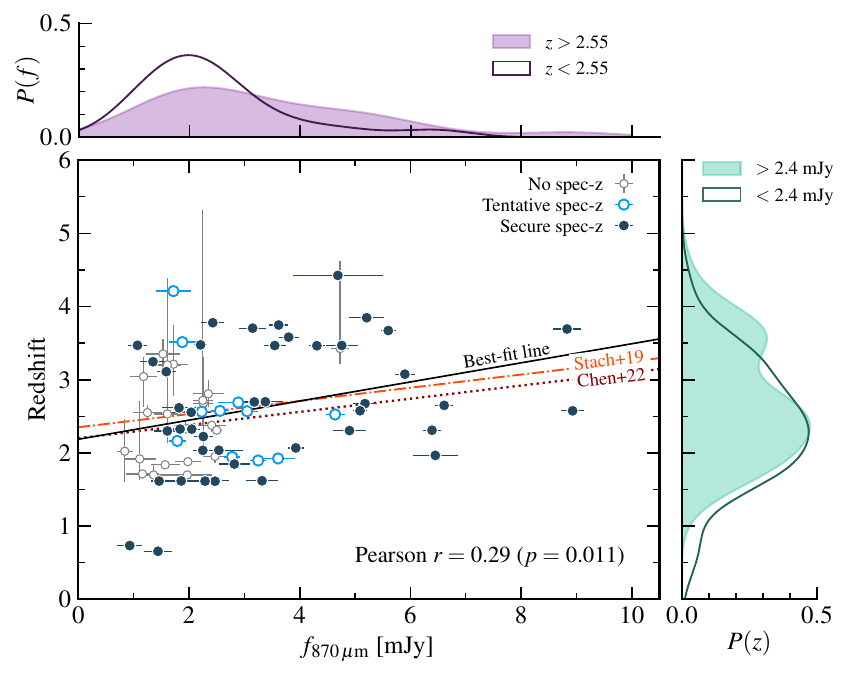}
    \caption{Center: Best redshift vs. $\falma$ for our DSFG sample. We plot photo-zs (gray), tentative spec-zs identified from a single line (light blue), and secure spec-zs (dark blue). {We find a statistically significant $z$--$\falma$ correlation (Pearson $r$ and $p$-value given in legend).} {Our best-fit line (black solid line) is steeper than but agrees with} the best-fit relations from the 708 AS2UDS DSFGs (\citealt{stach19}; orange dashed-dotted line) and the spectroscopic sample of \citet{chen22} (red dotted line) {within uncertainties.  } Top: Gaussian KDEs of the $\falma$ distribution, split above (filled purple) and below (dark purple line) the median redshift. Right: Gaussian KDEs of the redshift distribution, split above (filled green) and below (dark green line) the median $\falma$. {While at higher redshifts, DSFGs are distributed across a wide flux range, at lower redshifts, the population is weighted towards fainter submillimeter sources.} 
    }
    \label{fig:z_f870}
\end{figure*}

In Figure~\ref{fig:z_f870}, we plot the 870~$\mu$m fluxes of our sample vs. redshift (center panel).
We also show the $z$--$\falma$ relations from \citet{stach19}, representing a recent flux-complete but photo-z-based relation, and from \citet{chen22}, representing an incomplete but spectroscopically confirmed sample spanning $\falma = 2.5$--19~mJy.

{\citet{stach19} showed that for their sample of $\sim$700 DSFGs with photo-zs, a random subset of $\lesssim$100 galaxies would be unlikely to show a statistically significant trend. Our sample benefits from better depth and spectroscopic completeness but is limited by spatial coverage to a sample size below this threshold.
Still, we find a mild but statistically significant correlation of $z$ with $\falma$ using the Pearson correlation coefficient ($r=0.29$ with a $p$-value of 0.011). Our best-fit line is $z = (2.19 \pm 0.23) + (0.131 \pm 0.067)\times \falma$ (black line). This is slightly steeper than the literature relations, but agrees with them within the uncertainties. } 

In the top and right panels, we plot the distributions of $\falma$ split by median redshift and redshift split by median $\falma$, respectively. {A Komolgorov--Smirnov (K--S) test does not indicate any statistically significant difference in the redshift distributions above and below the median $\falma$ ($p$-value\,$=$\,0.16); however, there is a significant ($p$-value\,$=$\,0.019) difference between the $\falma$ distributions above and below the median redshift. While at higher redshifts, DSFGs are distributed across a wide flux range, at lower redshifts, the population is weighted towards fainter submillimeter sources.}

\subsection{Constraints on the high-redshift population}
\label{sec:z5fraction}

Next, we attempt to use our sample to constrain the fraction of 870~$\mu$m-selected DSFGs at high redshift ($z>4$--5). While it is known that a small subset of DSFGs populate the asymmetric high-redshift tail of the redshift distribution, the current constraints on the true prevalence and typical properties of dusty sources at $z>4$ are still uncertain (\citealp[e.g.,][]{gruppioni20, zavala21, algera23, sun25, barger26}). 
Furthermore, the observed fraction is dependent on the survey depth, selection wavelength, flux completeness, and quality of available photometric redshifts. 

We have only two spectroscopically confirmed $z\gtrsim4$ sources (one is tentative) and no confirmed $z>5$ sources in our sample. While several of the photo-z posterior distributions include $z>5$ solutions, no source has a best-fit photo-z at $z>5$. Based on the best available redshifts alone, we find a $z>4$ fraction of 2 out of 75 (2.7$^{+3.5}_{-1.7}$\%; errors are 1$\sigma$ Poisson errors; \citealt{gehrels86}) and a $z>5$ fraction of 0 out of 75 ($0^{+2.4}_{-0.0}$\%). 

As a further test, we can use the log-normal fit to the redshift distribution of our $\falma\geq2.25$ subset to make an estimate of the high-redshift fraction, simply by sampling the posterior and integrating the probability above each redshift threshold. This implies that 6.0\% of DSFGs lie at $z>4$ and 0.9\% lie at $z>5$, with 1$\sigma$ upper limits of 6.3\% at $z>4$ and 1.0\% at $z>5$. These upper limits fully account for the sampling of the log-normal and the full uncertainties on the fit parameters.

Comparing to the literature, our upper limits are {consistent with the fractions from AS2UDS (2.0$^{+0.6}_{-0.5}$\% at $z>4$ and 0.56$^{+0.45}_{-0.27}$\% at $z>5$; \citealp[][]{dudzeviciute20}) and from a combined ALMA sample in the A2744 field (4.9$^{+4.8}_{-2.7}$\% at $z>4$ and 0.0$^{+2.6}_{-0.0}$\% at $z>5$; \citealt{barger26}). Our estimates are slightly lower than the fraction from ALESS (16$^{+5.1}_{-4.0}$\% at $z>4$ and 2.0$^{+2.7}_{-1.3}$\% at $z>5$; \citealp[][]{dacunha15, danielson17}) and SCUBADive (9.7$^{+2.2}_{-1.8}$\% at $z>4$ and 2.1$^{+1.2}_{-0.8}$\% at $z>5$ using optical photo-zs; \citealp[][]{mckinney25}).
Note that these samples have brighter flux limits and thus lower completeness at the faint end of our sample's flux range. More importantly, the literature fractions are based largely on photo-zs, which may be particularly susceptible to outliers at the high redshift end. } 

To quantify the impact of photo-zs, we next ask whether the various photo-z estimates for our sample would have placed more objects at high redshifts if we did not have the spec-zs. 
Using the {\tt EAZY} fits from \citetalias{straatman16} (including photo-zs flagged as low quality, $Q>3$) and from \citetalias{eisenstein26}, or the {\tt MAGPHYS} fits (with or without the FIR data included), we would have {\it incorrectly} placed at least 3--6 sources at $z>4$ and 0--2~sources at $z>5$, depending on the method. 
The true redshifts for these sources tend to be at $3 < z < 4$; i.e., the redshifts are just moderately overestimated. Note that  {\tt MAGPHYS} fits that include the FIR photometry do not place any sources at $z>5$, and the two \citetalias{straatman16} fits with $z>5$ are flagged as low quality and therefore were known to be questionable {\it a priori}. By contrast, the {\tt MAGPHYS} fits without the FIR data and the \citetalias{eisenstein26} fits each place two sources at $z>5$, but they do not agree on which sources these are, reflecting the uncertainty in the fits.

Various explanations for the steep fall-off in the number of DSFGs at $z>4$ have been offered in the literature. From an observational standpoint, the difficulty of selecting and confirming these galaxies amidst the more common DSFGs around cosmic noon has been a limiting factor for a long time. Nevertheless, our results indicate that at least above $\falma = 2.25$~mJy, there is likely not a large population of $z>4$--5 DSFGs hidden in existing submillimeter surveys with underestimated photo-zs.

\begin{figure*}
    \centering
    \includegraphics[width=0.9\linewidth]{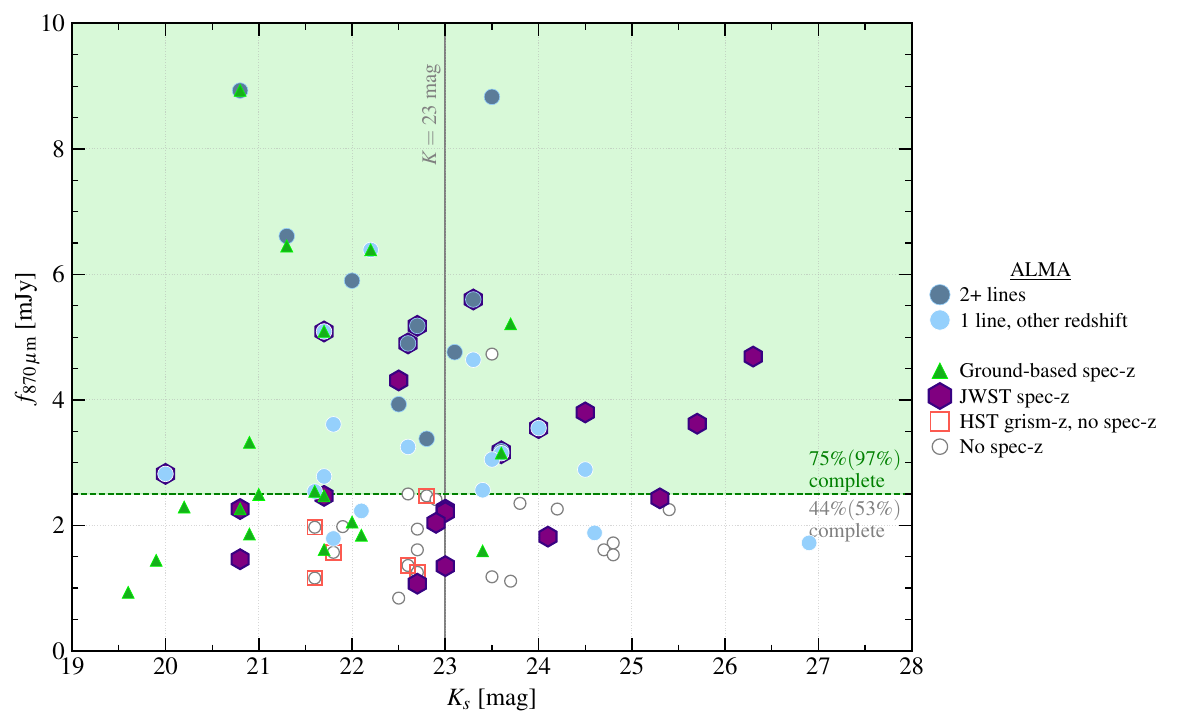}
    \caption{$\falma$ vs. $K_s$ magnitude for our DSFG sample, showing the different methods used to obtain redshifts (see legend). Sources may have redshifts from multiple instruments, as indicated by the stacked symbols. At $\falma > 2.5$~mJy (green dotted line and shaded region), the secure (tentative) spectroscopic completeness is 75\% (97\%); below, it is 44\% (53\%). We also mark $K_s=23$~mag (gray line) for illustrative purposes. 
    }
    \label{fig:f870_K}
\end{figure*}

\subsection{What sources are we still missing?}
\label{sec:missing_sources}

We now turn to the issue of spectroscopic completeness and which sources have eluded confirmation, despite benefiting from decades of spectroscopic campaigns, including significant investments of JWST and ALMA time. 

We first explore the relative completeness of our sample as a function of submillimeter and NIR flux.
In Figure~\ref{fig:f870_K}, we plot 870~$\mu$m flux versus $K_s$ magnitude, marking the methods used to obtain spec-zs. For clarity, we show the $\falma = 2.5$~mJy threshold (green dotted line and shaded region) above which the spectroscopic completeness for secure (tentative) spec-zs is 75\% (97\%); below, it is 44\% (53\%).
This figure demonstrates some trends that we might expect. For example, ALMA (dark and light blue circles) is very effective at the brightest 870~$\mu$m fluxes. Meanwhile, the distribution of ground-based spec-zs (green triangles) is agnostic to $\falma$ but is limited to $K_s < 24$~mag, while the JWST spec-zs (purple hexagons) are fairly evenly distributed across both $\falma$ and $K_s$ magnitude.

The success rate for sources targeted with JWST/NIRSpec appears to be much better than other methods. For example, for the 20 of our sources in the DJA NIRSpec catalog (version 4), only one has a NIRSpec PRISM observation that did not result in a secure spec-z (SG-ALMA-70, exposure time of $\sim$2400\,s). {However, it appears that the slit was not centered well on this source, which likely explains the non-detection.}
For the other 19 sources, all but one had PRISM exposure times ranging from $\sim$2400\,s to $\sim$8500\,s, while the last one was observed for $\sim$66500\,s (18.5\,hrs). This success rate suggests that, provided DSFGs can be selected and targeted on the NIRSpec MSA masks, the likelihood of obtaining a spec-z with the NIRSpec PRISM is extremely high even with relatively shallow programs \citep[e.g.,][]{maseda24}. Some of our sources were also observed in the higher-resolution NIRSpec gratings, although for these the success rate drops due to the relatively lower sensitivity and more limited redshift coverage. 

In total, only 29\% of our DSFG sample have JWST spec-zs. For $K_s<23$~mag ($K_s>23$~mag), this fraction is 23\% (32\%), and for $\falma>2.5$~mJy ($\falma<2.5$~mJy), it is 38\% (24\%). In other words, JWST has not yet provided particularly complete spectroscopic coverage for submillimeter-selected samples such as this one, likely due in part to their low space density ($\sim$1~arcmin$^{-2}$; hence the difficulty in targeting large samples efficiently with MSA pointings). {Unfortunately, it is difficult to quantify the selection biases involved in selecting targets for the various overlapping JWST/NIRSpec programs in GOODS-S, so we can only draw rough conclusions.}

Of the $\falma > 2.5$~mJy sources, only one object (SG-ALMA-12) still lacks a spec-z from any method. This source was only targeted in a single ALMA tuning due to a mistaken Keck/DEIMOS spec-z that turned out to correspond to a foreground galaxy. It is likely that a full ALMA spectral scan would identify a redshift, considering the photo-z ($z_{\rm phot} = 3.76$) and successful ALMA-based redshifts for sources at similar fluxes. A further eight $\falma > 2.5$~mJy sources have tentative spec-zs based off a single millimeter line. 

At the sensitivity of our ALMA scans, only four sources below $\falma = 2.5$~mJy have secure millimeter line detections, and none has 2 or more lines. Especially at $K_s>23$~mag, these faint DSFGs are very hard to confirm. Only two have JWST redshifts, despite JWST programs targeting optically faint or dark galaxies \citep[e.g.,][]{barrufet25}. {Moreover, since the completeness of our ALMA 870~$\mu$m survey begins to drop at fluxes below $\falma = 2.25$~mJy, we may be starting to miss fainter sources. To complete the spectroscopic survey of this population, which has been suggested to lie at the highest redshifts \citep[e.g.,][]{barrufet25, zavala26}, will require an additional concerted investment of ALMA and/or JWST time.}

\begin{figure}
    \centering
    \includegraphics[width=\linewidth]{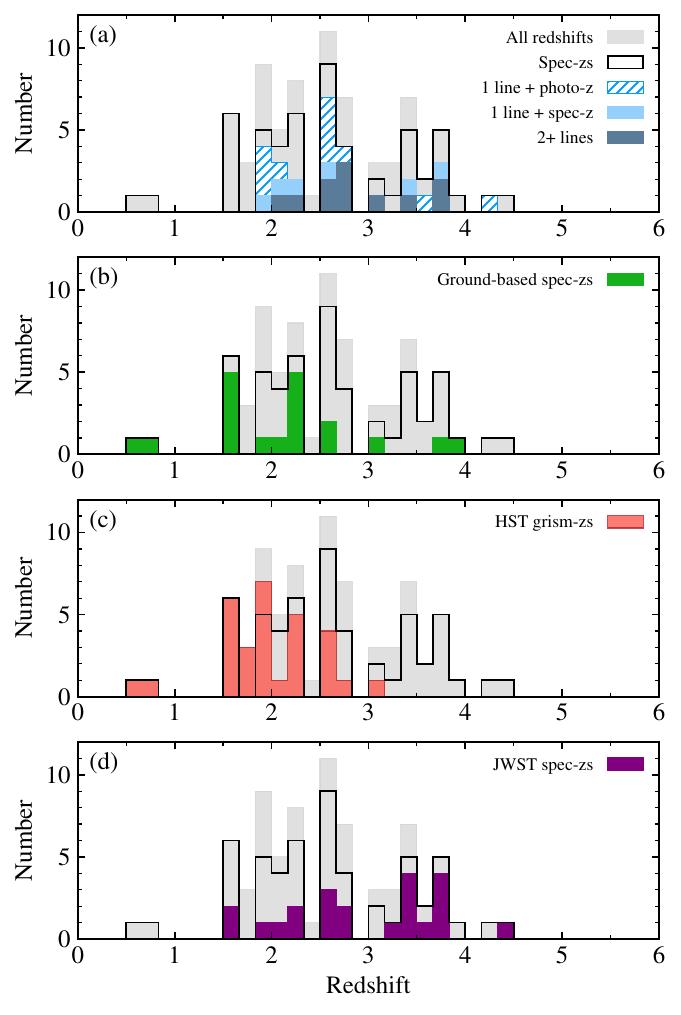}
    \caption{Redshift distribution from Figure~\ref{fig:zhist}, split by the source of the spec-zs. We show (a) ALMA spec-zs, including those based on two secure lines (dark blue) and those identified by one millimeter line and a literature spec-z or photo-z (light blue); (b) ground-based spec-zs (green), (c) HST grism-zs (red); and (d) JWST/NIRSpec spec-zs; purple).}
    \label{fig:zhist_types}
\end{figure}

In Figure~\ref{fig:zhist_types}, we investigate the relative success of each observing method/instrument as a function of redshift.  We plot the redshift distribution from Figure~\ref{fig:zhist}, highlighting in each panel, respectively, the distributions of ALMA spec-zs (including two ALMA spec-zs from the literature), ground-based spec-zs, HST grism-zs, and JWST spec-zs.

The ALMA-based redshifts are distributed between $1.8 \lesssim z\lesssim4$, although we note that just 12 sources have secure redshifts (2+ millimeter lines) from ALMA alone. Our observing program was designed to detect CO lines at $1\lesssim z \lesssim6$. Since we know the redshifts of most of the ALMA sources lie within this range, we can conclude that the non-detections are simply an issue of sensitivity. 

Ground-based spec-zs account for a smaller number of spec-zs in total, in spite of the numerous surveys that have been performed. This highlights the difficulty of detecting emission lines at optical/NIR wavelengths for DSFGs from the ground, hampered by their faintness and strong atmospheric absorption.
The HST grism-zs, while highly complete at $z<3$, fall off at higher redshifts. {This reflects the survey design of 3D-HST, which targeted galaxies at $0.7 < z < 3$ (\citetalias{momcheva16}), although beyond a redshift of $z\sim2.4$ the lack of strong emission lines in the G141 grism wavelength range may also limit the utility and accuracy of this method.}
As noted in Section~\ref{sec:photz_comp}, these are generally not as precise as spec-zs, and they can suffer from catastrophic outliers in some cases.

Thanks to its better wavelength coverage and sensitivity, JWST/NIRSpec is more effective at obtaining spec-zs at $z>3$ than ground-based instruments. Since nearly all the sources targeted with NIRSpec are confirmed, as discussed above, the observed distribution of the JWST spec-zs reflects the selection biases associated with the various NIRSpec programs.

As single-dish millimeter observations with instruments such as TolTEC and NIKA2 begin to be available for wider fields, efficient targeted spectroscopic campaigns along with more accurate photo-z estimation methods will become crucial for understanding the volume distribution and for determining the eventual descendants of DSFGs. The advent of the ALMA Wideband Sensitivity Upgrade \citep{carpenter23} will help with following up large, uniform samples with continuum imaging and spectroscopy down to fainter flux limits than are currently feasible, but it will likely be necessary to develop more efficient joint survey strategies with ALMA and JWST, or to utilize proxy selections based on other wavelengths or physical properties (\citealp[e.g.,][]{mckay25, barrufet25b, nicandro_rosenthal26, zavala26, barger26}) to build spectroscopically complete samples of faint DSFGs.

\section{Summary}
\label{sec:summary}

We presented new ALMA linescans of a large sample of DSFGs in the GOODS-S field. This sample is {nearly flux-complete} and unbiased above $\falma=2.25$~mJy, and it benefits from a wealth of deep multiwavelength observations and other spectroscopy.
\begin{itemize}
    \item We used the ALMA data to {obtain or confirm 26 spec-zs (15 secure, 11 tentative) between $2\lesssim z\lesssim4$}, primarily at the bright end ($\falma = 3$--9~mJy) of our sample. {We estimate that at least $\sim$80\% of the tentative spec-zs are correct.}
    
    \item Combining these redshifts (including tentative spec-zs) with others from the literature, our sample is now 72\% spectroscopically complete, while above $\falma=2.5$~mJy, it is as high as 97\% complete.
    
    \item We used this high spectroscopic completeness to test several photo-z codes on our sample. All photo-z methods had outlier fractions ($|\Delta z|/(1 + z_{\rm spec}) > 0.15$) greater than 20\%, including those using multi-band JWST data. While FIR-based photo-zs were not as accurate as those based on optical/NIR data, including the FIR data helped to limit high-redshift misidentifications and to reduce the overall outlier fraction.
        
    \item Our results are consistent with $\lesssim6$\% of $\falma\geq2.25$~mJy DSFGs lying at $z>4$ and $\lesssim1$\% lying at $z>5$. These fractions do not account for cosmic variance but they are not biased by incompleteness. Most of the photo-z methods tended to overestimate the number of DSFGs at $z>4$--5. 

    \item We found that nearly all DSFGs targeted with JWST/NIRSpec had successful spec-z identifications. However, the total JWST spectroscopic coverage is just 29\%, with no apparent variation with redshift, NIR magnitude, or submillimeter flux.
\end{itemize}

\section*{Acknowledgements}
{
We thank the anonymous referee for a careful report that helped to improve the manuscript.

We gratefully acknowledge support for this research from 
the William F. Vilas Estate and the North American ALMA Science Center through the ALMA Ambassadors program (S.~J.~M.), a WARF Named Professorship from the University of Wisconsin-Madison Office of the 
Vice Chancellor for Research and Graduate Education with funding from the Wisconsin Alumni Research Foundation (A.~J.~B.), NASA grant 80NSSC22K0483 (L.~L.~C.), and ANID-Chile BASAL CATA FB210003 and FONDECYT Regular 1241005 (F.~E.~B.). 

The National Radio Astronomy Observatory is a facility of the National Science Foundation operated under cooperative agreement by Associated Universities, Inc.
This paper makes use of the following ALMA data: 
ADS/JAO.ALMA\#2015.1.00242.S, \\ 
ADS/JAO.ALMA\#2021.1.00024.S.\\ 
and ADS/JAO.ALMA\#2024.1.01213.S. 
ALMA is a partnership of ESO (representing its member states), NSF (USA), and NINS (Japan), together with NRC (Canada), MOST and ASIAA (Taiwan), and KASI (Republic of Korea), in cooperation with the Republic of Chile. The Joint ALMA Observatory is operated by ESO, AUI/NRAO, and NAOJ.

Some of the data presented in this paper were obtained from the Mikulski Archive for Space Telescopes (MAST) at the Space Telescope Science Institute, which is operated by the Association of Universities for Research in Astronomy, Inc., under NASA contract NAS 5-03127 for JWST. The archival JWST data used in this work are from the JADES \citep{10.17909/8tdj-8n28}, JEMS \citep{10.17909/fsc4-dt61}, and FRESCO \citep{10.17909/gdyc-7g80} programs.

The James Clerk Maxwell Telescope is operated by the East Asian Observatory on behalf of The National Astronomical Observatory of Japan, Academia Sinica Institute of Astronomy and Astrophysics, the Korea Astronomy and Space Science Institute, the National Astronomical Observatories of China and the Chinese Academy of Sciences (grant No.~XDB09000000), with additional funding support from the Science and Technology Facilities Council of the United Kingdom and participating universities 
in the United Kingdom and Canada. 

We wish to recognize and acknowledge 
the very significant cultural role and reverence that the summit of Maunakea has always had within the indigenous Hawaiian community. We are most fortunate to have the opportunity to conduct observations from this mountain.
}

\facilities{ALMA, HST, JCMT, JWST}
\software{{\tt astropy} \citep{astropy:2013,astropy:2018,astropy:2022}, {\tt CASA} \citep[][]{mcmullin07},  {\tt emcee}} \citep{foreman13}, {\tt interferopy} \citep{boogaard21}, {\tt MAGPHYS} \citep[][]{dacunha08}, {\tt matplotlib} \citep{hunter07}, {\tt numpy} \citep{harris20}, {\tt pandas} \citep{mckinney10}, {\tt pymc} \citep{pymc23}, {\tt scipy} \citep{scipy20}

\bibliographystyle{aasjournalv7}
\bibliography{bib}

@string{june = {June}}

@misc{10.17909/8tdj-8n28,
 author = {{Rieke},  Marcia and {Robertson},  Brant and {Tacchella},  Sandro and {Willmer},  Christopher and {Johnson},  Ben and {Carniani},  Stefano and {Bunker},  Andy and {Willott},  Chris},
 doi = {10.17909/8TDJ-8N28},
 hideurl = {http://archive.stsci.edu/doi/resolve/resolve.html?doi=10.17909/8tdj-8n28},
 publisher = {STScI/MAST},
 title = {Data from the JWST Advanced Deep Extragalactic Survey (JADES)},
 year = {2023}
}

@misc{10.17909/fsc4-dt61,
 author = {{Williams},  Christina and {Tacchella},  Sandro and {Maseda},  Michael},
 doi = {10.17909/FSC4-DT61},
 hideurl = {http://archive.stsci.edu/doi/resolve/resolve.html?doi=10.17909/fsc4-dt61},
 publisher = {STScI/MAST},
 title = {Data from the JWST Extragalactic Medium-band Survey (JEMS)},
 year = {2023}
}

@misc{10.17909/gdyc-7g80,
 author = {{Oesch},  Pascal and {Magee},  Dan},
 doi = {10.17909/GDYC-7G80},
 hideurl = {http://archive.stsci.edu/doi/resolve/resolve.html?doi=10.17909/gdyc-7g80},
 publisher = {STScI/MAST},
 title = {The JWST FRESCO Survey},
 year = {2023}
}

@article{alberts24,
 adsurl = {https://ui.adsabs.harvard.edu/abs/2024ApJ...976..224A},
 archiveprefix = {arXiv},
 author = {{Alberts}, Stacey and {Lyu}, Jianwei and {Shivaei}, Irene and {Rieke}, George H. and {P{\'e}rez-Gonz{\'a}lez}, Pablo G. and {Bonaventura}, Nina and {Zhu}, Yongda and {Helton}, Jakob M. and {Ji}, Zhiyuan and {Morrison}, Jane and {Robertson}, Brant E. and {Stone}, Meredith A. and {Sun}, Yang and {Williams}, Christina C. and {Willmer}, Christopher N.~A.},
 doi = {10.3847/1538-4357/ad7396},
 eid = {224},
 hideeprint = {2405.15972},
 journal = {\apj},
 month = {December},
 number = {2},
 pages = {224},
 primaryclass = {astro-ph.GA},
 title = {{SMILES Initial Data Release: Unveiling the Obscured Universe with MIRI Multiband Imaging}},
 volume = {976},
 year = {2024}
}

@article{alberts26,
 adsurl = {https://ui.adsabs.harvard.edu/abs/2026arXiv260115955A},
 archiveprefix = {arXiv},
 author = {{Alberts}, Stacey and {Eisenstein}, Daniel J. and {Bunker}, Andrew J. and {Curtis-Lake}, Emma and {Duan}, Qiao and {Hainline}, Kevin and {Hausen}, Ryan and {Helton}, Jakob M. and {Ji}, Zhiyuan and {Johnson}, Benjamin D. and {Lyu}, Jianwei and {Morrison}, Jane and {Perez-Gonzalez}, Pablo G. and {Rieke}, George H. and {Rieke}, Marcia and {Rinaldi}, Pierluigi and {Robertson}, Brant and {Sun}, Yang and {Tacchella}, Sandro and {Williams}, Christina C. and {Willmer}, Christopher N.~A. and {Wu}, Zihao},
 doi = {10.48550/arXiv.2601.15955},
 eid = {arXiv:2601.15955},
 hideeprint = {2601.15955},
 journal = {arXiv e-prints},
 month = {January},
 pages = {arXiv:2601.15955},
 primaryclass = {astro-ph.GA},
 title = {{JWST Advanced Deep Extragalactic Survey (JADES) Data Release 5: MIRI Coordinated Parallels in GOODS-S and GOODS-N}},
 year = {2026}
}

@article{algera23,
 adsurl = {https://ui.adsabs.harvard.edu/abs/2023MNRAS.518.6142A},
 archiveprefix = {arXiv},
 author = {{Algera}, Hiddo S.~B. and {Inami}, Hanae and {Oesch}, Pascal A. and {Sommovigo}, Laura and {Bouwens}, Rychard J. and {Topping}, Michael W. and {Schouws}, Sander and {Stefanon}, Mauro and {Stark}, Daniel P. and {Aravena}, Manuel and {Barrufet}, Laia and {da Cunha}, Elisabete and {Dayal}, Pratika and {Endsley}, Ryan and {Ferrara}, Andrea and {Fudamoto}, Yoshinobu and {Gonzalez}, Valentino and {Graziani}, Luca and {Hodge}, Jacqueline A. and {Hygate}, Alexander P.~S. and {de Looze}, Ilse and {Nanayakkara}, Themiya and {Schneider}, Raffaella and {van der Werf}, Paul P.},
 doi = {10.1093/mnras/stac3195},
 eprint = {2208.08243},
 journal = {\mnras},
 month = {February},
 number = {4},
 pages = {6142-6157},
 primaryclass = {astro-ph.GA},
 title = {{The ALMA REBELS survey: the dust-obscured cosmic star formation rate density at redshift 7}},
 volume = {518},
 year = {2023}
}

@article{alvarez_crespo21,
 adsurl = {https://ui.adsabs.harvard.edu/abs/2021A&A...646A.174A},
 archiveprefix = {arXiv},
 author = {{{\'A}lvarez Crespo}, N. and {Smoli{\'c}}, V. and {Finoguenov}, A. and {Barrufet}, L. and {Aravena}, M.},
 doi = {10.1051/0004-6361/202039227},
 eid = {A174},
 hideeprint = {2101.02977},
 journal = {\aap},
 month = {February},
 pages = {A174},
 primaryclass = {astro-ph.GA},
 title = {{Environments of a sample of AzTEC submillimetre galaxies in the COSMOS field}},
 volume = {646},
 year = {2021}
}

@article{aravena20,
 adsurl = {https://ui.adsabs.harvard.edu/abs/2020ApJ...901...79A},
 archiveprefix = {arXiv},
 author = {{Aravena}, Manuel and {Boogaard}, Leindert and {G{\'o}nzalez-L{\'o}pez}, Jorge and {Decarli}, Roberto and {Walter}, Fabian and {Carilli}, Chris L. and {Smail}, Ian and {Weiss}, Axel and {Assef}, Roberto J. and {Bauer}, Franz Erik and {Bouwens}, Rychard J. and {Cortes}, Paulo C. and {Cox}, Pierre and {da Cunha}, Elisabete and {Daddi}, Emanuele and {D{\'\i}az-Santos}, Tanio and {Inami}, Hanae and {Ivison}, Rob and {Novak}, Mladen and {Popping}, Gerg{\"o} and {Riechers}, Dominik and {van der Werf}, Paul and {Wagg}, Jeff},
 doi = {10.3847/1538-4357/ab99a2},
 eid = {79},
 hideeprint = {2006.04284},
 journal = {\apj},
 month = {September},
 number = {1},
 pages = {79},
 primaryclass = {astro-ph.GA},
 title = {{The ALMA Spectroscopic Survey in the Hubble Ultra Deep Field: The Nature of the Faintest Dusty Star-forming Galaxies}},
 volume = {901},
 year = {2020}
}

@article{araya-araya26,
 adsurl = {https://ui.adsabs.harvard.edu/abs/2026A&A...707A.305A},
 archiveprefix = {arXiv},
 author = {{Araya-Araya}, Pablo and {Cochrane}, Rachel K. and {Sodr{\'e}}, Jr., Laerte and {Yates}, Robert M. and {Hayward}, Christopher C. and {van Daalen}, Marcel P. and {Vicentin}, Marcelo C. and {Gullberg}, Bitten and {Valentino}, Francesco},
 doi = {10.1051/0004-6361/202557426},
 eid = {A305},
 hideeprint = {2509.26646},
 journal = {\aap},
 month = {March},
 pages = {A305},
 primaryclass = {astro-ph.GA},
 title = {{The connection between dusty star-forming galaxies and the first massive quenched galaxies}},
 volume = {707},
 year = {2026}
}

@article{aretxaga03,
 adsurl = {https://ui.adsabs.harvard.edu/abs/2003MNRAS.342..759A},
 archiveprefix = {arXiv},
 author = {{Aretxaga}, Itziar and {Hughes}, David H. and {Chapin}, Edward L. and {Gazta{\~n}aga}, Enrique and {Dunlop}, James S. and {Ivison}, Rob J.},
 doi = {10.1046/j.1365-8711.2003.06560.x},
 hideeprint = {astro-ph/0205313},
 journal = {\mnras},
 month = {July},
 number = {3},
 pages = {759-801},
 primaryclass = {astro-ph},
 title = {{Breaking the `redshift deadlock'- II. The redshift distribution for the submillimetre population of galaxies}},
 volume = {342},
 year = {2003}
}

@article{aretxaga07,
 adsurl = {https://ui.adsabs.harvard.edu/abs/2007MNRAS.379.1571A},
 archiveprefix = {arXiv},
 author = {{Aretxaga}, Itziar and {Hughes}, David H. and {Coppin}, Kristen and {Mortier}, Angela M.~J. and {Wagg}, Jeff and {Dunlop}, James S. and {Chapin}, Edward L. and {Eales}, Stephen A. and {Gazta{\~n}aga}, Enrique and {Halpern}, Mark and {Ivison}, Rob J. and {van Kampen}, Eelco and {Scott}, Douglas and {Serjeant}, Stephen and {Smail}, Ian and {Babbedge}, Thomas and {Benson}, Andrew J. and {Chapman}, Scott and {Clements}, David L. and {Dunne}, Loretta and {Dye}, Simon and {Farrah}, Duncan and {Jarvis}, Matt J. and {Mann}, Robert G. and {Pope}, Alexandra and {Priddey}, Robert and {Rawlings}, Steve and {Seigar}, Marc and {Silva}, Laura and {Simpson}, Chris and {Vaccari}, Mattia},
 doi = {10.1111/j.1365-2966.2007.12036.x},
 eprint = {astro-ph/0702503},
 journal = {\mnras},
 month = {August},
 number = {4},
 pages = {1571-1588},
 primaryclass = {astro-ph},
 title = {{The SCUBA Half Degree Extragalactic Survey - IV. Radio-mm-FIR photometric redshifts}},
 volume = {379},
 year = {2007}
}

@article{astropy:2013,
 adsurl = {https://ui.adsabs.harvard.edu/abs/2013A&A...558A..33A},
 archiveprefix = {arXiv},
 author = {{Astropy Collaboration} and {Robitaille}, Thomas P. and {Tollerud}, Erik J. and {Greenfield}, Perry and {Droettboom}, Michael and {Bray}, Erik and {Aldcroft}, Tom and {Davis}, Matt and {Ginsburg}, Adam and {Price-Whelan}, Adrian M. and {Kerzendorf}, Wolfgang E. and {Conley}, Alexander and {Crighton}, Neil and {Barbary}, Kyle and {Muna}, Demitri and {Ferguson}, Henry and {Grollier}, Fr{\'e}d{\'e}ric and {Parikh}, Madhura M. and {Nair}, Prasanth H. and {Unther}, Hans M. and {Deil}, Christoph and {Woillez}, Julien and {Conseil}, Simon and {Kramer}, Roban and {Turner}, James E.~H. and {Singer}, Leo and {Fox}, Ryan and {Weaver}, Benjamin A. and {Zabalza}, Victor and {Edwards}, Zachary I. and {Azalee Bostroem}, K. and {Burke}, D.~J. and {Casey}, Andrew R. and {Crawford}, Steven M. and {Dencheva}, Nadia and {Ely}, Justin and {Jenness}, Tim and {Labrie}, Kathleen and {Lim}, Pey Lian and {Pierfederici}, Francesco and {Pontzen}, Andrew and {Ptak}, Andy and {Refsdal}, Brian and {Servillat}, Mathieu and {Streicher}, Ole},
 doi = {10.1051/0004-6361/201322068},
 eid = {A33},
 hideeprint = {1307.6212},
 journal = {\aap},
 month = {October},
 pages = {A33},
 primaryclass = {astro-ph.IM},
 title = {{Astropy: A community Python package for astronomy}},
 volume = {558},
 year = {2013}
}

@article{astropy:2018,
 adsurl = {https://ui.adsabs.harvard.edu/abs/2018AJ....156..123A},
 archiveprefix = {arXiv},
 author = {{Astropy Collaboration} and {Price-Whelan}, A.~M. and {Sip{\H{o}}cz}, B.~M. and {G{\"u}nther}, H.~M. and {Lim}, P.~L. and {Crawford}, S.~M. and {Conseil}, S. and {Shupe}, D.~L. and {Craig}, M.~W. and {Dencheva}, N. and {Ginsburg}, A. and {VanderPlas}, J.~T. and {Bradley}, L.~D. and {P{\'e}rez-Su{\'a}rez}, D. and {de Val-Borro}, M. and {Aldcroft}, T.~L. and {Cruz}, K.~L. and {Robitaille}, T.~P. and {Tollerud}, E.~J. and {Ardelean}, C. and {Babej}, T. and {Bach}, Y.~P. and {Bachetti}, M. and {Bakanov}, A.~V. and {Bamford}, S.~P. and {Barentsen}, G. and {Barmby}, P. and {Baumbach}, A. and {Berry}, K.~L. and {Biscani}, F. and {Boquien}, M. and {Bostroem}, K.~A. and {Bouma}, L.~G. and {Brammer}, G.~B. and {Bray}, E.~M. and {Breytenbach}, H. and {Buddelmeijer}, H. and {Burke}, D.~J. and {Calderone}, G. and {Cano Rodr{\'\i}guez}, J.~L. and {Cara}, M. and {Cardoso}, J.~V.~M. and {Cheedella}, S. and {Copin}, Y. and {Corrales}, L. and {Crichton}, D. and {D'Avella}, D. and {Deil}, C. and {Depagne}, {\'E}. and {Dietrich}, J.~P. and {Donath}, A. and {Droettboom}, M. and {Earl}, N. and {Erben}, T. and {Fabbro}, S. and {Ferreira}, L.~A. and {Finethy}, T. and {Fox}, R.~T. and {Garrison}, L.~H. and {Gibbons}, S.~L.~J. and {Goldstein}, D.~A. and {Gommers}, R. and {Greco}, J.~P. and {Greenfield}, P. and {Groener}, A.~M. and {Grollier}, F. and {Hagen}, A. and {Hirst}, P. and {Homeier}, D. and {Horton}, A.~J. and {Hosseinzadeh}, G. and {Hu}, L. and {Hunkeler}, J.~S. and {Ivezi{\'c}}, {\v{Z}}. and {Jain}, A. and {Jenness}, T. and {Kanarek}, G. and {Kendrew}, S. and {Kern}, N.~S. and {Kerzendorf}, W.~E. and {Khvalko}, A. and {King}, J. and {Kirkby}, D. and {Kulkarni}, A.~M. and {Kumar}, A. and {Lee}, A. and {Lenz}, D. and {Littlefair}, S.~P. and {Ma}, Z. and {Macleod}, D.~M. and {Mastropietro}, M. and {McCully}, C. and {Montagnac}, S. and {Morris}, B.~M. and {Mueller}, M. and {Mumford}, S.~J. and {Muna}, D. and {Murphy}, N.~A. and {Nelson}, S. and {Nguyen}, G.~H. and {Ninan}, J.~P. and {N{\"o}the}, M. and {Ogaz}, S. and {Oh}, S. and {Parejko}, J.~K. and {Parley}, N. and {Pascual}, S. and {Patil}, R. and {Patil}, A.~A. and {Plunkett}, A.~L. and {Prochaska}, J.~X. and {Rastogi}, T. and {Reddy Janga}, V. and {Sabater}, J. and {Sakurikar}, P. and {Seifert}, M. and {Sherbert}, L.~E. and {Sherwood-Taylor}, H. and {Shih}, A.~Y. and {Sick}, J. and {Silbiger}, M.~T. and {Singanamalla}, S. and {Singer}, L.~P. and {Sladen}, P.~H. and {Sooley}, K.~A. and {Sornarajah}, S. and {Streicher}, O. and {Teuben}, P. and {Thomas}, S.~W. and {Tremblay}, G.~R. and {Turner}, J.~E.~H. and {Terr{\'o}n}, V. and {van Kerkwijk}, M.~H. and {de la Vega}, A. and {Watkins}, L.~L. and {Weaver}, B.~A. and {Whitmore}, J.~B. and {Woillez}, J. and {Zabalza}, V. and {Astropy Contributors}},
 doi = {10.3847/1538-3881/aabc4f},
 eid = {123},
 hideeprint = {1801.02634},
 journal = {\aj},
 month = {September},
 number = {3},
 pages = {123},
 primaryclass = {astro-ph.IM},
 title = {{The Astropy Project: Building an Open-science Project and Status of the v2.0 Core Package}},
 volume = {156},
 year = {2018}
}

@article{astropy:2022,
 adsurl = {https://ui.adsabs.harvard.edu/abs/2022ApJ...935..167A},
 archiveprefix = {arXiv},
 author = {{Astropy Collaboration} and {Price-Whelan}, Adrian M. and {Lim}, Pey Lian and {Earl}, Nicholas and {Starkman}, Nathaniel and {Bradley}, Larry and {Shupe}, David L. and {Patil}, Aarya A. and {Corrales}, Lia and {Brasseur}, C.~E. and {N{"o}the}, Maximilian and {Donath}, Axel and {Tollerud}, Erik and {Morris}, Brett M. and {Ginsburg}, Adam and {Vaher}, Eero and {Weaver}, Benjamin A. and {Tocknell}, James and {Jamieson}, William and {van Kerkwijk}, Marten H. and {Robitaille}, Thomas P. and {Merry}, Bruce and {Bachetti}, Matteo and {G{"u}nther}, H. Moritz and {Aldcroft}, Thomas L. and {Alvarado-Montes}, Jaime A. and {Archibald}, Anne M. and {B{'o}di}, Attila and {Bapat}, Shreyas and {Barentsen}, Geert and {Baz{'a}n}, Juanjo and {Biswas}, Manish and {Boquien}, M{'e}d{'e}ric and {Burke}, D.~J. and {Cara}, Daria and {Cara}, Mihai and {Conroy}, Kyle E. and {Conseil}, Simon and {Craig}, Matthew W. and {Cross}, Robert M. and {Cruz}, Kelle L. and {D'Eugenio}, Francesco and {Dencheva}, Nadia and {Devillepoix}, Hadrien A.~R. and {Dietrich}, J{"o}rg P. and {Eigenbrot}, Arthur Davis and {Erben}, Thomas and {Ferreira}, Leonardo and {Foreman-Mackey}, Daniel and {Fox}, Ryan and {Freij}, Nabil and {Garg}, Suyog and {Geda}, Robel and {Glattly}, Lauren and {Gondhalekar}, Yash and {Gordon}, Karl D. and {Grant}, David and {Greenfield}, Perry and {Groener}, Austen M. and {Guest}, Steve and {Gurovich}, Sebastian and {Handberg}, Rasmus and {Hart}, Akeem and {Hatfield-Dodds}, Zac and {Homeier}, Derek and {Hosseinzadeh}, Griffin and {Jenness}, Tim and {Jones}, Craig K. and {Joseph}, Prajwel and {Kalmbach}, J. Bryce and {Karamehmetoglu}, Emir and {Ka{l}uszy{'n}ski}, Miko{l}aj and {Kelley}, Michael S.~P. and {Kern}, Nicholas and {Kerzendorf}, Wolfgang E. and {Koch}, Eric W. and {Kulumani}, Shankar and {Lee}, Antony and {Ly}, Chun and {Ma}, Zhiyuan and {MacBride}, Conor and {Maljaars}, Jakob M. and {Muna}, Demitri and {Murphy}, N.~A. and {Norman}, Henrik and {O'Steen}, Richard and {Oman}, Kyle A. and {Pacifici}, Camilla and {Pascual}, Sergio and {Pascual-Granado}, J. and {Patil}, Rohit R. and {Perren}, Gabriel I. and {Pickering}, Timothy E. and {Rastogi}, Tanuj and {Roulston}, Benjamin R. and {Ryan}, Daniel F. and {Rykoff}, Eli S. and {Sabater}, Jose and {Sakurikar}, Parikshit and {Salgado}, Jes{'u}s and {Sanghi}, Aniket and {Saunders}, Nicholas and {Savchenko}, Volodymyr and {Schwardt}, Ludwig and {Seifert-Eckert}, Michael and {Shih}, Albert Y. and {Jain}, Anany Shrey and {Shukla}, Gyanendra and {Sick}, Jonathan and {Simpson}, Chris and {Singanamalla}, Sudheesh and {Singer}, Leo P. and {Singhal}, Jaladh and {Sinha}, Manodeep and {Sip{H{o}}cz}, Brigitta M. and {Spitler}, Lee R. and {Stansby}, David and {Streicher}, Ole and {Sumak}, Jani and {Swinbank}, John D. and {Taranu}, Dan S. and {Tewary}, Nikita and {Tremblay}, Grant R. and {Val-Borro}, Miguel de and {Van Kooten}, Samuel J. and {Vasovi{'c}}, Zlatan and {Verma}, Shresth and {de Miranda Cardoso}, Jos{'e} Vin{'i}cius and {Williams}, Peter K.~G. and {Wilson}, Tom J. and {Winkel}, Benjamin and {Wood-Vasey}, W.~M. and {Xue}, Rui and {Yoachim}, Peter and {Zhang}, Chen and {Zonca}, Andrea and {Astropy Project Contributors}},
 doi = {10.3847/1538-4357/ac7c74},
 eid = {167},
 hideeprint = {2206.14220},
 journal = {\apj},
 month = {August},
 number = {2},
 pages = {167},
 primaryclass = {astro-ph.IM},
 title = {{The Astropy Project: Sustaining and Growing a Community-oriented Open-source Project and the Latest Major Release (v5.0) of the Core Package}},
 volume = {935},
 year = {2022}
}

@article{barger00,
 adsurl = {https://ui.adsabs.harvard.edu/abs/2000AJ....119.2092B},
 archiveprefix = {arXiv},
 author = {{Barger}, A.~J. and {Cowie}, L.~L. and {Richards}, E.~A.},
 doi = {10.1086/301341},
 hideeprint = {astro-ph/0001096},
 journal = {\aj},
 month = {May},
 number = {5},
 pages = {2092-2109},
 primaryclass = {astro-ph},
 title = {{Mapping the Evolution of High-Redshift Dusty Galaxies with Submillimeter Observations of a Radio-selected Sample}},
 volume = {119},
 year = {2000}
}

@article{barger22,
 adsurl = {https://ui.adsabs.harvard.edu/abs/2022ApJ...934...56B},
 archiveprefix = {arXiv},
 author = {{Barger}, A.~J. and {Cowie}, L.~L. and {Blair}, A.~H. and {Jones}, L.~H.},
 doi = {10.3847/1538-4357/ac67e7},
 eid = {56},
 hideeprint = {2205.01114},
 journal = {\apj},
 month = {July},
 number = {1},
 pages = {56},
 primaryclass = {astro-ph.GA},
 title = {{A Submillimeter Perspective on the GOODS Fields (SUPER GOODS). V. Deep 450 {\ensuremath{\mu}}m Imaging}},
 volume = {934},
 year = {2022}
}

@article{barger26,
doi = {10.3847/1538-4357/ae7101},
url = {https://doi.org/10.3847/1538-4357/ae7101},
year = {2026},
month = {jun},
publisher = {The American Astronomical Society},
volume = {1004},
number = {2},
pages = {241},
author = {Barger, A. J. and Cowie, L. L. and McKay, S. J. and Bauer, F. E.},
title = {A Redshift-based Red Selection of Dusty Star-forming Galaxies},
journal = {\apj},
}

@article{barger99,
 adsurl = {https://ui.adsabs.harvard.edu/abs/1999ApJ...518L...5B},
 archiveprefix = {arXiv},
 author = {{Barger}, A.~J. and {Cowie}, L.~L. and {Sanders}, D.~B.},
 doi = {10.1086/312054},
 eprint = {astro-ph/9904126},
 journal = {\apjl},
 month = {June},
 number = {1},
 pages = {L5-L8},
 primaryclass = {astro-ph},
 title = {{Resolving the Submillimeter Background: The 850 Micron Galaxy Counts}},
 volume = {518},
 year = {1999}
}

@article{barger99b,
 adsurl = {https://ui.adsabs.harvard.edu/abs/1999AJ....117.2656B},
 archiveprefix = {arXiv},
 author = {{Barger}, A.~J. and {Cowie}, L.~L. and {Smail}, I. and {Ivison}, R.~J. and {Blain}, A.~W. and {Kneib}, J.-P.},
 doi = {10.1086/300890},
 eprint = {astro-ph/9903142},
 journal = {\aj},
 month = {June},
 number = {6},
 pages = {2656-2665},
 primaryclass = {astro-ph},
 title = {{Redshift Distribution of the Faint Submillimeter Galaxy Population}},
 volume = {117},
 year = {1999}
}

@article{barrufet25,
 adsurl = {https://ui.adsabs.harvard.edu/abs/2025MNRAS.537.3453B},
 archiveprefix = {arXiv},
 author = {{Barrufet}, L. and {Oesch}, P.~A. and {Marques-Chaves}, R. and {Arellano-Cordova}, K. and {Baggen}, J.~F.~W. and {Carnall}, A.~C. and {Cullen}, F. and {Dunlop}, J.~S. and {Gottumukkala}, R. and {Fudamoto}, Y. and {Illingworth}, G.~D. and {Magee}, D. and {McLure}, R.~J. and {McLeod}, D.~J. and {Micha{\l}owski}, M.~J. and {Stefanon}, M. and {van Dokkum}, P.~G. and {Weibel}, A.},
 doi = {10.1093/mnras/staf013},
 hideeprint = {2404.08052},
 journal = {\mnras},
 month = {March},
 number = {4},
 pages = {3453-3469},
 primaryclass = {astro-ph.GA},
 title = {{Quiescent or dusty? Unveiling the nature of extremely red galaxies at z > 3}},
 volume = {537},
 year = {2025}
}

@article{barrufet25b,
 adsurl = {https://ui.adsabs.harvard.edu/abs/2026MNRAS.548ag770B},
 archiveprefix = {arXiv},
 author = {{Barrufet}, L. and {Dunlop}, J.~S. and {Begley}, R. and {Flury}, S. and {McLeod}, D.~J. and {Arellano-Cordova}, K. and {Carnall}, A. and {Cullen}, F. and {Donnan}, C.~T. and {Liu}, F. and {McLure}, R. and {Scholte}, D. and {Stanton}, T.~M. and {Cochrane}, R.~K. and {Conselice}, C. and {Ellis}, R. and {P{\'e}rez-Gonz{\'a}lez}, P.~G. and {Gottumukkala}, R. and {Grogin}, N.~A. and {Illingworth}, G.~D. and {Koekemoer}, A.~M. and {Magee}, D. and {Michalowski}, M.},
 doi = {10.1093/mnras/stag770},
 eid = {stag770},
 eprint = {2508.05740},
 journal = {\mnras},
 month = {June},
 number = {4},
 pages = {stag770},
 primaryclass = {astro-ph.GA},
 title = {{Strength in numbers: Red Galaxies Bolster the cosmic star formation rate density at z {\ensuremath{\gtrsim}} 3}},
 volume = {548},
 year = {2026}
}

@article{battisti+19,
 adsurl = {https://ui.adsabs.harvard.edu/abs/2019ApJ...882...61B},
 archiveprefix = {arXiv},
 author = {{Battisti}, A.~J. and {da Cunha}, E. and {Grasha}, K. and {Salvato}, M. and {Daddi}, E. and {Davies}, L. and {Jin}, S. and {Liu}, D. and {Schinnerer}, E. and {Vaccari}, M. and {COSMOS Collaboration}},
 doi = {10.3847/1538-4357/ab345d},
 eid = {61},
 hideeprint = {1908.00771},
 journal = {\apj},
 month = {September},
 number = {1},
 pages = {61},
 primaryclass = {astro-ph.GA},
 title = {{MAGPHYS+photo-z: Constraining the Physical Properties of Galaxies with Unknown Redshifts}},
 volume = {882},
 year = {2019}
}

@article{birkin21,
 adsurl = {https://ui.adsabs.harvard.edu/abs/2021MNRAS.501.3926B},
 archiveprefix = {arXiv},
 author = {{Birkin}, Jack E. and {Weiss}, Axel and {Wardlow}, J.~L. and {Smail}, Ian and {Swinbank}, A.~M. and {Dudzevi{\v{c}}i{\={u}}t{\.{e}}}, U. and {An}, Fang Xia and {Ao}, Y. and {Chapman}, S.~C. and {Chen}, Chian-Chou and {da Cunha}, E. and {Dannerbauer}, H. and {Gullberg}, B. and {Hodge}, J.~A. and {Ikarashi}, S. and {Ivison}, R.~J. and {Matsuda}, Y. and {Stach}, S.~M. and {Walter}, F. and {Wang}, W.-H. and {van der Werf}, P.},
 doi = {10.1093/mnras/staa3862},
 hideeprint = {2009.03341},
 journal = {\mnras},
 month = {March},
 number = {3},
 pages = {3926-3950},
 primaryclass = {astro-ph.GA},
 title = {{An ALMA/NOEMA survey of the molecular gas properties of high-redshift star-forming galaxies}},
 volume = {501},
 year = {2021}
}

@article{birkin24,
 adsurl = {https://ui.adsabs.harvard.edu/abs/2024MNRAS.531...61B},
 archiveprefix = {arXiv},
 author = {{Birkin}, Jack E. and {Puglisi}, A. and {Swinbank}, A.~M. and {Smail}, Ian and {An}, Fang Xia and {Chapman}, S.~C. and {Chen}, Chian-Chou and {Conselice}, C.~J. and {Dudzevi{\v{c}}i{\={u}}t{\.{e}}}, U. and {Farrah}, D. and {Gullberg}, B. and {Matsuda}, Y. and {Schinnerer}, E. and {Scott}, D. and {Wardlow}, J.~L. and {van der Werf}, P.},
 doi = {10.1093/mnras/stae1089},
 eprint = {2301.05720},
 journal = {\mnras},
 month = {June},
 number = {1},
 pages = {61-83},
 primaryclass = {astro-ph.GA},
 title = {{KAOSS: turbulent, but disc-like kinematics in dust-obscured star-forming galaxies at z   1.3-2.6}},
 volume = {531},
 year = {2024}
}

@article{bodansky26,
 adsurl = {https://ui.adsabs.harvard.edu/abs/2026ApJ..1001..235B},
 author = {{Bodansky}, Sarah and {Whitaker}, Katherine E. and {Abdullah}, Ayesha and {Lin}, Jamie and {Oesch}, Pascal A. and {Pope}, Alexandra and {Xiao}, Mengyuan and {Covelo-Paz}, Alba and {Cutler}, Sam and {Garcia Diaz}, Carlos and {Lee}, Minju M. and {Manning}, Sinclaire M. and {Meyer}, Romain A. and {Narayanan}, Desika and {Nelson}, Erica and {Shivaei}, Irene and {van Dokkum}, Pieter},
 doi = {10.3847/1538-4357/ae4f64},
 eid = {235},
 journal = {\apj},
 month = {April},
 number = {2},
 pages = {235},
 title = {{JWST+ALMA Reveal the Buildup of Stellar Mass in the Cores of Dusty Star-forming Galaxies at Cosmic Noon}},
 volume = {1001},
 year = {2026}
}

@software{boogaard21,
 adsurl = {https://ui.adsabs.harvard.edu/abs/2021zndo...5775604B},
 author = {{Boogaard}, Leindert and {Meyer}, Romain A. and {Novak}, Mladen},
 doi = {10.5281/zenodo.5775604},
 eid = {10.5281/zenodo.5775604},
 month = {December},
 publisher = {Zenodo},
 title = {{Interferopy: analysing datacubes from radio-to-submm observations}},
 version = {v1.0.0-beta},
 year = {2021}
}

@article{boquien19,
 adsurl = {https://ui.adsabs.harvard.edu/abs/2019A&A...622A.103B},
 archiveprefix = {arXiv},
 author = {{Boquien}, M. and {Burgarella}, D. and {Roehlly}, Y. and {Buat}, V. and {Ciesla}, L. and {Corre}, D. and {Inoue}, A.~K. and {Salas}, H.},
 doi = {10.1051/0004-6361/201834156},
 eid = {A103},
 eprint = {1811.03094},
 journal = {\aap},
 month = {February},
 pages = {A103},
 primaryclass = {astro-ph.GA},
 title = {{CIGALE: a python Code Investigating GALaxy Emission}},
 volume = {622},
 year = {2019}
}

@article{bothwell13,
 adsurl = {https://ui.adsabs.harvard.edu/abs/2013MNRAS.429.3047B},
 archiveprefix = {arXiv},
 author = {{Bothwell}, M.~S. and {Smail}, Ian and {Chapman}, S.~C. and {Genzel}, R. and {Ivison}, R.~J. and {Tacconi}, L.~J. and {Alaghband-Zadeh}, S. and {Bertoldi}, F. and {Blain}, A.~W. and {Casey}, C.~M. and {Cox}, P. and {Greve}, T.~R. and {Lutz}, D. and {Neri}, R. and {Omont}, A. and {Swinbank}, A.~M.},
 doi = {10.1093/mnras/sts562},
 eprint = {1205.1511},
 journal = {\mnras},
 month = {March},
 number = {4},
 pages = {3047-3067},
 primaryclass = {astro-ph.CO},
 title = {{A survey of molecular gas in luminous sub-millimetre galaxies}},
 volume = {429},
 year = {2013}
}

@article{brammer08,
 adsurl = {https://ui.adsabs.harvard.edu/abs/2008ApJ...686.1503B},
 archiveprefix = {arXiv},
 author = {{Brammer}, Gabriel B. and {van Dokkum}, Pieter G. and {Coppi}, Paolo},
 doi = {10.1086/591786},
 eprint = {0807.1533},
 journal = {\apj},
 month = {October},
 number = {2},
 pages = {1503-1513},
 primaryclass = {astro-ph},
 title = {{EAZY: A Fast, Public Photometric Redshift Code}},
 volume = {686},
 year = {2008}
}

@article{brammer12,
 adsurl = {https://ui.adsabs.harvard.edu/abs/2012ApJS..200...13B},
 archiveprefix = {arXiv},
 author = {{Brammer}, Gabriel B. and {van Dokkum}, Pieter G. and {Franx}, Marijn and {Fumagalli}, Mattia and {Patel}, Shannon and {Rix}, Hans-Walter and {Skelton}, Rosalind E. and {Kriek}, Mariska and {Nelson}, Erica and {Schmidt}, Kasper B. and {Bezanson}, Rachel and {da Cunha}, Elisabete and {Erb}, Dawn K. and {Fan}, Xiaohui and {F{\"o}rster Schreiber}, Natascha and {Illingworth}, Garth D. and {Labb{\'e}}, Ivo and {Leja}, Joel and {Lundgren}, Britt and {Magee}, Dan and {Marchesini}, Danilo and {McCarthy}, Patrick and {Momcheva}, Ivelina and {Muzzin}, Adam and {Quadri}, Ryan and {Steidel}, Charles C. and {Tal}, Tomer and {Wake}, David and {Whitaker}, Katherine E. and {Williams}, Anna},
 doi = {10.1088/0067-0049/200/2/13},
 eid = {13},
 hideeprint = {1204.2829},
 journal = {\apjs},
 month = {June},
 number = {2},
 pages = {13},
 primaryclass = {astro-ph.CO},
 title = {{3D-HST: A Wide-field Grism Spectroscopic Survey with the Hubble Space Telescope}},
 volume = {200},
 year = {2012}
}

@software{brammer23,
 adsurl = {https://ui.adsabs.harvard.edu/abs/2023zndo...8370018B},
 author = {{Brammer}, Gabriel},
 doi = {10.5281/zenodo.8370018},
 eid = {10.5281/zenodo.8370018},
 month = {September},
 publisher = {Zenodo},
 title = {{grizli}},
 version = {1.9.11},
 year = {2023}
}

@article{bunker24,
 adsurl = {https://ui.adsabs.harvard.edu/abs/2024A&A...690A.288B},
 archiveprefix = {arXiv},
 author = {{Bunker}, Andrew J. and {Cameron}, Alex J. and {Curtis-Lake}, Emma and {Jakobsen}, Peter and {Carniani}, Stefano and {Curti}, Mirko and {Witstok}, Joris and {Maiolino}, Roberto and {D'Eugenio}, Francesco and {Looser}, Tobias J. and {Willott}, Chris and {Bonaventura}, Nina and {Hainline}, Kevin and {{\"U}bler}, Hannah and {Willmer}, Christopher N.~A. and {Saxena}, Aayush and {Smit}, Renske and {Alberts}, Stacey and {Arribas}, Santiago and {Baker}, William M. and {Baum}, Stefi and {Bhatawdekar}, Rachana and {Bowler}, Rebecca A.~A. and {Boyett}, Kristan and {Charlot}, Stephane and {Chen}, Zuyi and {Chevallard}, Jacopo and {Circosta}, Chiara and {DeCoursey}, Christa and {de Graaff}, Anna and {Egami}, Eiichi and {Eisenstein}, Daniel J. and {Endsley}, Ryan and {Ferruit}, Pierre and {Giardino}, Giovanna and {Hausen}, Ryan and {Helton}, Jakob M. and {Hviding}, Raphael E. and {Ji}, Zhiyuan and {Johnson}, Benjamin D. and {Jones}, Gareth C. and {Kumari}, Nimisha and {Laseter}, Isaac and {L{\"u}tzgendorf}, Nora and {Maseda}, Michael V. and {Nelson}, Erica and {Parlanti}, Eleonora and {Perna}, Michele and {Rauscher}, Bernard J. and {Rawle}, Tim and {Rix}, Hans-Walter and {Rieke}, Marcia and {Robertson}, Brant and {Rodr{\'\i}guez Del Pino}, Bruno and {Sandles}, Lester and {Scholtz}, Jan and {Sharpe}, Katherine and {Skarbinski}, Maya and {Stark}, Daniel P. and {Sun}, Fengwu and {Tacchella}, Sandro and {Topping}, Michael W. and {Villanueva}, Natalia C. and {Wallace}, Imaan E.~B. and {Williams}, Christina C. and {Woodrum}, Charity},
 doi = {10.1051/0004-6361/202347094},
 eid = {A288},
 hideeprint = {2306.02467},
 journal = {\aap},
 month = {October},
 pages = {A288},
 primaryclass = {astro-ph.GA},
 title = {{JADES NIRSpec initial data release for the Hubble Ultra Deep Field: Redshifts and line fluxes of distant galaxies from the deepest JWST Cycle 1 NIRSpec multi-object spectroscopy}},
 volume = {690},
 year = {2024}
}

@inproceedings{carpenter23,
 adsurl = {https://ui.adsabs.harvard.edu/abs/2023pcsf.conf..304C},
 archiveprefix = {arXiv},
 author = {{Carpenter}, John and {Brogan}, Crystal and {Iono}, Daisuke and {Mroczkowski}, Tony},
 booktitle = {Physics and Chemistry of Star Formation: The Dynamical ISM Across Time and Spatial Scales},
 doi = {10.48550/arXiv.2211.00195},
 editor = {{Ossenkopf-Okada}, V. and {Schaaf}, R. and {Breloy}, I. and {Stutzki}, J.},
 eprint = {2211.00195},
 month = {February},
 pages = {304},
 primaryclass = {astro-ph.IM},
 title = {{The ALMA Wideband Sensitivity Upgrade}},
 year = {2023}
}

@article{casey11,
 adsurl = {https://ui.adsabs.harvard.edu/abs/2011MNRAS.411.2739C},
 archiveprefix = {arXiv},
 author = {{Casey}, C.~M. and {Chapman}, S.~C. and {Smail}, Ian and {Alaghband-Zadeh}, S. and {Bothwell}, M.~S. and {Swinbank}, A.~M.},
 doi = {10.1111/j.1365-2966.2010.17876.x},
 hideeprint = {1009.5709},
 journal = {\mnras},
 month = {March},
 number = {4},
 pages = {2739-2749},
 primaryclass = {astro-ph.CO},
 title = {{Spectroscopic characterization of 250-{\ensuremath{\mu}}m-selected hyper-luminous star-forming galaxies}},
 volume = {411},
 year = {2011}
}

@article{casey13,
 adsurl = {https://ui.adsabs.harvard.edu/abs/2013MNRAS.436.1919C},
 archiveprefix = {arXiv},
 author = {{Casey}, Caitlin M. and {Chen}, Chian-Chou and {Cowie}, Lennox L. and {Barger}, Amy J. and {Capak}, Peter and {Ilbert}, Olivier and {Koss}, Michael and {Lee}, Nicholas and {Le Floc'h}, Emeric and {Sanders}, David B. and {Williams}, Jonathan P.},
 doi = {10.1093/mnras/stt1673},
 hideeprint = {1302.2619},
 journal = {\mnras},
 month = {December},
 number = {3},
 pages = {1919-1954},
 primaryclass = {astro-ph.CO},
 title = {{Characterization of SCUBA-2 450 {\ensuremath{\mu}}m and 850 {\ensuremath{\mu}}m selected galaxies in the COSMOS field}},
 volume = {436},
 year = {2013}
}

@article{casey14,
 adsurl = {https://ui.adsabs.harvard.edu/abs/2014PhR...541...45C},
 archiveprefix = {arXiv},
 author = {{Casey}, Caitlin M. and {Narayanan}, Desika and {Cooray}, Asantha},
 doi = {10.1016/j.physrep.2014.02.009},
 eprint = {1402.1456},
 journal = {\physrep},
 month = {August},
 number = {2},
 pages = {45-161},
 primaryclass = {astro-ph.CO},
 title = {{Dusty star-forming galaxies at high redshift}},
 volume = {541},
 year = {2014}
}

@article{casey16,
 adsurl = {https://ui.adsabs.harvard.edu/abs/2016ApJ...824...36C},
 archiveprefix = {arXiv},
 author = {{Casey}, Caitlin M.},
 doi = {10.3847/0004-637X/824/1/36},
 eid = {36},
 hideeprint = {1603.04437},
 journal = {\apj},
 month = {June},
 number = {1},
 pages = {36},
 primaryclass = {astro-ph.GA},
 title = {{The Ubiquity of Coeval Starbursts in Massive Galaxy Cluster Progenitors}},
 volume = {824},
 year = {2016}
}

@article{casey17,
 adsurl = {https://ui.adsabs.harvard.edu/abs/2017ApJ...840..101C},
 archiveprefix = {arXiv},
 author = {{Casey}, Caitlin M. and {Cooray}, Asantha and {Killi}, Meghana and {Capak}, Peter and {Chen}, Chian-Chou and {Hung}, Chao-Ling and {Kartaltepe}, Jeyhan and {Sanders}, D.~B. and {Scoville}, N.~Z.},
 doi = {10.3847/1538-4357/aa6cb1},
 eid = {101},
 eprint = {1703.10168},
 journal = {\apj},
 month = {May},
 number = {2},
 pages = {101},
 primaryclass = {astro-ph.GA},
 title = {{Near-infrared MOSFIRE Spectra of Dusty Star-forming Galaxies at 0.2 < z < 4}},
 volume = {840},
 year = {2017}
}

@article{casey20,
 adsurl = {https://ui.adsabs.harvard.edu/abs/2020ApJ...900...68C},
 archiveprefix = {arXiv},
 author = {{Casey}, Caitlin M.},
 doi = {10.3847/1538-4357/aba528},
 eid = {68},
 hideeprint = {2007.11012},
 journal = {\apj},
 month = {September},
 number = {1},
 pages = {68},
 primaryclass = {astro-ph.GA},
 title = {{Far-infrared Photometric Redshifts: A New Approach to a Highly Uncertain Enterprise}},
 volume = {900},
 year = {2020}
}

@article{castellano07,
 adsurl = {https://ui.adsabs.harvard.edu/abs/2007ApJ...671.1497C},
 archiveprefix = {arXiv},
 author = {{Castellano}, M. and {Salimbeni}, S. and {Trevese}, D. and {Grazian}, A. and {Pentericci}, L. and {Fiore}, F. and {Fontana}, A. and {Giallongo}, E. and {Santini}, P. and {Cristiani}, S. and {Nonino}, M. and {Vanzella}, E.},
 doi = {10.1086/521595},
 hideeprint = {0707.1783},
 journal = {\apj},
 month = {December},
 number = {2},
 pages = {1497-1502},
 primaryclass = {astro-ph},
 title = {{A Photometrically Detected Forming Cluster of Galaxies at Redshift 1.6 in the GOODS Field}},
 volume = {671},
 year = {2007}
}

@article{chapin09,
 adsurl = {https://ui.adsabs.harvard.edu/abs/2009MNRAS.398.1793C},
 archiveprefix = {arXiv},
 author = {{Chapin}, Edward L. and {Pope}, Alexandra and {Scott}, Douglas and {Aretxaga}, Itziar and {Austermann}, Jason E. and {Chary}, Ranga-Ram and {Coppin}, Kristen and {Halpern}, Mark and {Hughes}, David H. and {Lowenthal}, James D. and {Morrison}, Glenn E. and {Perera}, Thushara A. and {Scott}, Kimberly S. and {Wilson}, Grant W. and {Yun}, Min S.},
 doi = {10.1111/j.1365-2966.2009.15267.x},
 eprint = {0906.4561},
 journal = {\mnras},
 month = {October},
 number = {4},
 pages = {1793-1808},
 primaryclass = {astro-ph.CO},
 title = {{An AzTEC 1.1mm survey of the GOODS-N field - II. Multiwavelength identifications and redshift distribution}},
 volume = {398},
 year = {2009}
}

@article{chapman01,
 adsurl = {https://ui.adsabs.harvard.edu/abs/2001ApJ...548L..17C},
 archiveprefix = {arXiv},
 author = {{Chapman}, S.~C. and {Lewis}, G.~F. and {Scott}, D. and {Richards}, E. and {Borys}, C. and {Steidel}, C.~C. and {Adelberger}, K.~L. and {Shapley}, A.~E.},
 doi = {10.1086/318919},
 hideeprint = {astro-ph/0010101},
 journal = {\apjl},
 month = {February},
 number = {1},
 pages = {L17-L21},
 primaryclass = {astro-ph},
 title = {{Submillimeter Imaging of a Protocluster Region at Z=3.09}},
 volume = {548},
 year = {2001}
}

@article{chapman05,
 adsurl = {https://ui.adsabs.harvard.edu/abs/2005ApJ...622..772C},
 archiveprefix = {arXiv},
 author = {{Chapman}, S.~C. and {Blain}, A.~W. and {Smail}, Ian and {Ivison}, R.~J.},
 doi = {10.1086/428082},
 hideeprint = {astro-ph/0412573},
 journal = {\apj},
 month = {April},
 number = {2},
 pages = {772-796},
 primaryclass = {astro-ph},
 title = {{A Redshift Survey of the Submillimeter Galaxy Population}},
 volume = {622},
 year = {2005}
}

@article{chapman09,
 adsurl = {https://ui.adsabs.harvard.edu/abs/2009ApJ...691..560C},
 archiveprefix = {arXiv},
 author = {{Chapman}, S.~C. and {Blain}, A. and {Ibata}, R. and {Ivison}, R.~J. and {Smail}, I. and {Morrison}, G.},
 doi = {10.1088/0004-637X/691/1/560},
 hideeprint = {0809.1159},
 journal = {\apj},
 month = {January},
 number = {1},
 pages = {560-568},
 primaryclass = {astro-ph},
 title = {{Do Submillimeter Galaxies Really Trace the Most Massive Dark-Matter Halos? Discovery of a High-z Cluster in a Highly Active Phase of Evolution}},
 volume = {691},
 year = {2009}
}

@article{chen22,
 adsurl = {https://ui.adsabs.harvard.edu/abs/2022ApJ...929..159C},
 archiveprefix = {arXiv},
 author = {{Chen}, Chian-Chou and {Liao}, Cheng-Lin and {Smail}, Ian and {Swinbank}, A.~M. and {Ao}, Y. and {Bunker}, A.~J. and {Chapman}, S.~C. and {Hatsukade}, B. and {Ivison}, R.~J. and {Lee}, Minju M. and {Serjeant}, Stephen and {Umehata}, Hideki and {Wang}, Wei-Hao and {Zhao}, Y.},
 doi = {10.3847/1538-4357/ac61df},
 eid = {159},
 hideeprint = {2112.07430},
 journal = {\apj},
 month = {April},
 number = {2},
 pages = {159},
 primaryclass = {astro-ph.GA},
 title = {{An ALMA Spectroscopic Survey of the Brightest Submillimeter Galaxies in the SCUBA-2-COSMOS Field (AS2COSPEC): Survey Description and First Results}},
 volume = {929},
 year = {2022}
}

@article{cooper25,
 adsurl = {https://ui.adsabs.harvard.edu/abs/2025ApJ...982..125C},
 archiveprefix = {arXiv},
 author = {{Cooper}, Olivia R. and {Brammer}, Gabriel and {Heintz}, Kasper E. and {Toft}, Sune and {Casey}, Caitlin M. and {Setton}, David J. and {de Graaff}, Anna and {Boogaard}, Leindert and {Cleri}, Nikko J. and {Gillman}, Steven and {Gottumukkala}, Rashmi and {Greene}, Jenny E. and {Gullberg}, Bitten and {Hirschmann}, Michaela and {Hviding}, Raphael E. and {Lambrides}, Erini and {Leja}, Joel and {Long}, Arianna S. and {Manning}, Sinclaire M. and {Maseda}, Michael V. and {McConachie}, Ian and {McKinney}, Jed and {Narayanan}, Desika and {Price}, Sedona H. and {Strait}, Victoria and {Suess}, Katherine A. and {Weibel}, Andrea and {Williams}, Christina C.},
 doi = {10.3847/1538-4357/adb8e1},
 eid = {125},
 eprint = {2410.08387},
 journal = {\apj},
 month = {April},
 number = {2},
 pages = {125},
 primaryclass = {astro-ph.GA},
 title = {{RUBIES: JWST/NIRSpec Resolves Evolutionary Phases of Dusty Star-forming Galaxies at z {\ensuremath{\sim}} 2}},
 volume = {982},
 year = {2025}
}

@article{cowie17,
 adsurl = {https://ui.adsabs.harvard.edu/abs/2017ApJ...837..139C},
 archiveprefix = {arXiv},
 author = {{Cowie}, L.~L. and {Barger}, A.~J. and {Hsu}, L.-Y. and {Chen}, Chian-Chou and {Owen}, F.~N. and {Wang}, W.-H.},
 doi = {10.3847/1538-4357/aa60bb},
 eid = {139},
 hideeprint = {1702.03002},
 journal = {\apj},
 month = {March},
 number = {2},
 pages = {139},
 primaryclass = {astro-ph.GA},
 title = {{A Submillimeter Perspective on the GOODS Fields (SUPER GOODS). I. An Ultradeep SCUBA-2 Survey of the GOODS-N}},
 volume = {837},
 year = {2017}
}

@article{cowie18,
 adsurl = {https://ui.adsabs.harvard.edu/abs/2018ApJ...865..106C},
 archiveprefix = {arXiv},
 author = {{Cowie}, L.~L. and {Gonz{\'a}lez-L{\'o}pez}, J. and {Barger}, A.~J. and {Bauer}, F.~E. and {Hsu}, L.-Y. and {Wang}, W.-H.},
 doi = {10.3847/1538-4357/aadc63},
 eid = {106},
 hideeprint = {1805.09424},
 journal = {\apj},
 month = {October},
 number = {2},
 pages = {106},
 primaryclass = {astro-ph.GA},
 title = {{A Submillimeter Perspective on the GOODS Fields (SUPER GOODS). III. A Large Sample of ALMA Sources in the GOODS-S}},
 volume = {865},
 year = {2018}
}

@article{cowie23,
 adsurl = {https://ui.adsabs.harvard.edu/abs/2023ApJ...952...28C},
 archiveprefix = {arXiv},
 author = {{Cowie}, L.~L. and {Barger}, A.~J. and {Bauer}, F.~E.},
 doi = {10.3847/1538-4357/acd763},
 eid = {28},
 hideeprint = {2305.17167},
 journal = {\apj},
 month = {July},
 number = {1},
 pages = {28},
 primaryclass = {astro-ph.GA},
 title = {{2 mm Observations and the Search for High-redshift Dusty Star-forming Galaxies}},
 volume = {952},
 year = {2023}
}

@article{curtis-lake25,
 adsurl = {https://ui.adsabs.harvard.edu/abs/2026MNRAS.tmp..935C},
 archiveprefix = {arXiv},
 author = {{Curtis-Lake}, Emma and {Cameron}, Alex J. and {Bunker}, Andrew J. and {Scholtz}, Jan and {Carniani}, Stefano and {Parlanti}, Eleonora and {D'Eugenio}, Francesco and {Jakobsen}, Peter and {Willmer}, Christopher N.~A. and {Arribas}, Santiago and {Baker}, William M. and {Charlot}, St{\'e}phane and {Chevallard}, Jacopo and {Circosta}, Chiara and {Curti}, Mirko and {Duan}, Qiao and {Eisenstein}, Daniel J. and {Hainline}, Kevin and {Ji}, Zhiyuan and {Johnson}, Benjamin D. and {Jones}, Gareth C. and {Maiolino}, Roberto and {Maseda}, Michael V. and {Perna}, Michele and {P{\'e}rez-Gonz{\'a}lez}, Pablo G. and {Rawle}, Tim and {Rieke}, Marcia and {Rinaldi}, Pierluigi and {Robertson}, Brant and {Rodr{\'\i}guez Del Pino}, Bruno and {Saxena}, Aayush and {Shivaei}, Irene and {Smit}, Renske and {Tacchella}, Sandro and {{\"U}bler}, Hannah and {Venturi}, Giacomo and {Williams}, Christina C. and {Willott}, Chris},
 doi = {10.1093/mnras/stag836},
 eprint = {2510.01033},
 journal = {\mnras},
 month = {May},
 primaryclass = {astro-ph.GA},
 title = {{JADES Data Release 4 Paper I: Sample Selection, Observing Strategy and Redshifts of the complete spectroscopic sample}},
 year = {2026}
}

@article{dacunha08,
 adsurl = {https://ui.adsabs.harvard.edu/abs/2008MNRAS.388.1595D},
 archiveprefix = {arXiv},
 author = {{da Cunha}, Elisabete and {Charlot}, St{\'e}phane and {Elbaz}, David},
 doi = {10.1111/j.1365-2966.2008.13535.x},
 hideeprint = {0806.1020},
 journal = {\mnras},
 month = {August},
 number = {4},
 pages = {1595-1617},
 primaryclass = {astro-ph},
 title = {{A simple model to interpret the ultraviolet, optical and infrared emission from galaxies}},
 volume = {388},
 year = {2008}
}

@article{dacunha15,
 adsurl = {https://ui.adsabs.harvard.edu/abs/2015ApJ...806..110D},
 archiveprefix = {arXiv},
 author = {{da Cunha}, E. and {Walter}, F. and {Smail}, I.~R. and {Swinbank}, A.~M. and {Simpson}, J.~M. and {Decarli}, R. and {Hodge}, J.~A. and {Weiss}, A. and {van der Werf}, P.~P. and {Bertoldi}, F. and {Chapman}, S.~C. and {Cox}, P. and {Danielson}, A.~L.~R. and {Dannerbauer}, H. and {Greve}, T.~R. and {Ivison}, R.~J. and {Karim}, A. and {Thomson}, A.},
 doi = {10.1088/0004-637X/806/1/110},
 eid = {110},
 hideeprint = {1504.04376},
 journal = {\apj},
 month = {June},
 number = {1},
 pages = {110},
 primaryclass = {astro-ph.GA},
 title = {{An ALMA Survey of Sub-millimeter Galaxies in the Extended Chandra Deep Field South: Physical Properties Derived from Ultraviolet-to-radio Modeling}},
 volume = {806},
 year = {2015}
}

@article{daddi15,
 adsurl = {https://ui.adsabs.harvard.edu/abs/2015A&A...577A..46D},
 archiveprefix = {arXiv},
 author = {{Daddi}, E. and {Dannerbauer}, H. and {Liu}, D. and {Aravena}, M. and {Bournaud}, F. and {Walter}, F. and {Riechers}, D. and {Magdis}, G. and {Sargent}, M. and {B{\'e}thermin}, M. and {Carilli}, C. and {Cibinel}, A. and {Dickinson}, M. and {Elbaz}, D. and {Gao}, Y. and {Gobat}, R. and {Hodge}, J. and {Krips}, M.},
 doi = {10.1051/0004-6361/201425043},
 eid = {A46},
 eprint = {1409.8158},
 journal = {\aap},
 month = {May},
 pages = {A46},
 primaryclass = {astro-ph.GA},
 title = {{CO excitation of normal star-forming galaxies out to z = 1.5 as regulated by the properties of their interstellar medium}},
 volume = {577},
 year = {2015}
}

@article{danielson17,
 adsurl = {https://ui.adsabs.harvard.edu/abs/2017ApJ...840...78D},
 archiveprefix = {arXiv},
 author = {{Danielson}, A.~L.~R. and {Swinbank}, A.~M. and {Smail}, Ian and {Simpson}, J.~M. and {Casey}, C.~M. and {Chapman}, S.~C. and {da Cunha}, E. and {Hodge}, J.~A. and {Walter}, F. and {Wardlow}, J.~L. and {Alexander}, D.~M. and {Brandt}, W.~N. and {de Breuck}, C. and {Coppin}, K.~E.~K. and {Dannerbauer}, H. and {Dickinson}, M. and {Edge}, A.~C. and {Gawiser}, E. and {Ivison}, R.~J. and {Karim}, A. and {Kovacs}, A. and {Lutz}, D. and {Menten}, K. and {Schinnerer}, E. and {Wei{\ss}}, A. and {van der Werf}, P.},
 doi = {10.3847/1538-4357/aa6caf},
 eid = {78},
 hideeprint = {1705.03503},
 journal = {\apj},
 month = {May},
 number = {2},
 pages = {78},
 primaryclass = {astro-ph.GA},
 title = {{An ALMA Survey of Submillimeter Galaxies in the Extended Chandra  Deep Field South: Spectroscopic Redshifts}},
 volume = {840},
 year = {2017}
}

@article{degraaff25,
 adsurl = {https://ui.adsabs.harvard.edu/abs/2025A&A...697A.189D},
 archiveprefix = {arXiv},
 author = {{de Graaff}, Anna and {Brammer}, Gabriel and {Weibel}, Andrea and {Lewis}, Zach and {Maseda}, Michael V. and {Oesch}, Pascal A. and {Bezanson}, Rachel and {Boogaard}, Leindert A. and {Cleri}, Nikko J. and {Cooper}, Olivia R. and {Gottumukkala}, Rashmi and {Greene}, Jenny E. and {Hirschmann}, Michaela and {Hviding}, Raphael E. and {Katz}, Harley and {Labb{\'e}}, Ivo and {Leja}, Joel and {Matthee}, Jorryt and {McConachie}, Ian and {Miller}, Tim B. and {Naidu}, Rohan P. and {Price}, Sedona H. and {Rix}, Hans-Walter and {Setton}, David J. and {Suess}, Katherine A. and {Wang}, Bingjie and {Whitaker}, Katherine E. and {Williams}, Christina C.},
 doi = {10.1051/0004-6361/202452186},
 eid = {A189},
 hideeprint = {2409.05948},
 journal = {\aap},
 month = {May},
 pages = {A189},
 primaryclass = {astro-ph.GA},
 title = {{RUBIES: A complete census of the bright and red distant Universe with JWST/NIRSpec}},
 volume = {697},
 year = {2025}
}

@article{deugenio25,
 adsurl = {https://ui.adsabs.harvard.edu/abs/2025ApJS..277....4D},
 archiveprefix = {arXiv},
 author = {{D'Eugenio}, Francesco and {Cameron}, Alex J. and {Scholtz}, Jan and {Carniani}, Stefano and {Willott}, Chris J. and {Curtis-Lake}, Emma and {Bunker}, Andrew J. and {Parlanti}, Eleonora and {Maiolino}, Roberto and {Willmer}, Christopher N.~A. and {Jakobsen}, Peter and {Robertson}, Brant E. and {Johnson}, Benjamin D. and {Tacchella}, Sandro and {Cargile}, Phillip A. and {Rawle}, Tim and {Arribas}, Santiago and {Chevallard}, Jacopo and {Curti}, Mirko and {Egami}, Eiichi and {Eisenstein}, Daniel J. and {Kumari}, Nimisha and {Looser}, Tobias J. and {Rieke}, Marcia J. and {Rodr{\'\i}guez Del Pino}, Bruno and {Saxena}, Aayush and {{\"U}bler}, Hannah and {Venturi}, Giacomo and {Witstok}, Joris and {Baker}, William M. and {Bhatawdekar}, Rachana and {Bonaventura}, Nina and {Boyett}, Kristan and {Charlot}, Stephane and {Danhaive}, A. Lola and {Hainline}, Kevin N. and {Hausen}, Ryan and {Helton}, Jakob M. and {Ji}, Xihan and {Ji}, Zhiyuan and {Jones}, Gareth C. and {Juod{\v{z}}balis}, Ignas and {Maseda}, Michael V. and {P{\'e}rez-Gonz{\'a}lez}, Pablo G. and {Perna}, Michele and {Pusk{\'a}s}, D{\'a}vid and {Shivaei}, Irene and {Silcock}, Maddie S. and {Simmonds}, Charlotte and {Smit}, Renske and {Sun}, Fengwu and {Villanueva}, Natalia C. and {Williams}, Christina C. and {Zhu}, Yongda},
 doi = {10.3847/1538-4365/ada148},
 eid = {4},
 hideeprint = {2404.06531},
 journal = {\apjs},
 month = {March},
 number = {1},
 pages = {4},
 primaryclass = {astro-ph.GA},
 title = {{JADES Data Release 3: NIRSpec/Microshutter Assembly Spectroscopy for 4000 Galaxies in the GOODS Fields}},
 volume = {277},
 year = {2025}
}

@article{drew22,
 adsurl = {https://ui.adsabs.harvard.edu/abs/2022ApJ...930..142D},
 archiveprefix = {arXiv},
 author = {{Drew}, Patrick M. and {Casey}, Caitlin M.},
 doi = {10.3847/1538-4357/ac6270},
 eid = {142},
 hideeprint = {2203.16655},
 journal = {\apj},
 month = {May},
 number = {2},
 pages = {142},
 primaryclass = {astro-ph.GA},
 title = {{No Redshift Evolution of Galaxies' Dust Temperatures Seen from 0 < z < 2}},
 volume = {930},
 year = {2022}
}

@article{dudzeviciute20,
 adsurl = {https://ui.adsabs.harvard.edu/abs/2020MNRAS.494.3828D},
 archiveprefix = {arXiv},
 author = {{Dudzevi{\v{c}}i{\={u}}t{\.{e}}}, U. and {Smail}, Ian and {Swinbank}, A.~M. and {Stach}, S.~M. and {Almaini}, O. and {da Cunha}, E. and {An}, Fang Xia and {Arumugam}, V. and {Birkin}, J. and {Blain}, A.~W. and {Chapman}, S.~C. and {Chen}, C.-C. and {Conselice}, C.~J. and {Coppin}, K.~E.~K. and {Dunlop}, J.~S. and {Farrah}, D. and {Geach}, J.~E. and {Gullberg}, B. and {Hartley}, W.~G. and {Hodge}, J.~A. and {Ivison}, R.~J. and {Maltby}, D.~T. and {Scott}, D. and {Simpson}, C.~J. and {Simpson}, J.~M. and {Thomson}, A.~P. and {Walter}, F. and {Wardlow}, J.~L. and {Weiss}, A. and {van der Werf}, P.},
 doi = {10.1093/mnras/staa769},
 hideeprint = {1910.07524},
 journal = {\mnras},
 month = {May},
 number = {3},
 pages = {3828-3860},
 primaryclass = {astro-ph.GA},
 title = {{An ALMA survey of the SCUBA-2 CLS UDS field: physical properties of 707 sub-millimetre galaxies}},
 volume = {494},
 year = {2020}
}

@article{eisenstein26,
 adsurl = {https://ui.adsabs.harvard.edu/abs/2026ApJS..283....6E},
 archiveprefix = {arXiv},
 author = {{Eisenstein}, Daniel J. and {Willott}, Chris and {Alberts}, Stacey and {Arribas}, Santiago and {Bonaventura}, Nina and {Bunker}, Andrew J. and {Cameron}, Alex J. and {Carniani}, Stefano and {Charlot}, Stephane and {Curtis-Lake}, Emma and {D'Eugenio}, Francesco and {Ferruit}, Pierre and {Giardino}, Giovanna and {Hainline}, Kevin and {Hausen}, Ryan and {Jakobsen}, Peter and {Johnson}, Benjamin D. and {Maiolino}, Roberto and {Rauscher}, Bernard J. and {Rieke}, Marcia and {Rieke}, George and {Rix}, Hans-Walter and {Robertson}, Brant and {Stark}, Daniel P. and {Tacchella}, Sandro and {Williams}, Christina C. and {Willmer}, Christopher N.~A. and {Baker}, William M. and {Baum}, Stefi and {Bhatawdekar}, Rachana and {Boyett}, Kristan and {Chen}, Zuyi and {Chevallard}, Jacopo and {Circosta}, Chiara and {Curti}, Mirko and {Danhaive}, A. Lola and {DeCoursey}, Christa and {Endsley}, Ryan and {de Graaff}, Anna and {Dressler}, Alan and {Egami}, Eiichi and {Helton}, Jakob M. and {Hviding}, Raphael E. and {Ji}, Zhiyuan and {Jones}, Gareth C. and {Kumari}, Nimisha and {L{\"u}tzgendorf}, Nora and {Laseter}, Isaac and {Looser}, Tobias J. and {Lyu}, Jianwei and {Maseda}, Michael V. and {Nelson}, Erica and {Parlanti}, Eleonora and {Perna}, Michele and {Pusk{\'a}s}, D{\'a}vid and {Rawle}, Tim and {Rodr{\'\i}guez Del Pino}, Bruno and {Rujopakarn}, Wiphu and {Sandles}, Lester and {Saxena}, Aayush and {Scholtz}, Jan and {Sharpe}, Katherine and {Shivaei}, Irene and {Silcock}, Maddie S. and {Simmonds}, Charlotte and {Skarbinski}, Maya and {Smit}, Renske and {Stone}, Meredith and {Suess}, Katherine A. and {Sun}, Fengwu and {Tang}, Mengtao and {Topping}, Michael W. and {{\"U}bler}, Hannah and {Villanueva}, Natalia C. and {Wallace}, Imaan E.~B. and {Whitler}, Lily and {Witstok}, Joris and {Woodrum}, Charity},
 doi = {10.3847/1538-4365/ae3163},
 eid = {6},
 hideeprint = {2306.02465},
 journal = {\apjs},
 month = {March},
 number = {1},
 pages = {6},
 primaryclass = {astro-ph.GA},
 title = {{Overview of the JWST Advanced Deep Extragalactic Survey (JADES)}},
 volume = {283},
 year = {2026}
}

@article{elbaz11,
 adsurl = {https://ui.adsabs.harvard.edu/abs/2011A&A...533A.119E},
 archiveprefix = {arXiv},
 author = {{Elbaz}, D. and {Dickinson}, M. and {Hwang}, H.~S. and {D{\'\i}az-Santos}, T. and {Magdis}, G. and {Magnelli}, B. and {Le Borgne}, D. and {Galliano}, F. and {Pannella}, M. and {Chanial}, P. and {Armus}, L. and {Charmandaris}, V. and {Daddi}, E. and {Aussel}, H. and {Popesso}, P. and {Kartaltepe}, J. and {Altieri}, B. and {Valtchanov}, I. and {Coia}, D. and {Dannerbauer}, H. and {Dasyra}, K. and {Leiton}, R. and {Mazzarella}, J. and {Alexander}, D.~M. and {Buat}, V. and {Burgarella}, D. and {Chary}, R.-R. and {Gilli}, R. and {Ivison}, R.~J. and {Juneau}, S. and {Le Floc'h}, E. and {Lutz}, D. and {Morrison}, G.~E. and {Mullaney}, J.~R. and {Murphy}, E. and {Pope}, A. and {Scott}, D. and {Brodwin}, M. and {Calzetti}, D. and {Cesarsky}, C. and {Charlot}, S. and {Dole}, H. and {Eisenhardt}, P. and {Ferguson}, H.~C. and {F{\"o}rster Schreiber}, N. and {Frayer}, D. and {Giavalisco}, M. and {Huynh}, M. and {Koekemoer}, A.~M. and {Papovich}, C. and {Reddy}, N. and {Surace}, C. and {Teplitz}, H. and {Yun}, M.~S. and {Wilson}, G.},
 doi = {10.1051/0004-6361/201117239},
 eid = {A119},
 hideeprint = {1105.2537},
 journal = {\aap},
 month = {September},
 pages = {A119},
 primaryclass = {astro-ph.CO},
 title = {{GOODS-Herschel: an infrared main sequence for star-forming galaxies}},
 volume = {533},
 year = {2011}
}

@article{elliott25,
 adsurl = {https://ui.adsabs.harvard.edu/abs/2026MNRAS.548ag564E},
 archiveprefix = {arXiv},
 author = {{Elliott}, Edward J. and {Baugh}, Carlton M. and {Lacey}, Cedric G.},
 doi = {10.1093/mnras/stag564},
 eid = {stag564},
 eprint = {2511.12550},
 journal = {\mnras},
 month = {May},
 number = {2},
 pages = {stag564},
 primaryclass = {astro-ph.GA},
 title = {{Sub-millimetre galaxies in hierarchical models: revisiting the need for a top-heavy stellar initial mass function with Bayesian optimization}},
 volume = {548},
 year = {2026}
}

@article{foreman13,
 adsurl = {https://ui.adsabs.harvard.edu/abs/2013PASP..125..306F},
 archiveprefix = {arXiv},
 author = {{Foreman-Mackey}, Daniel and {Hogg}, David W. and {Lang}, Dustin and {Goodman}, Jonathan},
 doi = {10.1086/670067},
 hideeprint = {1202.3665},
 journal = {\pasp},
 month = {March},
 number = {925},
 pages = {306},
 primaryclass = {astro-ph.IM},
 title = {{emcee: The MCMC Hammer}},
 volume = {125},
 year = {2013}
}

@article{franco18,
 adsurl = {https://ui.adsabs.harvard.edu/abs/2018A&A...620A.152F},
 archiveprefix = {arXiv},
 author = {{Franco}, M. and {Elbaz}, D. and {B{\'e}thermin}, M. and {Magnelli}, B. and {Schreiber}, C. and {Ciesla}, L. and {Dickinson}, M. and {Nagar}, N. and {Silverman}, J. and {Daddi}, E. and {Alexander}, D.~M. and {Wang}, T. and {Pannella}, M. and {Le Floc'h}, E. and {Pope}, A. and {Giavalisco}, M. and {Maury}, A.~J. and {Bournaud}, F. and {Chary}, R. and {Demarco}, R. and {Ferguson}, H. and {Finkelstein}, S.~L. and {Inami}, H. and {Iono}, D. and {Juneau}, S. and {Lagache}, G. and {Leiton}, R. and {Lin}, L. and {Magdis}, G. and {Messias}, H. and {Motohara}, K. and {Mullaney}, J. and {Okumura}, K. and {Papovich}, C. and {Pforr}, J. and {Rujopakarn}, W. and {Sargent}, M. and {Shu}, X. and {Zhou}, L.},
 doi = {10.1051/0004-6361/201832928},
 eid = {A152},
 eprint = {1803.00157},
 journal = {\aap},
 month = {December},
 pages = {A152},
 primaryclass = {astro-ph.GA},
 title = {{GOODS-ALMA: 1.1 mm galaxy survey. I. Source catalog and optically dark galaxies}},
 volume = {620},
 year = {2018}
}

@article{garilli21,
 adsurl = {https://ui.adsabs.harvard.edu/abs/2021A&A...647A.150G},
 archiveprefix = {arXiv},
 author = {{Garilli}, B. and {McLure}, R. and {Pentericci}, L. and {Franzetti}, P. and {Gargiulo}, A. and {Carnall}, A. and {Cucciati}, O. and {Iovino}, A. and {Amorin}, R. and {Bolzonella}, M. and {Bongiorno}, A. and {Castellano}, M. and {Cimatti}, A. and {Cirasuolo}, M. and {Cullen}, F. and {Dunlop}, J. and {Elbaz}, D. and {Finkelstein}, S. and {Fontana}, A. and {Fontanot}, F. and {Fumana}, M. and {Guaita}, L. and {Hartley}, W. and {Jarvis}, M. and {Juneau}, S. and {Maccagni}, D. and {McLeod}, D. and {Nandra}, K. and {Pompei}, E. and {Pozzetti}, L. and {Scodeggio}, M. and {Talia}, M. and {Calabr{\`o}}, A. and {Cresci}, G. and {Fynbo}, J.~P.~U. and {Hathi}, N.~P. and {Hibon}, P. and {Koekemoer}, A.~M. and {Magliocchetti}, M. and {Salvato}, M. and {Vietri}, G. and {Zamorani}, G. and {Almaini}, O. and {Balestra}, I. and {Bardelli}, S. and {Begley}, R. and {Brammer}, G. and {Bell}, E.~F. and {Bowler}, R.~A.~A. and {Brusa}, M. and {Buitrago}, F. and {Caputi}, C. and {Cassata}, P. and {Charlot}, S. and {Citro}, A. and {Cristiani}, S. and {Curtis-Lake}, E. and {Dickinson}, M. and {Fazio}, G. and {Ferguson}, H.~C. and {Fiore}, F. and {Franco}, M. and {Georgakakis}, A. and {Giavalisco}, M. and {Grazian}, A. and {Hamadouche}, M. and {Jung}, I. and {Kim}, S. and {Khusanova}, Y. and {Le F{\`e}vre}, O. and {Longhetti}, M. and {Lotz}, J. and {Mannucci}, F. and {Maltby}, D. and {Matsuoka}, K. and {Mendez-Hernandez}, H. and {Mendez-Abreu}, J. and {Mignoli}, M. and {Moresco}, M. and {Nonino}, M. and {Pannella}, M. and {Papovich}, C. and {Popesso}, P. and {Roberts-Borsani}, G. and {Rosario}, D.~J. and {Saldana-Lopez}, A. and {Santini}, P. and {Saxena}, A. and {Schaerer}, D. and {Schreiber}, C. and {Stark}, D. and {Tasca}, L.~A.~M. and {Thomas}, R. and {Vanzella}, E. and {Wild}, V. and {Williams}, C. and {Zucca}, E.},
 doi = {10.1051/0004-6361/202040059},
 eid = {A150},
 eprint = {2101.07645},
 journal = {\aap},
 month = {March},
 pages = {A150},
 primaryclass = {astro-ph.GA},
 title = {{The VANDELS ESO public spectroscopic survey. Final data release of 2087 spectra and spectroscopic measurements}},
 volume = {647},
 year = {2021}
}

@article{gehrels86,
 adsurl = {https://ui.adsabs.harvard.edu/abs/1986ApJ...303..336G},
 author = {{Gehrels}, N.},
 doi = {10.1086/164079},
 journal = {\apj},
 month = {April},
 pages = {336},
 title = {{Confidence Limits for Small Numbers of Events in Astrophysical Data}},
 volume = {303},
 year = {1986}
}

@article{giavalisco04,
 adsurl = {https://ui.adsabs.harvard.edu/abs/2004ApJ...600L..93G},
 archiveprefix = {arXiv},
 author = {{Giavalisco}, M. and {Ferguson}, H.~C. and {Koekemoer}, A.~M. and {Dickinson}, M. and {Alexander}, D.~M. and {Bauer}, F.~E. and {Bergeron}, J. and {Biagetti}, C. and {Brandt}, W.~N. and {Casertano}, S. and {Cesarsky}, C. and {Chatzichristou}, E. and {Conselice}, C. and {Cristiani}, S. and {Da Costa}, L. and {Dahlen}, T. and {de Mello}, D. and {Eisenhardt}, P. and {Erben}, T. and {Fall}, S.~M. and {Fassnacht}, C. and {Fosbury}, R. and {Fruchter}, A. and {Gardner}, J.~P. and {Grogin}, N. and {Hook}, R.~N. and {Hornschemeier}, A.~E. and {Idzi}, R. and {Jogee}, S. and {Kretchmer}, C. and {Laidler}, V. and {Lee}, K.~S. and {Livio}, M. and {Lucas}, R. and {Madau}, P. and {Mobasher}, B. and {Moustakas}, L.~A. and {Nonino}, M. and {Padovani}, P. and {Papovich}, C. and {Park}, Y. and {Ravindranath}, S. and {Renzini}, A. and {Richardson}, M. and {Riess}, A. and {Rosati}, P. and {Schirmer}, M. and {Schreier}, E. and {Somerville}, R.~S. and {Spinrad}, H. and {Stern}, D. and {Stiavelli}, M. and {Strolger}, L. and {Urry}, C.~M. and {Vandame}, B. and {Williams}, R. and {Wolf}, C.},
 doi = {10.1086/379232},
 eprint = {astro-ph/0309105},
 journal = {\apjl},
 month = {January},
 number = {2},
 pages = {L93-L98},
 primaryclass = {astro-ph},
 title = {{The Great Observatories Origins Deep Survey: Initial Results from Optical and Near-Infrared Imaging}},
 volume = {600},
 year = {2004}
}

@article{gillman23,
 adsurl = {https://ui.adsabs.harvard.edu/abs/2023A&A...676A..26G},
 archiveprefix = {arXiv},
 author = {{Gillman}, Steven and {Gullberg}, Bitten and {Brammer}, Gabe and {Vijayan}, Aswin P. and {Lee}, Minju and {Bl{\'a}nquez}, David and {Brinch}, Malte and {Greve}, Thomas R. and {Jermann}, Iris and {Jin}, Shuowen and {Kokorev}, Vasily and {Liu}, Lijie and {Magdis}, Georgios and {Rizzo}, Francesca and {Valentino}, Francesco},
 doi = {10.1051/0004-6361/202346531},
 eid = {A26},
 eprint = {2303.17246},
 journal = {\aap},
 month = {August},
 pages = {A26},
 primaryclass = {astro-ph.GA},
 title = {{Sub-millimetre galaxies with Webb. Near-infrared counterparts and multi-wavelength morphology}},
 volume = {676},
 year = {2023}
}

@article{gillman24,
 adsurl = {https://ui.adsabs.harvard.edu/abs/2024A&A...691A.299G},
 archiveprefix = {arXiv},
 author = {{Gillman}, Steven and {Smail}, Ian and {Gullberg}, Bitten and {Swinbank}, A.~M. and {Vijayan}, Aswin P. and {Lee}, Minju and {Brammer}, Gabe and {Dudzevi{\v{c}}i{\={u}}t{\.{e}}}, Ugn{\.{e}} and {Greve}, Thomas R. and {Almaini}, Omar and {Brinch}, Malte and {Chapman}, Scott C. and {Chen}, Chian-Chou and {Ikarashi}, Soh and {Matsuda}, Yuichi and {Wang}, Wei-Hao and {Walter}, Fabian and {van der Werf}, Paul P.},
 doi = {10.1051/0004-6361/202451006},
 eid = {A299},
 eprint = {2406.03544},
 journal = {\aap},
 month = {November},
 pages = {A299},
 primaryclass = {astro-ph.GA},
 title = {{The structure of massive star-forming galaxies from JWST and ALMA: Dusty, high-redshift disc galaxies}},
 volume = {691},
 year = {2024}
}

@article{gomez-guijarro22,
 adsurl = {https://ui.adsabs.harvard.edu/abs/2022A&A...658A..43G},
 archiveprefix = {arXiv},
 author = {{G{\'o}mez-Guijarro}, C. and {Elbaz}, D. and {Xiao}, M. and {B{\'e}thermin}, M. and {Franco}, M. and {Magnelli}, B. and {Daddi}, E. and {Dickinson}, M. and {Demarco}, R. and {Inami}, H. and {Rujopakarn}, W. and {Magdis}, G.~E. and {Shu}, X. and {Chary}, R. and {Zhou}, L. and {Alexander}, D.~M. and {Bournaud}, F. and {Ciesla}, L. and {Ferguson}, H.~C. and {Finkelstein}, S.~L. and {Giavalisco}, M. and {Iono}, D. and {Juneau}, S. and {Kartaltepe}, J.~S. and {Lagache}, G. and {Le Floc'h}, E. and {Leiton}, R. and {Lin}, L. and {Motohara}, K. and {Mullaney}, J. and {Okumura}, K. and {Pannella}, M. and {Papovich}, C. and {Pope}, A. and {Sargent}, M.~T. and {Silverman}, J.~D. and {Treister}, E. and {Wang}, T.},
 doi = {10.1051/0004-6361/202141615},
 eid = {A43},
 hideeprint = {2106.13246},
 journal = {\aap},
 month = {February},
 pages = {A43},
 primaryclass = {astro-ph.GA},
 title = {{GOODS-ALMA 2.0: Source catalog, number counts, and prevailing compact sizes in 1.1 mm galaxies}},
 volume = {658},
 year = {2022}
}

@article{gonzalez-lopez19,
 adsurl = {https://ui.adsabs.harvard.edu/abs/2019ApJ...882..139G},
 archiveprefix = {arXiv},
 author = {{Gonz{\'a}lez-L{\'o}pez}, Jorge and {Decarli}, Roberto and {Pavesi}, Riccardo and {Walter}, Fabian and {Aravena}, Manuel and {Carilli}, Chris and {Boogaard}, Leindert and {Popping}, Gerg{\"o} and {Weiss}, Axel and {Assef}, Roberto J. and {Bauer}, Franz Erik and {Bertoldi}, Frank and {Bouwens}, Richard and {Contini}, Thierry and {Cortes}, Paulo C. and {Cox}, Pierre and {da Cunha}, Elisabete and {Daddi}, Emanuele and {D{\'\i}az-Santos}, Tanio and {Inami}, Hanae and {Hodge}, Jacqueline and {Ivison}, Rob and {Le F{\`e}vre}, Olivier and {Magnelli}, Benjamin and {Oesch}, Pascal and {Riechers}, Dominik and {Rix}, Hans-Walter and {Smail}, Ian and {Swinbank}, A.~M. and {Somerville}, Rachel S. and {Uzgil}, Bade and {van der Werf}, Paul},
 doi = {10.3847/1538-4357/ab3105},
 eid = {139},
 hideeprint = {1903.09161},
 journal = {\apj},
 month = {September},
 number = {2},
 pages = {139},
 primaryclass = {astro-ph.GA},
 title = {{The Atacama Large Millimeter/submillimeter Array Spectroscopic Survey in the Hubble Ultra Deep Field: CO Emission Lines and 3 mm Continuum Sources}},
 volume = {882},
 year = {2019}
}

@article{gruppioni20,
 adsurl = {https://ui.adsabs.harvard.edu/abs/2020A&A...643A...8G},
 archiveprefix = {arXiv},
 author = {{Gruppioni}, C. and {B{\'e}thermin}, M. and {Loiacono}, F. and {Le F{\`e}vre}, O. and {Capak}, P. and {Cassata}, P. and {Faisst}, A.~L. and {Schaerer}, D. and {Silverman}, J. and {Yan}, L. and {Bardelli}, S. and {Boquien}, M. and {Carraro}, R. and {Cimatti}, A. and {Dessauges-Zavadsky}, M. and {Ginolfi}, M. and {Fujimoto}, S. and {Hathi}, N.~P. and {Jones}, G.~C. and {Khusanova}, Y. and {Koekemoer}, A.~M. and {Lagache}, G. and {Lemaux}, B.~C. and {Oesch}, P.~A. and {Pozzi}, F. and {Riechers}, D.~A. and {Rodighiero}, G. and {Romano}, M. and {Talia}, M. and {Vallini}, L. and {Vergani}, D. and {Zamorani}, G. and {Zucca}, E.},
 doi = {10.1051/0004-6361/202038487},
 eid = {A8},
 eprint = {2006.04974},
 journal = {\aap},
 month = {November},
 pages = {A8},
 primaryclass = {astro-ph.GA},
 title = {{The ALPINE-ALMA [CII] survey. The nature, luminosity function, and star formation history of dusty galaxies up to z ≃ 6}},
 volume = {643},
 year = {2020}
}

@article{guaita20,
 adsurl = {https://ui.adsabs.harvard.edu/abs/2020A&A...640A.107G},
 archiveprefix = {arXiv},
 author = {{Guaita}, L. and {Pompei}, E. and {Castellano}, M. and {Pentericci}, L. and {Cucciati}, O. and {Zamorani}, G. and {Zoldan}, A. and {Fontanot}, F. and {Bauer}, F.~E. and {Amorin}, R. and {Bolzonella}, M. and {de Lucia}, G. and {Gargiulo}, A. and {Hathi}, N.~P. and {Hibon}, P. and {Hirschmann}, M. and {Koekemoer}, A.~M. and {McLure}, R. and {Pozzetti}, L. and {Talia}, M. and {Thomas}, R. and {Xie}, L.},
 doi = {10.1051/0004-6361/201935855},
 eid = {A107},
 hideeprint = {2007.12314},
 journal = {\aap},
 month = {August},
 pages = {A107},
 primaryclass = {astro-ph.GA},
 title = {{The VANDELS survey: Discovery of massive overdensities of galaxies at z > 2. Location of Ly{\ensuremath{\alpha}}-emitting galaxies with respect to environment}},
 volume = {640},
 year = {2020}
}

@article{guo13,
 adsurl = {https://ui.adsabs.harvard.edu/abs/2013ApJS..207...24G},
 archiveprefix = {arXiv},
 author = {{Guo}, Yicheng and {Ferguson}, Henry C. and {Giavalisco}, Mauro and {Barro}, Guillermo and {Willner}, S.~P. and {Ashby}, Matthew L.~N. and {Dahlen}, Tomas and {Donley}, Jennifer L. and {Faber}, Sandra M. and {Fontana}, Adriano and {Galametz}, Audrey and {Grazian}, Andrea and {Huang}, Kuang-Han and {Kocevski}, Dale D. and {Koekemoer}, Anton M. and {Koo}, David C. and {McGrath}, Elizabeth J. and {Peth}, Michael and {Salvato}, Mara and {Wuyts}, Stijn and {Castellano}, Marco and {Cooray}, Asantha R. and {Dickinson}, Mark E. and {Dunlop}, James S. and {Fazio}, G.~G. and {Gardner}, Jonathan P. and {Gawiser}, Eric and {Grogin}, Norman A. and {Hathi}, Nimish P. and {Hsu}, Li-Ting and {Lee}, Kyoung-Soo and {Lucas}, Ray A. and {Mobasher}, Bahram and {Nandra}, Kirpal and {Newman}, Jeffery A. and {van der Wel}, Arjen},
 doi = {10.1088/0067-0049/207/2/24},
 eid = {24},
 hideeprint = {1308.4405},
 journal = {\apjs},
 month = {August},
 number = {2},
 pages = {24},
 primaryclass = {astro-ph.CO},
 title = {{CANDELS Multi-wavelength Catalogs: Source Detection and Photometry in the GOODS-South Field}},
 volume = {207},
 year = {2013}
}

@article{harris20,
 adsurl = {https://ui.adsabs.harvard.edu/abs/2020Natur.585..357H},
 archiveprefix = {arXiv},
 author = {{Harris}, Charles R. and {Millman}, K. Jarrod and {van der Walt}, St{\'e}fan J. and {Gommers}, Ralf and {Virtanen}, Pauli and {Cournapeau}, David and {Wieser}, Eric and {Taylor}, Julian and {Berg}, Sebastian and {Smith}, Nathaniel J. and {Kern}, Robert and {Picus}, Matti and {Hoyer}, Stephan and {van Kerkwijk}, Marten H. and {Brett}, Matthew and {Haldane}, Allan and {del R{\'\i}o}, Jaime Fern{\'a}ndez and {Wiebe}, Mark and {Peterson}, Pearu and {G{\'e}rard-Marchant}, Pierre and {Sheppard}, Kevin and {Reddy}, Tyler and {Weckesser}, Warren and {Abbasi}, Hameer and {Gohlke}, Christoph and {Oliphant}, Travis E.},
 doi = {10.1038/s41586-020-2649-2},
 eprint = {2006.10256},
 journal = {\nat},
 month = {September},
 number = {7825},
 pages = {357-362},
 primaryclass = {cs.MS},
 title = {{Array programming with NumPy}},
 volume = {585},
 year = {2020}
}

@article{hayward21,
 adsurl = {https://ui.adsabs.harvard.edu/abs/2021MNRAS.502.2922H},
 archiveprefix = {arXiv},
 author = {{Hayward}, Christopher C. and {Sparre}, Martin and {Chapman}, Scott C. and {Hernquist}, Lars and {Nelson}, Dylan and {Pakmor}, R{\"u}diger and {Pillepich}, Annalisa and {Springel}, Volker and {Torrey}, Paul and {Vogelsberger}, Mark and {Weinberger}, Rainer},
 doi = {10.1093/mnras/stab246},
 hideeprint = {2007.01885},
 journal = {\mnras},
 month = {April},
 number = {2},
 pages = {2922-2933},
 primaryclass = {astro-ph.GA},
 title = {{Submillimetre galaxies in cosmological hydrodynamical simulations - an opportunity for constraining feedback models}},
 volume = {502},
 year = {2021}
}

@article{heintz25,
 adsurl = {https://ui.adsabs.harvard.edu/abs/2025A&A...693A..60H},
 archiveprefix = {arXiv},
 author = {{Heintz}, K.~E. and {Brammer}, G.~B. and {Watson}, D. and {Oesch}, P.~A. and {Keating}, L.~C. and {Hayes}, M.~J. and {Abdurro'uf} and {Arellano-C{\'o}rdova}, K.~Z. and {Carnall}, A.~C. and {Christiansen}, C.~R. and {Cullen}, F. and {Dav{\'e}}, R. and {Dayal}, P. and {Ferrara}, A. and {Finlator}, K. and {Fynbo}, J.~P.~U. and {Flury}, S.~R. and {Gelli}, V. and {Gillman}, S. and {Gottumukkala}, R. and {Gould}, K. and {Greve}, T.~R. and {Hardin}, S.~E. and {Hsiao}, T.~Y.-Y. and {Hutter}, A. and {Jakobsson}, P. and {Killi}, M. and {Khosravaninezhad}, N. and {Laursen}, P. and {Lee}, M.~M. and {Magdis}, G.~E. and {Matthee}, J. and {Naidu}, R.~P. and {Narayanan}, D. and {Pollock}, C. and {Prescott}, M.~K.~M. and {Rusakov}, V. and {Shuntov}, M. and {Sneppen}, A. and {Smit}, R. and {Tanvir}, N.~R. and {Terp}, C. and {Toft}, S. and {Valentino}, F. and {Vijayan}, A.~P. and {Weaver}, J.~R. and {Wise}, J.~H. and {Witstok}, J.},
 doi = {10.1051/0004-6361/202450243},
 eid = {A60},
 hideeprint = {2404.02211},
 journal = {\aap},
 month = {January},
 pages = {A60},
 primaryclass = {astro-ph.GA},
 title = {{The JWST-PRIMAL archival survey: A JWST/NIRSpec reference sample for the physical properties and Lyman-{\ensuremath{\alpha}} absorption and emission of {\ensuremath{\sim}}600 galaxies at z = 5.0 ‑ 13.4}},
 volume = {693},
 year = {2025}
}

@article{hodge20,
 adsurl = {https://ui.adsabs.harvard.edu/abs/2020RSOS....700556H},
 archiveprefix = {arXiv},
 author = {{Hodge}, J.~A. and {da Cunha}, E.},
 doi = {10.1098/rsos.200556},
 eid = {200556},
 eprint = {2004.00934},
 journal = {Royal Society Open Science},
 month = {December},
 number = {12},
 pages = {200556},
 primaryclass = {astro-ph.GA},
 title = {{High-redshift star formation in the Atacama large millimetre/submillimetre array era}},
 volume = {7},
 year = {2020}
}

@article{hughes97,
 adsurl = {https://ui.adsabs.harvard.edu/abs/1997MNRAS.289..766H},
 archiveprefix = {arXiv},
 author = {{Hughes}, D.~H. and {Dunlop}, J.~S. and {Rawlings}, S.},
 doi = {10.1093/mnras/289.4.766},
 hideeprint = {astro-ph/9705094},
 journal = {\mnras},
 month = {August},
 number = {4},
 pages = {766-782},
 primaryclass = {astro-ph},
 title = {{High-redshift radio galaxies and quasars at submillimetre wavelengths: assessing their evolutionary status}},
 volume = {289},
 year = {1997}
}

@article{hunter07,
 adsurl = {https://ui.adsabs.harvard.edu/abs/2007CSE.....9...90H},
 author = {{Hunter}, John D.},
 doi = {10.1109/MCSE.2007.55},
 journal = {Computing in Science and Engineering},
 month = {January},
 number = {3},
 pages = {90-95},
 title = {{Matplotlib: A 2D Graphics Environment}},
 volume = {9},
 year = {2007}
}

@article{ikarashi15,
 adsurl = {https://ui.adsabs.harvard.edu/abs/2015ApJ...810..133I},
 archiveprefix = {arXiv},
 author = {{Ikarashi}, Soh and {Ivison}, R.~J. and {Caputi}, Karina I. and {Aretxaga}, Itziar and {Dunlop}, James S. and {Hatsukade}, Bunyo and {Hughes}, David H. and {Iono}, Daisuke and {Izumi}, Takuma and {Kawabe}, Ryohei and {Kohno}, Kotaro and {Lagos}, Claudia D.~P. and {Motohara}, Kentaro and {Nakanishi}, Kouichiro and {Ohta}, Kouji and {Tamura}, Yoichi and {Umehata}, Hideki and {Wilson}, Grant W. and {Yabe}, Kiyoto and {Yun}, Min S.},
 doi = {10.1088/0004-637X/810/2/133},
 eid = {133},
 eprint = {1411.5038},
 journal = {\apj},
 month = {September},
 number = {2},
 pages = {133},
 primaryclass = {astro-ph.GA},
 title = {{Compact Starbursts in z {\ensuremath{\sim}} 3-6 Submillimeter Galaxies Revealed by ALMA}},
 volume = {810},
 year = {2015}
}

@article{illingworth16,
 adsurl = {https://ui.adsabs.harvard.edu/abs/2016arXiv160600841I},
 archiveprefix = {arXiv},
 author = {{Illingworth}, Garth and {Magee}, Daniel and {Bouwens}, Rychard and {Oesch}, Pascal and {Labbe}, Ivo and {van Dokkum}, Pieter and {Whitaker}, Katherine and {Holden}, Bradford and {Franx}, Marijn and {Gonzalez}, Valentino},
 doi = {10.48550/arXiv.1606.00841},
 eid = {arXiv:1606.00841},
 hideeprint = {1606.00841},
 journal = {arXiv e-prints},
 month = {June},
 pages = {arXiv:1606.00841},
 primaryclass = {astro-ph.GA},
 title = {{The Hubble Legacy Fields (HLF-GOODS-S) v1.5 Data Products: Combining 2442 Orbits of GOODS-S/CDF-S Region ACS and WFC3/IR Images}},
 year = {2016}
}

@article{ivison16,
 adsurl = {https://ui.adsabs.harvard.edu/abs/2016ApJ...832...78I},
 archiveprefix = {arXiv},
 author = {{Ivison}, R.~J. and {Lewis}, A.~J.~R. and {Weiss}, A. and {Arumugam}, V. and {Simpson}, J.~M. and {Holland}, W.~S. and {Maddox}, S. and {Dunne}, L. and {Valiante}, E. and {van der Werf}, P. and {Omont}, A. and {Dannerbauer}, H. and {Smail}, Ian and {Bertoldi}, F. and {Bremer}, M. and {Bussmann}, R.~S. and {Cai}, Z.-Y. and {Clements}, D.~L. and {Cooray}, A. and {De Zotti}, G. and {Eales}, S.~A. and {Fuller}, C. and {Gonzalez-Nuevo}, J. and {Ibar}, E. and {Negrello}, M. and {Oteo}, I. and {P{\'e}rez-Fournon}, I. and {Riechers}, D. and {Stevens}, J.~A. and {Swinbank}, A.~M. and {Wardlow}, J.},
 doi = {10.3847/0004-637X/832/1/78},
 eid = {78},
 hideeprint = {1611.00762},
 journal = {\apj},
 month = {November},
 number = {1},
 pages = {78},
 primaryclass = {astro-ph.GA},
 title = {{The Space Density of Luminous Dusty Star-forming Galaxies at z > 4: SCUBA-2 and LABOCA Imaging of Ultrared Galaxies from Herschel-ATLAS}},
 volume = {832},
 year = {2016}
}

@article{jin24,
 adsurl = {https://ui.adsabs.harvard.edu/abs/2024A&A...690L..16J},
 archiveprefix = {arXiv},
 author = {{Jin}, Shuowen and {Sillassen}, Nikolaj B. and {Hodge}, Jacqueline and {Magdis}, Georgios E. and {Rizzo}, Francesca and {Casey}, Caitlin and {Koekemoer}, Anton M. and {Valentino}, Francesco and {Kokorev}, Vasily and {Magnelli}, Benjamin and {Gobat}, Raphael and {Gillman}, Steven and {Franco}, Maximilien and {Faisst}, Andreas and {Kartaltepe}, Jeyhan and {Schinnerer}, Eva and {Toft}, Sune and {Algera}, Hiddo S.~B. and {Harish}, Santosh and {Lee}, Minju and {Liu}, Daizhong and {Shuntov}, Marko and {Talia}, Margherita and {Vijayan}, Aswin},
 doi = {10.1051/0004-6361/202451445},
 eid = {L16},
 eprint = {2407.07585},
 journal = {\aap},
 month = {October},
 pages = {L16},
 primaryclass = {astro-ph.GA},
 title = {{A photo-z cautionary tale: Redshift confirmation of COSBO-7 at z = 2.625}},
 volume = {690},
 year = {2024}
}

@article{kriek15,
 adsurl = {https://ui.adsabs.harvard.edu/abs/2015ApJS..218...15K},
 archiveprefix = {arXiv},
 author = {{Kriek}, Mariska and {Shapley}, Alice E. and {Reddy}, Naveen A. and {Siana}, Brian and {Coil}, Alison L. and {Mobasher}, Bahram and {Freeman}, William R. and {de Groot}, Laura and {Price}, Sedona H. and {Sanders}, Ryan and {Shivaei}, Irene and {Brammer}, Gabriel B. and {Momcheva}, Ivelina G. and {Skelton}, Rosalind E. and {van Dokkum}, Pieter G. and {Whitaker}, Katherine E. and {Aird}, James and {Azadi}, Mojegan and {Kassis}, Marc and {Bullock}, James S. and {Conroy}, Charlie and {Dav{\'e}}, Romeel and {Kere{\v{s}}}, Du{\v{s}}an and {Krumholz}, Mark},
 doi = {10.1088/0067-0049/218/2/15},
 eid = {15},
 eprint = {1412.1835},
 journal = {\apjs},
 month = {June},
 number = {2},
 pages = {15},
 primaryclass = {astro-ph.GA},
 title = {{The MOSFIRE Deep Evolution Field (MOSDEF) Survey: Rest-frame Optical Spectroscopy for \raisebox{-0.5ex}\textasciitilde1500 H-selected Galaxies at 1.37 < z < 3.8}},
 volume = {218},
 year = {2015}
}

@article{kumar25,
 adsurl = {https://ui.adsabs.harvard.edu/abs/2025A&A...698A.236K},
 archiveprefix = {arXiv},
 author = {{Kumar}, Ankit and {Artale}, M. Celeste and {Montero-Dorta}, Antonio D. and {Guaita}, Lucia and {Lee}, Kyoung-Soo and {Pope}, Alexandra and {Schaye}, Joop and {Schaller}, Matthieu and {Gawiser}, Eric and {Hwang}, Ho Seong and {Jeong}, Woong-Seob and {Lee}, Jaehyun and {Padilla}, Nelson and {Park}, Changbom and {Ramakrishnan}, Vandana and {Singh}, Akriti and {Yang}, Yujin},
 doi = {10.1051/0004-6361/202553981},
 eid = {A236},
 hideeprint = {2501.19327},
 journal = {\aap},
 month = {June},
 pages = {A236},
 primaryclass = {astro-ph.CO},
 title = {{Modeling submillimeter galaxies in cosmological simulations: Contribution to the cosmic star formation density and predictions for future surveys}},
 volume = {698},
 year = {2025}
}

@article{kumar26,
 adsurl = {https://ui.adsabs.harvard.edu/abs/2026arXiv260211751K},
 archiveprefix = {arXiv},
 author = {{Kumar}, Ankit and {Artale}, M. Celeste and {Montero-Dorta}, Antonio D. and {Guaita}, Lucia and {Schaye}, Joop and {Lee}, Kyoung-Soo and {Pope}, Alexandra and {Rodriguez}, Facundo and {Gawiser}, Eric and {Hwang}, Ho Seong and {Troncoso Iribarren}, Paulina and {Lee}, Jaehyun and {Lee}, Seong-Kook and {Park}, Changbom and {Yang}, Yujin},
 doi = {10.48550/arXiv.2602.11751},
 eid = {arXiv:2602.11751},
 hideeprint = {2602.11751},
 journal = {arXiv e-prints},
 month = {February},
 pages = {arXiv:2602.11751},
 primaryclass = {astro-ph.CO},
 title = {{Evolution of submillimeter galaxies across cosmic-web environments}},
 year = {2026}
}

@article{kurk13,
 adsurl = {https://ui.adsabs.harvard.edu/abs/2013A&A...549A..63K},
 archiveprefix = {arXiv},
 author = {{Kurk}, J. and {Cimatti}, A. and {Daddi}, E. and {Mignoli}, M. and {Pozzetti}, L. and {Dickinson}, M. and {Bolzonella}, M. and {Zamorani}, G. and {Cassata}, P. and {Rodighiero}, G. and {Franceschini}, A. and {Renzini}, A. and {Rosati}, P. and {Halliday}, C. and {Berta}, S.},
 doi = {10.1051/0004-6361/201117847},
 eid = {A63},
 eprint = {1209.1561},
 journal = {\aap},
 month = {January},
 pages = {A63},
 primaryclass = {astro-ph.CO},
 title = {{GMASS ultradeep spectroscopy of galaxies at z \raisebox{-0.5ex}\textasciitilde 2. VII. Sample selection and spectroscopy}},
 volume = {549},
 year = {2013}
}

@article{lacey16,
 adsurl = {https://ui.adsabs.harvard.edu/abs/2016MNRAS.462.3854L},
 archiveprefix = {arXiv},
 author = {{Lacey}, Cedric G. and {Baugh}, Carlton M. and {Frenk}, Carlos S. and {Benson}, Andrew J. and {Bower}, Richard G. and {Cole}, Shaun and {Gonzalez-Perez}, Violeta and {Helly}, John C. and {Lagos}, Claudia D.~P. and {Mitchell}, Peter D.},
 doi = {10.1093/mnras/stw1888},
 eprint = {1509.08473},
 journal = {\mnras},
 month = {November},
 number = {4},
 pages = {3854-3911},
 primaryclass = {astro-ph.GA},
 title = {{A unified multiwavelength model of galaxy formation}},
 volume = {462},
 year = {2016}
}

@article{lapasia25,
 adsurl = {https://ui.adsabs.harvard.edu/abs/2026arXiv260108693L},
 archiveprefix = {arXiv},
 author = {{Lapasia}, Yash and {Tacchella}, Sandro and {D'Eugenio}, Francesco and {Pusk{\'a}s}, D{\'a}vid and {Bunker}, Andrew J. and {Danhaive}, A. Lola and {Johnson}, Benjamin D. and {Maiolino}, Roberto and {Robertson}, Brant and {Simmonds}, Charlotte and {Shivaei}, Irene and {Williams}, Christina C. and {Willmer}, Christopher},
 doi = {10.48550/arXiv.2601.08693},
 eid = {arXiv:2601.08693},
 hideeprint = {2601.08693},
 journal = {arXiv e-prints},
 month = {January},
 pages = {arXiv:2601.08693},
 primaryclass = {astro-ph.GA},
 title = {{Stellar masses of optically dark galaxies: uncertainty introduced by the attenuation law and star-formation histories}},
 year = {2026}
}

@article{lilly99,
 adsurl = {https://ui.adsabs.harvard.edu/abs/1999ApJ...518..641L},
 archiveprefix = {arXiv},
 author = {{Lilly}, Simon J. and {Eales}, Stephen A. and {Gear}, Walter K.~P. and {Hammer}, Fran{\c{c}}ois and {Le F{\`e}vre}, Olivier and {Crampton}, David and {Bond}, J. Richard and {Dunne}, Loretta},
 doi = {10.1086/307310},
 eprint = {astro-ph/9901047},
 journal = {\apj},
 month = {June},
 number = {2},
 pages = {641-655},
 primaryclass = {astro-ph},
 title = {{The Canada-United Kingdom Deep Submillimeter Survey. II. First Identifications, Redshifts, and Implications for Galaxy Evolution}},
 volume = {518},
 year = {1999}
}

@article{liu15,
 adsurl = {https://ui.adsabs.harvard.edu/abs/2015ApJ...810L..14L},
 archiveprefix = {arXiv},
 author = {{Liu}, Daizhong and {Gao}, Yu and {Isaak}, Kate and {Daddi}, Emanuele and {Yang}, Chentao and {Lu}, Nanyao and {van der Werf}, Paul},
 doi = {10.1088/2041-8205/810/2/L14},
 eid = {L14},
 eprint = {1504.05897},
 journal = {\apjl},
 month = {September},
 number = {2},
 pages = {L14},
 primaryclass = {astro-ph.GA},
 title = {{High-J CO versus Far-infrared Relations in Normal and Starburst Galaxies}},
 volume = {810},
 year = {2015}
}

@article{lovell21,
 adsurl = {https://ui.adsabs.harvard.edu/abs/2021MNRAS.502..772L},
 archiveprefix = {arXiv},
 author = {{Lovell}, Christopher C. and {Geach}, James E. and {Dav{\'e}}, Romeel and {Narayanan}, Desika and {Li}, Qi},
 doi = {10.1093/mnras/staa4043},
 hideeprint = {2006.15156},
 journal = {\mnras},
 month = {March},
 number = {1},
 pages = {772-793},
 primaryclass = {astro-ph.GA},
 title = {{Reproducing submillimetre galaxy number counts with cosmological hydrodynamic simulations}},
 volume = {502},
 year = {2021}
}

@article{maseda24,
 adsurl = {https://ui.adsabs.harvard.edu/abs/2024A&A...689A..73M},
 archiveprefix = {arXiv},
 author = {{Maseda}, Michael V. and {de Graaff}, Anna and {Franx}, Marijn and {Rix}, Hans-Walter and {Carniani}, Stefano and {Laseter}, Isaac and {Dudzevi{\v{c}}i{\={u}}t{\.{e}}}, Ugn{\.{e}} and {Rawle}, Tim and {Parlanti}, Eleonora and {Arribas}, Santiago and {Bunker}, Andrew J. and {Cameron}, Alex J. and {Charlot}, Stephane and {Curti}, Mirko and {D'Eugenio}, Francesco and {Jones}, Gareth C. and {Kumari}, Nimisha and {Maiolino}, Roberto and {{\"U}bler}, Hannah and {Saxena}, Aayush and {Smit}, Renske and {Willott}, Chris and {Witstok}, Joris},
 doi = {10.1051/0004-6361/202449914},
 eid = {A73},
 hideeprint = {2403.05506},
 journal = {\aap},
 month = {September},
 pages = {A73},
 primaryclass = {astro-ph.GA},
 title = {{The NIRSpec Wide GTO Survey}},
 volume = {689},
 year = {2024}
}

@article{mcalpine19,
 adsurl = {https://ui.adsabs.harvard.edu/abs/2019MNRAS.488.2440M},
 archiveprefix = {arXiv},
 author = {{McAlpine}, Stuart and {Smail}, Ian and {Bower}, Richard G. and {Swinbank}, A.~M. and {Trayford}, James W. and {Theuns}, Tom and {Baes}, Maarten and {Camps}, Peter and {Crain}, Robert A. and {Schaye}, Joop},
 doi = {10.1093/mnras/stz1692},
 hideeprint = {1901.05467},
 journal = {\mnras},
 month = {September},
 number = {2},
 pages = {2440-2454},
 primaryclass = {astro-ph.GA},
 title = {{The nature of submillimetre and highly star-forming galaxies in the EAGLE simulation}},
 volume = {488},
 year = {2019}
}

@article{mckay23,
 adsurl = {https://ui.adsabs.harvard.edu/abs/2023ApJ...951...48M},
 archiveprefix = {arXiv},
 author = {{McKay}, S.~J. and {Barger}, A.~J. and {Cowie}, L.~L. and {Bauer}, F.~E. and {Rosenthal}, M.~J. Nicandro},
 doi = {10.3847/1538-4357/acd1e5},
 eid = {48},
 hideeprint = {2305.06388},
 journal = {\apj},
 month = {July},
 number = {1},
 pages = {48},
 primaryclass = {astro-ph.GA},
 title = {{Dust Properties of 870 {\ensuremath{\mu}}m-selected Galaxies in GOODS-S}},
 volume = {951},
 year = {2023}
}

@article{mckay25,
 adsurl = {https://ui.adsabs.harvard.edu/abs/2025ApJ...988..135M},
 archiveprefix = {arXiv},
 author = {{McKay}, S.~J. and {Barger}, A.~J. and {Cowie}, L.~L. and {Nicandro Rosenthal}, M.~J.},
 doi = {10.3847/1538-4357/ade394},
 eid = {135},
 hideeprint = {2503.00102},
 journal = {\apj},
 month = {July},
 number = {1},
 pages = {135},
 primaryclass = {astro-ph.GA},
 title = {{The Physical Properties and Morphologies of Faint Dusty Star-forming Galaxies Identified with JWST}},
 volume = {988},
 year = {2025}
}

@software{mckay_mbb,
 author = {Stephen McKay},
 doi = {10.5281/zenodo.18842404},
 month = {March},
 publisher = {Zenodo},
 swhid = {swh:1:dir:a6f9767c262316dde00d159d376c7bdf2544f55e
;origin=https:// doi.org/10.5281/zenodo.17781466;vi
sit=swh:1:snp:9132e922511399ef0eb81074066988ef7273
acb0;anchor=swh:1:rel:724b4a843b54c896fed8c7e3bfc0
424c3a13a54f;path=sjmckay-mbb-1f52be3},
 title = {sjmckay/mbb: mbb v0.5.2},
 url = {https://doi.org/10.5281/zenodo.18842404},
 version = {v0.5.2},
 year = {2026}
}

@software{mckinney10,
 adsurl = {https://ui.adsabs.harvard.edu/abs/2010scpy.soft.....M},
 author = {{McKinney}, Wes},
 doi = {10.25080/Majora-92bf1922-00a},
 howpublished = {Proceedings of the 9th Python in Science Conference},
 month = {January},
 title = {{Data Structures for Statistical Computing in Python}},
 year = {2010}
}

@article{mckinney25,
 adsurl = {https://ui.adsabs.harvard.edu/abs/2025ApJ...979..229M},
 archiveprefix = {arXiv},
 author = {{McKinney}, Jed and {Casey}, Caitlin M. and {Long}, Arianna S. and {Cooper}, Olivia R. and {Manning}, Sinclaire M. and {Franco}, Maximilien and {Akins}, Hollis and {Lambrides}, Erini and {Gammon}, Elaine and {Silva}, Camila and {Gentile}, Fabrizio and {Zavala}, Jorge A. and {Amvrosiadis}, Aristeidis and {Andika}, Irham and {Brinch}, Malte and {Champagne}, Jaclyn B. and {Chartab}, Nima and {Drakos}, Nicole E. and {Faisst}, Andreas L. and {Fujimoto}, Seiji and {Gillman}, Steven and {Gozaliasl}, Ghassem and {Greve}, Thomas R. and {Harish}, Santosh and {Hayward}, Christopher C. and {Hirschmann}, Michaela and {Ilbert}, Olivier and {Kalita}, Boris S. and {Kartaltepe}, Jeyhan S. and {Koekemoer}, Anton M. and {Kokorev}, Vasily and {Liu}, Daizhong and {Magdis}, Georgios and {McCracken}, Henry Joy and {Rhodes}, Jason and {Robertson}, Brant E. and {Talia}, Margherita and {Valentino}, Francesco and {Vijayan}, Aswin P.},
 doi = {10.3847/1538-4357/ada357},
 eid = {229},
 hideeprint = {2408.08346},
 journal = {\apj},
 month = {February},
 number = {2},
 pages = {229},
 primaryclass = {astro-ph.GA},
 title = {{SCUBADive. I. JWST+ALMA Analysis of 289 Submillimeter Galaxies in COSMOS-web}},
 volume = {979},
 year = {2025}
}

@article{mclure18,
 adsurl = {https://ui.adsabs.harvard.edu/abs/2018MNRAS.479...25M},
 archiveprefix = {arXiv},
 author = {{McLure}, R.~J. and {Pentericci}, L. and {Cimatti}, A. and {Dunlop}, J.~S. and {Elbaz}, D. and {Fontana}, A. and {Nandra}, K. and {Amorin}, R. and {Bolzonella}, M. and {Bongiorno}, A. and {Carnall}, A.~C. and {Castellano}, M. and {Cirasuolo}, M. and {Cucciati}, O. and {Cullen}, F. and {De Barros}, S. and {Finkelstein}, S.~L. and {Fontanot}, F. and {Franzetti}, P. and {Fumana}, M. and {Gargiulo}, A. and {Garilli}, B. and {Guaita}, L. and {Hartley}, W.~G. and {Iovino}, A. and {Jarvis}, M.~J. and {Juneau}, S. and {Karman}, W. and {Maccagni}, D. and {Marchi}, F. and {M{\'a}rmol-Queralt{\'o}}, E. and {Pompei}, E. and {Pozzetti}, L. and {Scodeggio}, M. and {Sommariva}, V. and {Talia}, M. and {Almaini}, O. and {Balestra}, I. and {Bardelli}, S. and {Bell}, E.~F. and {Bourne}, N. and {Bowler}, R.~A.~A. and {Brusa}, M. and {Buitrago}, F. and {Caputi}, K.~I. and {Cassata}, P. and {Charlot}, S. and {Citro}, A. and {Cresci}, G. and {Cristiani}, S. and {Curtis-Lake}, E. and {Dickinson}, M. and {Fazio}, G.~G. and {Ferguson}, H.~C. and {Fiore}, F. and {Franco}, M. and {Fynbo}, J.~P.~U. and {Galametz}, A. and {Georgakakis}, A. and {Giavalisco}, M. and {Grazian}, A. and {Hathi}, N.~P. and {Jung}, I. and {Kim}, S. and {Koekemoer}, A.~M. and {Khusanova}, Y. and {Le F{\`e}vre}, O. and {Lotz}, J.~M. and {Mannucci}, F. and {Maltby}, D.~T. and {Matsuoka}, K. and {McLeod}, D.~J. and {Mendez-Hernandez}, H. and {Mendez-Abreu}, J. and {Mignoli}, M. and {Moresco}, M. and {Mortlock}, A. and {Nonino}, M. and {Pannella}, M. and {Papovich}, C. and {Popesso}, P. and {Rosario}, D.~P. and {Salvato}, M. and {Santini}, P. and {Schaerer}, D. and {Schreiber}, C. and {Stark}, D.~P. and {Tasca}, L.~A.~M. and {Thomas}, R. and {Treu}, T. and {Vanzella}, E. and {Wild}, V. and {Williams}, C.~C. and {Zamorani}, G. and {Zucca}, E.},
 doi = {10.1093/mnras/sty1213},
 eprint = {1803.07414},
 journal = {\mnras},
 month = {September},
 number = {1},
 pages = {25-42},
 primaryclass = {astro-ph.GA},
 title = {{The VANDELS ESO public spectroscopic survey}},
 volume = {479},
 year = {2018}
}

@inproceedings{mcmullin07,
 adsurl = {https://ui.adsabs.harvard.edu/abs/2007ASPC..376..127M},
 author = {{McMullin}, J.~P. and {Waters}, B. and {Schiebel}, D. and {Young}, W. and {Golap}, K.},
 booktitle = {Astronomical Data Analysis Software and Systems XVI},
 editor = {{Shaw}, R.~A. and {Hill}, F. and {Bell}, D.~J.},
 month = {October},
 pages = {127},
 series = {Astronomical Society of the Pacific Conference Series},
 title = {{CASA Architecture and Applications}},
 volume = {376},
 year = {2007}
}

@article{momcheva16,
 adsurl = {https://ui.adsabs.harvard.edu/abs/2016ApJS..225...27M},
 archiveprefix = {arXiv},
 author = {{Momcheva}, Ivelina G. and {Brammer}, Gabriel B. and {van Dokkum}, Pieter G. and {Skelton}, Rosalind E. and {Whitaker}, Katherine E. and {Nelson}, Erica J. and {Fumagalli}, Mattia and {Maseda}, Michael V. and {Leja}, Joel and {Franx}, Marijn and {Rix}, Hans-Walter and {Bezanson}, Rachel and {Da Cunha}, Elisabete and {Dickey}, Claire and {F{\"o}rster Schreiber}, Natascha M. and {Illingworth}, Garth and {Kriek}, Mariska and {Labb{\'e}}, Ivo and {Ulf Lange}, Johannes and {Lundgren}, Britt F. and {Magee}, Daniel and {Marchesini}, Danilo and {Oesch}, Pascal and {Pacifici}, Camilla and {Patel}, Shannon G. and {Price}, Sedona and {Tal}, Tomer and {Wake}, David A. and {van der Wel}, Arjen and {Wuyts}, Stijn},
 doi = {10.3847/0067-0049/225/2/27},
 eid = {27},
 hideeprint = {1510.02106},
 journal = {\apjs},
 month = {August},
 number = {2},
 pages = {27},
 primaryclass = {astro-ph.GA},
 title = {{The 3D-HST Survey: Hubble Space Telescope WFC3/G141 Grism Spectra, Redshifts, and Emission Line Measurements for \raisebox{-0.5ex}\textasciitilde 100,000 Galaxies}},
 volume = {225},
 year = {2016}
}

@article{narayanan15,
 adsurl = {https://ui.adsabs.harvard.edu/abs/2015Natur.525..496N},
 archiveprefix = {arXiv},
 author = {{Narayanan}, Desika and {Turk}, Matthew and {Feldmann}, Robert and {Robitaille}, Thomas and {Hopkins}, Philip and {Thompson}, Robert and {Hayward}, Christopher and {Ball}, David and {Faucher-Gigu{\`e}re}, Claude-Andr{\'e} and {Kere{\v{s}}}, Du{\v{s}}an},
 doi = {10.1038/nature15383},
 hideeprint = {1509.06377},
 journal = {\nat},
 month = {September},
 number = {7570},
 pages = {496-499},
 primaryclass = {astro-ph.GA},
 title = {{The formation of submillimetre-bright galaxies from gas infall over a billion years}},
 volume = {525},
 year = {2015}
}

@article{nicandro_rosenthal25,
 adsurl = {https://ui.adsabs.harvard.edu/abs/2025ApJ...979..247N},
 archiveprefix = {arXiv},
 author = {{Nicandro Rosenthal}, Michael J. and {Barger}, Amy J. and {Cowie}, Lennox L. and {Jones}, Logan H. and {McKay}, Stephen J. and {Taylor}, Anthony J.},
 doi = {10.3847/1538-4357/ad9c67},
 eid = {247},
 eprint = {2411.07291},
 journal = {\apj},
 month = {February},
 number = {2},
 pages = {247},
 primaryclass = {astro-ph.GA},
 title = {{Spectroscopic Confirmation of a Massive Protocluster with Two Substructures at z ≃ 3.1}},
 volume = {979},
 year = {2025}
}

@article{nicandro_rosenthal26,
 adsurl = {https://ui.adsabs.harvard.edu/abs/2026arXiv260120862N},
 archiveprefix = {arXiv},
 author = {{Nicandro Rosenthal}, Michael J. and {McKay}, Stephen J. and {Barger}, Amy J. and {Cowie}, Lennox L.},
 doi = {10.48550/arXiv.2601.20862},
 eid = {arXiv:2601.20862},
 eprint = {2601.20862},
 journal = {arXiv e-prints},
 month = {January},
 pages = {arXiv:2601.20862},
 primaryclass = {astro-ph.GA},
 title = {{Molecular Gas Detections in Eight Faint DSFGs with Red NIR Colors at z = 1.2-2.5}},
 year = {2026}
}

@article{oesch23,
 adsurl = {https://ui.adsabs.harvard.edu/abs/2023MNRAS.525.2864O},
 archiveprefix = {arXiv},
 author = {{Oesch}, P.~A. and {Brammer}, G. and {Naidu}, R.~P. and {Bouwens}, R.~J. and {Chisholm}, J. and {Illingworth}, G.~D. and {Matthee}, J. and {Nelson}, E. and {Qin}, Y. and {Reddy}, N. and {Shapley}, A. and {Shivaei}, I. and {van Dokkum}, P. and {Weibel}, A. and {Whitaker}, K. and {Wuyts}, S. and {Covelo-Paz}, A. and {Endsley}, R. and {Fudamoto}, Y. and {Giovinazzo}, E. and {Herard-Demanche}, T. and {Kerutt}, J. and {Kramarenko}, I. and {Labbe}, I. and {Leonova}, E. and {Lin}, J. and {Magee}, D. and {Marchesini}, D. and {Maseda}, M. and {Mason}, C. and {Matharu}, J. and {Meyer}, R.~A. and {Neufeld}, C. and {Prieto Lyon}, G. and {Schaerer}, D. and {Sharma}, R. and {Shuntov}, M. and {Smit}, R. and {Stefanon}, M. and {Wyithe}, J.~S.~B. and {Xiao}, M.},
 doi = {10.1093/mnras/stad2411},
 hideeprint = {2304.02026},
 journal = {\mnras},
 month = {October},
 number = {2},
 pages = {2864-2874},
 primaryclass = {astro-ph.GA},
 title = {{The JWST FRESCO survey: legacy NIRCam/grism spectroscopy and imaging in the two GOODS fields}},
 volume = {525},
 year = {2023}
}

@article{oke83,
 adsurl = {https://ui.adsabs.harvard.edu/abs/1983ApJ...266..713O},
 author = {{Oke}, J.~B. and {Gunn}, J.~E.},
 doi = {10.1086/160817},
 journal = {\apj},
 month = {March},
 pages = {713-717},
 title = {{Secondary standard stars for absolute spectrophotometry.}},
 volume = {266},
 year = {1983}
}

@article{oliver12,
 adsurl = {https://ui.adsabs.harvard.edu/abs/2012MNRAS.424.1614O},
 archiveprefix = {arXiv},
 author = {{Oliver}, S.~J. and {Bock}, J. and {Altieri}, B. and {Amblard}, A. and {Arumugam}, V. and {Aussel}, H. and {Babbedge}, T. and {Beelen}, A. and {B{\'e}thermin}, M. and {Blain}, A. and {Boselli}, A. and {Bridge}, C. and {Brisbin}, D. and {Buat}, V. and {Burgarella}, D. and {Castro-Rodr{\'\i}guez}, N. and {Cava}, A. and {Chanial}, P. and {Cirasuolo}, M. and {Clements}, D.~L. and {Conley}, A. and {Conversi}, L. and {Cooray}, A. and {Dowell}, C.~D. and {Dubois}, E.~N. and {Dwek}, E. and {Dye}, S. and {Eales}, S. and {Elbaz}, D. and {Farrah}, D. and {Feltre}, A. and {Ferrero}, P. and {Fiolet}, N. and {Fox}, M. and {Franceschini}, A. and {Gear}, W. and {Giovannoli}, E. and {Glenn}, J. and {Gong}, Y. and {Gonz{\'a}lez Solares}, E.~A. and {Griffin}, M. and {Halpern}, M. and {Harwit}, M. and {Hatziminaoglou}, E. and {Heinis}, S. and {Hurley}, P. and {Hwang}, H.~S. and {Hyde}, A. and {Ibar}, E. and {Ilbert}, O. and {Isaak}, K. and {Ivison}, R.~J. and {Lagache}, G. and {Le Floc'h}, E. and {Levenson}, L. and {Faro}, B. Lo and {Lu}, N. and {Madden}, S. and {Maffei}, B. and {Magdis}, G. and {Mainetti}, G. and {Marchetti}, L. and {Marsden}, G. and {Marshall}, J. and {Mortier}, A.~M.~J. and {Nguyen}, H.~T. and {O'Halloran}, B. and {Omont}, A. and {Page}, M.~J. and {Panuzzo}, P. and {Papageorgiou}, A. and {Patel}, H. and {Pearson}, C.~P. and {P{\'e}rez-Fournon}, I. and {Pohlen}, M. and {Rawlings}, J.~I. and {Raymond}, G. and {Rigopoulou}, D. and {Riguccini}, L. and {Rizzo}, D. and {Rodighiero}, G. and {Roseboom}, I.~G. and {Rowan-Robinson}, M. and {S{\'a}nchez Portal}, M. and {Schulz}, B. and {Scott}, Douglas and {Seymour}, N. and {Shupe}, D.~L. and {Smith}, A.~J. and {Stevens}, J.~A. and {Symeonidis}, M. and {Trichas}, M. and {Tugwell}, K.~E. and {Vaccari}, M. and {Valtchanov}, I. and {Vieira}, J.~D. and {Viero}, M. and {Vigroux}, L. and {Wang}, L. and {Ward}, R. and {Wardlow}, J. and {Wright}, G. and {Xu}, C.~K. and {Zemcov}, M.},
 doi = {10.1111/j.1365-2966.2012.20912.x},
 hideeprint = {1203.2562},
 journal = {\mnras},
 month = {August},
 number = {3},
 pages = {1614-1635},
 primaryclass = {astro-ph.CO},
 title = {{The Herschel Multi-tiered Extragalactic Survey: HerMES}},
 volume = {424},
 year = {2012}
}

@article{pentericci18,
 adsurl = {https://ui.adsabs.harvard.edu/abs/2018A&A...616A.174P},
 archiveprefix = {arXiv},
 author = {{Pentericci}, L. and {McLure}, R.~J. and {Garilli}, B. and {Cucciati}, O. and {Franzetti}, P. and {Iovino}, A. and {Amorin}, R. and {Bolzonella}, M. and {Bongiorno}, A. and {Carnall}, A.~C. and {Castellano}, M. and {Cimatti}, A. and {Cirasuolo}, M. and {Cullen}, F. and {De Barros}, S. and {Dunlop}, J.~S. and {Elbaz}, D. and {Finkelstein}, S.~L. and {Fontana}, A. and {Fontanot}, F. and {Fumana}, M. and {Gargiulo}, A. and {Guaita}, L. and {Hartley}, W.~G. and {Jarvis}, M.~J. and {Juneau}, S. and {Karman}, W. and {Maccagni}, D. and {Marchi}, F. and {Marmol-Queralto}, E. and {Nandra}, K. and {Pompei}, E. and {Pozzetti}, L. and {Scodeggio}, M. and {Sommariva}, V. and {Talia}, M. and {Almaini}, O. and {Balestra}, I. and {Bardelli}, S. and {Bell}, E.~F. and {Bourne}, N. and {Bowler}, R.~A.~A. and {Brusa}, M. and {Buitrago}, F. and {Caputi}, K.~I. and {Cassata}, P. and {Charlot}, S. and {Citro}, A. and {Cresci}, G. and {Cristiani}, S. and {Curtis-Lake}, E. and {Dickinson}, M. and {Fazio}, G.~G. and {Ferguson}, H.~C. and {Fiore}, F. and {Franco}, M. and {Fynbo}, J.~P.~U. and {Galametz}, A. and {Georgakakis}, A. and {Giavalisco}, M. and {Grazian}, A. and {Hathi}, N.~P. and {Jung}, I. and {Kim}, S. and {Koekemoer}, A.~M. and {Khusanova}, Y. and {Le F{\`e}vre}, O. and {Lotz}, J.~M. and {Mannucci}, F. and {Maltby}, D.~T. and {Matsuoka}, K. and {McLeod}, D.~J. and {Mendez-Hernandez}, H. and {Mendez-Abreu}, J. and {Mignoli}, M. and {Moresco}, M. and {Mortlock}, A. and {Nonino}, M. and {Pannella}, M. and {Papovich}, C. and {Popesso}, P. and {Rosario}, D.~P. and {Salvato}, M. and {Santini}, P. and {Schaerer}, D. and {Schreiber}, C. and {Stark}, D.~P. and {Tasca}, L.~A.~M. and {Thomas}, R. and {Treu}, T. and {Vanzella}, E. and {Wild}, V. and {Williams}, C.~C. and {Zamorani}, G. and {Zucca}, E.},
 doi = {10.1051/0004-6361/201833047},
 eid = {A174},
 eprint = {1803.07373},
 journal = {\aap},
 month = {September},
 pages = {A174},
 primaryclass = {astro-ph.GA},
 title = {{The VANDELS ESO public spectroscopic survey: Observations and first data release}},
 volume = {616},
 year = {2018}
}

\newpage
\begin{longrotatetable}
\begin{deluxetable*}{ccccccccccccccccc}
\tablewidth{\textwidth}
\tabletypesize{\scriptsize}
\tablecaption{\label{tab:redshifts}
Redshifts for the ALMA Sample}
\tablehead{
\colhead{Source}&\colhead{R.A.}&\colhead{Decl.}&\colhead{$f_{870 \rm \mu m}$}&\colhead{$\log(L_{\rm IR}/{\rm L}_\odot)$}&\colhead{$z_{\rm spec, ALMA}$}&\colhead{Flag}&\colhead{$z_{\rm spec, other}$}&\colhead{Ref.}&\colhead{Best $z$}&\colhead{$z$ Type}
&\colhead{$z_{\rm phot, {\tt MAGPHYS}}$}&\colhead{$z_{\rm phot, {\tt MBB}}$}&\colhead{$z_{\rm phot, {\tt MMPZ}}$}
&\colhead{$z_{\rm phot, S16}$}&\colhead{$z_{\rm phot, E26}$}&\colhead{$z_{\rm grism, M16}$}\\
&(J2000)&(J2000)&(mJy)&&&&&&&&&&&&}
\colnumbers

\startdata
SG-ALMA-1&53.030371&-27.855778&8.93\,$\pm$\,0.21&$12.80^{+0.02}_{-0.02}$&2.574&4&\nodata&\nodata&2.574&0&$2.42^{+0.42}_{-0.35}$&$3.19^{+0.06}_{-0.05}$&$2.92^{+3.58}_{-1.25}$&$1.50^{+0.07}_{-0.00}$&$1.67^{+0.12}_{-0.13}$&\nodata\\
SG-ALMA-2&53.047208&-27.870000&8.83\,$\pm$\,0.26&$12.97^{+0.03}_{-0.04}$&3.693&4&\nodata&\nodata&3.693&0&$3.19^{+1.27}_{-0.32}$&$3.68^{+0.08}_{-0.07}$&$3.23^{+2.81}_{-1.20}$&$4.45^{+0.13}_{-0.28}$&$3.53^{+0.03}_{-0.06}$&\nodata\\
SG-ALMA-3&53.063875&-27.843778&6.61\,$\pm$\,0.16&$12.86^{+0.02}_{-0.02}$&2.649&4&\nodata&\nodata&2.649&0&$3.17^{+0.37}_{-0.24}$&$2.82^{+0.05}_{-0.04}$&$2.17^{+1.61}_{-1.00}$&$3.12^{+0.03}_{-0.03}$&$2.82^{+0.13}_{-0.16}$&$2.92^{+0.00}_{-0.11}$\\
SG-ALMA-4&53.020371&-27.779917&6.45\,$\pm$\,0.41&$12.71^{+0.01}_{-0.01}$&1.965&3&\nodata&9&1.965&0&$2.83^{+0.18}_{-0.25}$&$2.31^{+0.04}_{-0.03}$&$2.23^{+4.89}_{-1.30}$&$1.95^{+0.03}_{-0.02}$&$1.85^{+0.11}_{-0.17}$&$1.80^{+0.08}_{-0.02}$\\
SG-ALMA-5&53.118750&-27.782861&6.39\,$\pm$\,0.16&$12.85^{+0.02}_{-0.02}$&2.309&3&2.309&3&2.309&0&$2.89^{+0.25}_{-0.25}$&$2.45^{+0.04}_{-0.04}$&$1.82^{+0.91}_{-0.90}$&$2.29^{+0.26}_{-0.02}$&$2.35^{+0.07}_{-0.12}$&$2.52^{+0.00}_{-0.26}$\\
SG-ALMA-6&53.195125&-27.855778&5.90\,$\pm$\,0.18&$12.85^{+0.03}_{-0.03}$&3.075&4&\nodata&\nodata&3.075&0&$4.16^{+0.26}_{-0.42}$&$3.05^{+0.06}_{-0.06}$&$2.52^{+3.79}_{-1.15}$&$2.00^{+0.01}_{-0.07}$&$2.39^{+1.06}_{-0.06}$&$1.10^{+0.00}_{-0.00}$\\
SG-ALMA-7&53.158333&-27.733611&5.60\,$\pm$\,0.14&$12.91^{+0.03}_{-0.04}$&3.672&4&3.675&14&3.672&0&$3.36^{+0.33}_{-0.24}$&$3.65^{+0.10}_{-0.09}$&$2.73^{+1.66}_{-0.85}$&$3.49^{+0.04}_{-0.14}$&$2.97^{+0.04}_{-0.06}$&\nodata\\
SG-ALMA-8&53.105208&-27.875194&5.18\,$\pm$\,0.22&$12.59^{+0.04}_{-0.04}$&2.673&4&2.678&14&2.673&0&$2.70^{+0.27}_{-0.12}$&$3.06^{+0.06}_{-0.06}$&$2.67^{+4.60}_{-1.25}$&$2.69^{+0.04}_{-0.07}$&$2.68^{+0.14}_{-0.26}$&\nodata\\
SG-ALMA-9&53.148875&-27.821167&5.09\,$\pm$\,0.12&$12.83^{+0.06}_{-0.05}$&2.575&3&2.576&2&2.575&0&$2.31^{+0.29}_{-0.29}$&$2.24^{+0.04}_{-0.04}$&$2.12^{+4.79}_{-1.50}$&$2.55^{+0.07}_{-0.01}$&$2.59^{+0.03}_{-0.14}$&$2.44^{+0.02}_{-0.01}$\\
SG-ALMA-10&53.082083&-27.767278&4.90\,$\pm$\,0.29&$12.59^{+0.03}_{-0.02}$&2.304&4&2.290&12&2.304&0&$1.93^{+0.57}_{-0.28}$&$2.32^{+0.04}_{-0.05}$&$2.17^{+5.84}_{-1.30}$&$2.41^{+0.05}_{-0.07}$&$2.36^{+0.12}_{-0.18}$&\nodata\\
SG-ALMA-11&53.079375&-27.870806&4.76\,$\pm$\,0.29&$13.00^{+0.02}_{-0.03}$&3.468&4&\nodata&\nodata&3.468&0&$2.79^{+0.27}_{-0.36}$&$2.77^{+0.06}_{-0.05}$&$2.27^{+4.54}_{-1.15}$&$1.86^{+0.04}_{-0.09}$&$1.92^{+0.07}_{-0.05}$&\nodata\\
SG-ALMA-12&53.142792&-27.827861&4.73\,$\pm$\,0.16&$12.92^{+0.03}_{-0.03}$&\nodata&\nodata&\nodata&\nodata&3.43&2&$3.43^{+1.20}_{-0.21}$&$3.26^{+0.08}_{-0.08}$&$2.62^{+3.30}_{-1.10}$&$3.76^{+0.05}_{-0.04}$&$3.74^{+0.02}_{-0.10}$&\nodata\\
SG-ALMA-13&53.074833&-27.875889&4.69\,$\pm$\,0.81&$12.97^{+0.05}_{-0.05}$&4.426&3&4.440&10&4.426&0&$4.25^{+0.95}_{-1.78}$&$2.67^{+0.13}_{-0.11}$&$2.58^{+5.77}_{-0.85}$&$2.73^{+1.71}_{-0.05}$&$4.51^{+0.11}_{-0.10}$&\nodata\\
SG-ALMA-14&53.092333&-27.826833&4.64\,$\pm$\,0.17&$12.98^{+0.02}_{-0.02}$&2.524&2&\nodata&\nodata&2.524&0&$2.79^{+0.46}_{-0.15}$&$1.87^{+0.04}_{-0.04}$&$1.48^{+0.66}_{-0.55}$&$2.73^{+0.08}_{-0.09}$&$2.36^{+0.18}_{-0.08}$&\nodata\\
SG-ALMA-15&53.024292&-27.805694&3.93\,$\pm$\,0.15&$12.36^{+0.05}_{-0.04}$&2.067&4&\nodata&\nodata&2.067&0&$2.41^{+0.29}_{-0.33}$&$2.26^{+0.06}_{-0.06}$&$1.88^{+5.79}_{-1.25}$&$2.15^{+0.08}_{-0.08}$&$2.09^{+0.03}_{-0.29}$&\nodata\\
SG-ALMA-16&53.082750&-27.866583&4.31\,$\pm$\,0.15&$12.54^{+0.07}_{-0.13}$&\nodata&\nodata&3.464&13&3.464&0&$3.31^{+0.31}_{-0.25}$&$3.56^{+0.24}_{-0.20}$&$3.08^{+1.27}_{-0.95}$&$3.38^{+0.04}_{-0.21}$&$3.53^{+0.06}_{-0.27}$&\nodata\\
SG-ALMA-17&53.146625&-27.871028&3.80\,$\pm$\,0.18&$12.40^{+0.07}_{-0.09}$&3.581&3&3.583&14&3.581&0&$3.38^{+0.42}_{-0.16}$&$4.01^{+0.17}_{-0.17}$&$3.12^{+2.47}_{-1.20}$&$3.58^{+0.24}_{-0.12}$&$3.57^{+0.14}_{-0.17}$&\nodata\\
SG-ALMA-18&53.092833&-27.801331&5.21\,$\pm$\,0.32&$12.86^{+0.04}_{-0.06}$&\nodata&\nodata&3.847&7&3.847&0&$3.45^{+0.37}_{-0.31}$&$3.76^{+0.16}_{-0.13}$&$2.62^{+2.28}_{-0.95}$&$2.96^{+0.06}_{-0.10}$&$2.82^{+0.13}_{-0.10}$&\nodata\\
SG-ALMA-19&53.108792&-27.869028&3.62\,$\pm$\,0.17&$12.44^{+0.14}_{-0.09}$&\nodata&\nodata&3.747&13&3.747&0&$4.40^{+0.51}_{-1.25}$&$4.50^{+0.36}_{-0.29}$&$3.38^{+2.81}_{-1.15}$&$4.48^{+1.17}_{-0.40}$&$17.24^{+3.74}_{-0.60}$&\nodata\\
SG-ALMA-20&53.198292&-27.747861&3.61\,$\pm$\,0.30&$12.63^{+0.02}_{-0.02}$&1.923&2&\nodata&\nodata&1.923&0&$1.91^{+0.00}_{-0.00}$&$1.80^{+0.04}_{-0.04}$&$1.07^{+1.03}_{-1.05}$&$1.94^{+0.03}_{-0.04}$&$1.83^{+0.10}_{-0.12}$&$1.57^{+0.09}_{-0.10}$\\
SG-ALMA-21&53.178333&-27.870194&3.55\,$\pm$\,0.20&$12.53^{+0.08}_{-0.09}$&3.467&3&3.478&10&3.467&0&$3.71^{+1.13}_{-0.53}$&$3.88^{+0.20}_{-0.17}$&$2.67^{+1.46}_{-0.75}$&$3.79^{+0.05}_{-0.09}$&$4.23^{+0.42}_{-0.52}$&\nodata\\
SG-ALMA-22&53.183458&-27.776611&3.38\,$\pm$\,0.32&$12.65^{+0.03}_{-0.04}$&\nodata&\nodata&2.698&8&2.698&0&$3.24^{+0.10}_{-0.12}$&$2.74^{+0.08}_{-0.08}$&$2.17^{+5.06}_{-1.75}$&$2.86^{+0.03}_{-0.12}$&$2.90^{+0.04}_{-0.08}$&$2.70^{+0.01}_{-0.01}$\\
SG-ALMA-23&53.157167&-27.833497&3.32\,$\pm$\,0.29&$12.44^{+0.02}_{-0.02}$&\nodata&\nodata&1.618&1&1.618&0&$1.59^{+0.07}_{-0.07}$&$1.95^{+0.04}_{-0.04}$&$1.27^{+1.13}_{-1.25}$&$1.58^{+0.04}_{-0.01}$&$1.62^{+0.06}_{-0.05}$&$1.42^{+0.00}_{-0.06}$\\
SG-ALMA-24&53.102750&-27.892833&3.25\,$\pm$\,0.14&$12.55^{+0.01}_{-0.02}$&1.895&2&\nodata&\nodata&1.895&0&$2.26^{+0.41}_{-0.62}$&$2.16^{+0.04}_{-0.04}$&$1.38^{+0.91}_{-0.85}$&$1.97^{+0.03}_{-0.07}$&$1.85^{+0.08}_{-0.09}$&$1.92^{+0.00}_{-0.01}$\\
SG-ALMA-25&53.181375&-27.777556&3.18\,$\pm$\,0.23&$12.43^{+0.04}_{-0.05}$&\nodata&\nodata&2.696&8&2.696&0&$2.94^{+0.16}_{-0.23}$&$3.09^{+0.10}_{-0.09}$&$2.48^{+5.30}_{-1.20}$&$2.92^{+0.06}_{-0.07}$&$2.81^{+0.16}_{-0.08}$&\nodata\\
SG-ALMA-26&53.070250&-27.845611&3.15\,$\pm$\,0.25&$12.70^{+0.06}_{-0.08}$&3.704&3&3.688&5&3.704&0&$3.46^{+0.28}_{-0.24}$&$2.98^{+0.24}_{-0.17}$&$2.48^{+4.72}_{-1.00}$&$3.79^{+0.04}_{-0.08}$&$3.70^{+0.01}_{-0.10}$&\nodata\\
SG-ALMA-27&53.014583&-27.844389&3.05\,$\pm$\,0.19&$12.14^{+0.08}_{-0.14}$&2.569&2&\nodata&\nodata&2.569&0&$2.55^{+0.49}_{-0.40}$&$3.42^{+0.26}_{-0.21}$&$2.83^{+4.00}_{-1.10}$&$1.78^{+0.96}_{-0.11}$&$2.46^{+0.11}_{-0.09}$&\nodata\\
SG-ALMA-28&53.139250&-27.890722&2.89\,$\pm$\,0.37&$12.47^{+0.07}_{-0.10}$&2.688&1&\nodata&\nodata&2.688&0&$2.51^{+0.32}_{-0.59}$&$2.94^{+0.13}_{-0.11}$&$2.38^{+5.59}_{-1.20}$&$3.33^{+1.05}_{-0.13}$&$2.90^{+6.20}_{-0.03}$&\nodata\\
SG-ALMA-29&53.137125&-27.761361&2.82\,$\pm$\,0.28&$12.13^{+0.05}_{-0.04}$&1.847&3&1.848&14&1.847&0&$1.94^{+0.33}_{-0.19}$&$2.31^{+0.06}_{-0.06}$&$2.38^{+6.18}_{-1.25}$&$0.52^{+0.02}_{-0.01}$&$1.68^{+0.16}_{-0.05}$&$1.97^{+0.14}_{-0.02}$\\
SG-ALMA-30&53.071708&-27.843667&2.78\,$\pm$\,0.15&$12.33^{+0.03}_{-0.05}$&1.944&2&\nodata&\nodata&1.944&0&$1.99^{+0.27}_{-0.37}$&$2.18^{+0.05}_{-0.05}$&$1.62^{+4.73}_{-1.60}$&$1.86^{+0.06}_{-0.05}$&$1.77^{+0.08}_{-0.10}$&$1.72^{+0.03}_{-0.01}$\\
SG-ALMA-31&53.077375&-27.859611&2.54\,$\pm$\,0.43&$12.35^{+0.02}_{-0.04}$&2.035&3&2.036&9&2.035&0&$1.58^{+0.03}_{-0.20}$&$1.90^{+0.06}_{-0.05}$&$1.88^{+6.25}_{-1.35}$&$1.95^{+0.06}_{-0.05}$&$2.04^{+0.00}_{-0.35}$&$2.25^{+0.02}_{-0.30}$\\
SG-ALMA-32&53.049750&-27.770944&2.56\,$\pm$\,0.16&$12.70^{+0.03}_{-0.03}$&2.574&2&\nodata&\nodata&2.574&0&$3.18^{+0.23}_{-0.33}$&$2.91^{+0.09}_{-0.08}$&$1.62^{+0.53}_{-0.55}$&$2.75^{+0.04}_{-0.18}$&$2.56^{+0.10}_{-0.19}$&\nodata\\
SG-ALMA-33&53.072708&-27.834278&2.49\,$\pm$\,0.23&$12.44^{+0.02}_{-0.02}$&\nodata&\nodata&1.617&1&1.617&0&$1.44^{+0.19}_{-0.25}$&$1.74^{+0.04}_{-0.04}$&$1.23^{+4.21}_{-1.20}$&$1.58^{+0.03}_{-0.05}$&$1.68^{+0.05}_{-0.12}$&$1.48^{+0.03}_{-0.03}$\\
SG-ALMA-34&53.090750&-27.782472&2.47\,$\pm$\,0.21&$12.35^{+0.04}_{-0.04}$&\nodata&\nodata&\nodata&\nodata&1.95&1&$1.76^{+0.31}_{-0.27}$&$1.93^{+0.06}_{-0.07}$&$1.38^{+1.44}_{-1.10}$&$1.95^{+0.02}_{-0.09}$&$1.75^{+0.11}_{-0.13}$&$1.93^{+0.02}_{-0.10}$\\
SG-ALMA-35&53.091708&-27.712139&2.47\,$\pm$\,0.13&$12.22^{+0.02}_{-0.04}$&\nodata&\nodata&1.612&3&1.612&0&$1.58^{+0.15}_{-0.10}$&$2.00^{+0.06}_{-0.05}$&$1.57^{+4.82}_{-1.55}$&$1.71^{+0.01}_{-0.06}$&$1.63^{+0.03}_{-0.09}$&$1.71^{+0.09}_{-0.04}$\\
SG-ALMA-36&53.086583&-27.810222&2.41\,$\pm$\,0.25&$12.20^{+0.06}_{-0.06}$&\nodata&\nodata&\nodata&\nodata&2.38&1&$2.68^{+0.41}_{-0.50}$&$2.23^{+0.07}_{-0.07}$&$2.12^{+6.18}_{-1.30}$&$2.38^{+0.05}_{-0.09}$&$2.36^{+0.24}_{-0.05}$&\nodata\\
SG-ALMA-37&53.146375&-27.888806&2.35\,$\pm$\,0.28&$12.46^{+0.06}_{-0.08}$&\nodata&\nodata&\nodata&\nodata&2.81&2&$2.81^{+0.19}_{-0.19}$&$2.59^{+0.11}_{-0.11}$&$1.88^{+4.74}_{-1.85}$&$2.96^{+0.09}_{-0.08}$&$4.01^{+0.06}_{-0.98}$&\nodata\\
SG-ALMA-38&53.092333&-27.803222&2.50\,$\pm$\,0.10&$12.36^{+0.04}_{-0.05}$&\nodata&\nodata&\nodata&\nodata&2.31&1&$2.74^{+0.23}_{-0.31}$&$2.40^{+0.05}_{-0.05}$&$2.23^{+5.49}_{-1.25}$&$2.31^{+0.07}_{-0.04}$&$2.35^{+0.08}_{-0.11}$&\nodata\\
SG-ALMA-39&53.124292&-27.882694&2.26\,$\pm$\,0.18&$12.73^{+0.04}_{-0.03}$&\nodata&\nodata&\nodata&\nodata&2.68&2&$2.68^{+0.64}_{-0.67}$&$2.41^{+0.08}_{-0.08}$&$1.62^{+3.69}_{-1.60}$&$3.04^{+0.17}_{-0.05}$&$2.81^{+0.08}_{-0.07}$&\nodata\\
SG-ALMA-40&53.131083&-27.773194&2.26\,$\pm$\,0.17&$12.48^{+0.03}_{-0.03}$&\nodata&\nodata&2.223&4&2.223&0&$2.14^{+0.38}_{-0.13}$&$1.95^{+0.05}_{-0.05}$&$1.82^{+5.82}_{-1.45}$&$2.23^{+0.02}_{-0.03}$&$2.30^{+0.01}_{-0.16}$&$2.20^{+0.00}_{-0.00}$\\
SG-ALMA-41&53.172792&-27.858833&2.25\,$\pm$\,0.18&$12.67^{+0.07}_{-0.10}$&\nodata&\nodata&\nodata&\nodata&2.72&2&$2.72^{+2.60}_{-0.33}$&$3.46^{+0.46}_{-0.34}$&$2.83^{+1.01}_{-0.75}$&$4.13^{+0.33}_{-0.68}$&$2.97^{+0.21}_{-0.11}$&\nodata\\
SG-ALMA-42&53.091625&-27.853389&2.25\,$\pm$\,0.18&$12.33^{+0.03}_{-0.04}$&\nodata&\nodata&2.034&13&2.034&0&$1.99^{+0.11}_{-0.07}$&$2.08^{+0.05}_{-0.06}$&$1.88^{+5.86}_{-1.45}$&$2.34^{+0.08}_{-0.04}$&$3.09^{+0.21}_{-0.03}$&\nodata\\
SG-ALMA-43&53.068833&-27.879722&2.23\,$\pm$\,0.41&$12.46^{+0.04}_{-0.06}$&2.562&2&\nodata&\nodata&2.562&0&$2.55^{+0.21}_{-0.31}$&$2.60^{+0.07}_{-0.06}$&$2.17^{+5.73}_{-1.15}$&$2.39^{+0.23}_{-0.07}$&$2.46^{+0.11}_{-0.04}$&\nodata\\
SG-ALMA-44&53.087125&-27.840194&2.21\,$\pm$\,0.12&$12.27^{+0.11}_{-0.11}$&\nodata&\nodata&3.476&14&3.476&0&$3.52^{+0.18}_{-0.14}$&$3.84^{+0.38}_{-0.32}$&$3.08^{+2.29}_{-1.00}$&\nodata&$15.03^{+0.08}_{-0.34}$&\nodata\\
SG-ALMA-45&53.041083&-27.837694&2.43\,$\pm$\,0.21&$12.68^{+0.06}_{-0.07}$&\nodata&\nodata&3.779&14&3.779&0&$2.69^{+0.28}_{-0.47}$&$2.66^{+0.09}_{-0.08}$&$2.23^{+5.30}_{-1.30}$&$7.62^{+0.31}_{-0.47}$&$3.06^{+0.27}_{-0.12}$&\nodata\\
SG-ALMA-46&53.104875&-27.705278&2.29\,$\pm$\,0.11&$12.47^{+0.02}_{-0.02}$&\nodata&\nodata&1.613&6&1.613&0&$1.92^{+0.65}_{-0.43}$&$1.52^{+0.05}_{-0.04}$&$0.97^{+3.93}_{-0.95}$&$0.10^{+0.01}_{-0.01}$&\nodata&$0.99^{+0.00}_{-0.00}$\\
SG-ALMA-47&53.163500&-27.890556&2.05\,$\pm$\,0.15&$12.55^{+0.02}_{-0.02}$&\nodata&\nodata&2.324&9&2.324&0&$2.47^{+0.25}_{-0.22}$&$1.73^{+0.05}_{-0.06}$&$1.62^{+5.62}_{-1.60}$&$2.20^{+0.03}_{-0.08}$&$2.18^{+0.14}_{-0.10}$&$2.28^{+0.00}_{-0.01}$\\
SG-ALMA-48&53.160625&-27.776250&2.04\,$\pm$\,0.36&$12.87^{+0.02}_{-0.03}$&\nodata&\nodata&2.554&12&2.554&0&$2.19^{+0.30}_{-0.05}$&$1.99^{+0.06}_{-0.06}$&$1.27^{+1.78}_{-1.25}$&$2.58^{+0.10}_{-0.02}$&$2.71^{+0.06}_{-0.14}$&$2.54^{+0.00}_{-0.00}$\\
SG-ALMA-49&53.053667&-27.869278&1.98\,$\pm$\,0.23&$12.08^{+0.05}_{-0.05}$&\nodata&\nodata&\nodata&\nodata&1.88&1&$2.32^{+0.38}_{-0.67}$&$1.98^{+0.09}_{-0.09}$&$1.62^{+5.43}_{-1.60}$&$1.88^{+0.06}_{-0.03}$&$1.94^{+0.07}_{-0.21}$&\nodata\\
SG-ALMA-50&53.089542&-27.711639&1.97\,$\pm$\,0.45&$12.19^{+0.04}_{-0.04}$&\nodata&\nodata&\nodata&\nodata&1.70&1&$1.74^{+0.31}_{-0.09}$&$1.76^{+0.07}_{-0.07}$&$1.62^{+5.66}_{-1.60}$&$1.70^{+0.01}_{-0.06}$&$1.76^{+0.16}_{-0.06}$&$1.69^{+0.04}_{-0.13}$\\
SG-ALMA-51&53.067792&-27.728889&1.94\,$\pm$\,0.22&$12.14^{+0.06}_{-0.08}$&\nodata&\nodata&\nodata&\nodata&2.33&1&$2.23^{+0.22}_{-0.33}$&$2.15^{+0.07}_{-0.08}$&$2.23^{+6.25}_{-1.25}$&$2.33^{+0.10}_{-0.03}$&$2.56^{+0.15}_{-0.18}$&\nodata\\
SG-ALMA-52&53.064792&-27.862611&1.88\,$\pm$\,0.24&$12.33^{+0.09}_{-0.16}$&3.514&1&\nodata&\nodata&3.514&0&$4.42^{+0.73}_{-1.57}$&$3.53^{+0.27}_{-0.26}$&$2.48^{+3.86}_{-1.10}$&$4.79^{+0.31}_{-0.43}$&$3.57^{+0.00}_{-0.04}$&\nodata\\
SG-ALMA-53&53.198875&-27.843944&1.86\,$\pm$\,0.32&$12.00^{+0.03}_{-0.03}$&\nodata&\nodata&1.616&9&1.616&0&$2.07^{+0.25}_{-0.37}$&$1.90^{+0.08}_{-0.07}$&$1.62^{+5.55}_{-1.60}$&$1.57^{+0.04}_{-0.06}$&$2.48^{+0.22}_{-0.89}$&$1.41^{+0.02}_{-0.02}$\\
SG-ALMA-54&53.181958&-27.814194&1.82\,$\pm$\,0.30&$12.60^{+0.03}_{-0.04}$&\nodata&\nodata&2.616&12&2.616&0&$1.15^{+0.11}_{-0.08}$&$1.65^{+0.08}_{-0.09}$&$0.97^{+4.96}_{-0.95}$&$9.42^{+0.41}_{-0.07}$&$2.63^{+0.10}_{-0.10}$&\nodata\\
SG-ALMA-55&53.048375&-27.770306&1.79\,$\pm$\,0.15&$12.89^{+0.04}_{-0.04}$&2.163&1&\nodata&\nodata&2.163&0&$2.59^{+0.25}_{-0.27}$&$2.21^{+0.07}_{-0.06}$&$1.73^{+5.05}_{-1.70}$&$1.06^{+0.02}_{-0.01}$&$2.48^{+0.19}_{-0.12}$&\nodata\\
SG-ALMA-56&53.107042&-27.718333&1.61\,$\pm$\,0.25&$12.38^{+0.04}_{-0.04}$&\nodata&\nodata&2.299&4&2.299&0&$3.04^{+0.26}_{-0.38}$&$0.11^{+0.02}_{-0.01}$&$1.73^{+5.47}_{-1.70}$&$1.55^{+0.01}_{-0.02}$&$0.78^{+1.93}_{-0.13}$&$2.11^{+0.01}_{-0.07}$\\
SG-ALMA-57&53.033125&-27.816778&1.72\,$\pm$\,0.26&$11.57^{+0.33}_{-0.33}$&\nodata&\nodata&\nodata&\nodata&3.21&2&$3.21^{+0.53}_{-0.44}$&$8.59^{+2.40}_{-2.90}$&$6.47^{+4.04}_{-1.35}$&$3.08^{+0.61}_{-0.08}$&$2.80^{+0.18}_{-0.07}$&\nodata\\
SG-ALMA-58&53.183625&-27.836500&1.72\,$\pm$\,0.31&$12.00^{+0.43}_{-0.26}$&4.211&2&\nodata&\nodata&4.211&0&$4.57^{+0.50}_{-0.96}$&$5.57^{+1.12}_{-0.97}$&$4.28^{+0.70}_{-0.80}$&$4.73^{+0.17}_{-0.34}$&$3.76^{+0.83}_{-0.50}$&\nodata\\
SG-ALMA-59&53.094042&-27.804194&1.84\,$\pm$\,0.13&$11.96^{+0.06}_{-0.06}$&\nodata&\nodata&2.325&4&2.325&0&$2.71^{+0.26}_{-0.17}$&$2.32^{+0.08}_{-0.07}$&$2.23^{+5.98}_{-1.30}$&$2.31^{+0.09}_{-0.06}$&$2.61^{+0.02}_{-0.34}$&$2.37^{+0.01}_{-0.08}$\\
SG-ALMA-60&53.124583&-27.893278&1.61\,$\pm$\,0.25&$12.18^{+0.06}_{-0.06}$&\nodata&\nodata&\nodata&\nodata&2.53&1&$2.64^{+0.23}_{-0.30}$&$2.61^{+0.14}_{-0.16}$&$2.27^{+5.34}_{-2.25}$&$2.53^{+0.07}_{-0.12}$&$2.34^{+0.39}_{-0.07}$&\nodata\\
SG-ALMA-61&53.132750&-27.720278&1.61\,$\pm$\,0.25&$11.95^{+0.41}_{-0.27}$&\nodata&\nodata&\nodata&\nodata&3.20&2&$3.20^{+1.20}_{-1.07}$&$4.48^{+4.06}_{-1.24}$&$2.92^{+1.82}_{-0.95}$&$4.67^{+0.56}_{-0.19}$&$4.33^{+12.01}_{-0.08}$&\nodata\\
SG-ALMA-62&53.080667&-27.720861&1.59\,$\pm$\,0.17&$12.32^{+0.07}_{-0.13}$&\nodata&\nodata&3.110&5&3.110&0&$3.30^{+0.46}_{-0.42}$&$2.93^{+0.36}_{-0.25}$&$2.23^{+1.10}_{-0.80}$&$2.94^{+0.09}_{-0.06}$&$2.90^{+0.10}_{-0.13}$&\nodata\\
SG-ALMA-63&53.120000&-27.808250&1.57\,$\pm$\,0.26&$12.09^{+0.03}_{-0.03}$&\nodata&\nodata&\nodata&\nodata&1.84&1&$1.83^{+0.07}_{-0.08}$&$1.84^{+0.08}_{-0.08}$&$1.73^{+5.57}_{-1.70}$&$1.84^{+0.05}_{-0.06}$&$1.81^{+0.09}_{-0.06}$&$1.69^{+0.05}_{-0.06}$\\
SG-ALMA-64&53.117083&-27.874917&1.53\,$\pm$\,0.31&$12.72^{+0.03}_{-0.03}$&\nodata&\nodata&\nodata&\nodata&3.35&2&$3.35^{+0.21}_{-0.20}$&$3.02^{+0.22}_{-0.24}$&$1.32^{+0.62}_{-0.65}$&$3.27^{+0.13}_{-0.06}$&$3.61^{+0.22}_{-0.45}$&\nodata\\
SG-ALMA-65&53.131458&-27.841361&1.46\,$\pm$\,0.14&$12.01^{+0.04}_{-0.04}$&\nodata&\nodata&1.614&13&1.614&0&$1.56^{+0.06}_{-0.05}$&$2.03^{+0.07}_{-0.06}$&$1.73^{+5.76}_{-1.70}$&$1.58^{+0.05}_{-0.02}$&$1.72^{+0.05}_{-0.07}$&$1.61^{+0.03}_{-0.01}$\\
SG-ALMA-66&53.044708&-27.802025&1.44\,$\pm$\,0.26&$11.64^{+0.02}_{-0.02}$&\nodata&\nodata&0.654&6&0.654&0&$1.09^{+0.26}_{-0.29}$&$1.29^{+0.05}_{-0.05}$&$0.42^{+0.48}_{-0.40}$&$0.68^{+0.01}_{-0.03}$&$0.68^{+0.06}_{-0.06}$&$0.66^{+0.00}_{-0.00}$\\
SG-ALMA-67&53.072000&-27.819000&1.36\,$\pm$\,0.19&$11.90^{+0.04}_{-0.06}$&\nodata&\nodata&\nodata&\nodata&1.70&1&$1.63^{+0.51}_{-0.22}$&$2.30^{+0.11}_{-0.11}$&$1.32^{+3.10}_{-1.30}$&$1.70^{+0.15}_{-0.03}$&$1.74^{+0.07}_{-0.13}$&$1.56^{+0.11}_{-0.02}$\\
SG-ALMA-68&53.120458&-27.742056&1.35\,$\pm$\,0.24&$13.74^{+0.03}_{-0.02}$&\nodata&\nodata&3.246&11&3.246&0&$1.76^{+0.44}_{-0.42}$&$1.55^{+0.05}_{-0.05}$&$0.03^{+2.76}_{-0.00}$&\nodata&$3.64^{+0.06}_{-0.36}$&\nodata\\
SG-ALMA-69&53.113125&-27.886611&1.25\,$\pm$\,0.27&$12.13^{+0.09}_{-0.11}$&\nodata&\nodata&\nodata&\nodata&2.55&1&$2.91^{+0.21}_{-0.22}$&$2.77^{+0.23}_{-0.21}$&$1.73^{+5.33}_{-1.70}$&$2.55^{+0.10}_{-0.07}$&$2.67^{+0.11}_{-0.21}$&$2.43^{+0.08}_{-0.12}$\\
SG-ALMA-70&53.141250&-27.872833&1.18\,$\pm$\,0.25&$12.49^{+0.08}_{-0.07}$&\nodata&\nodata&\nodata&\nodata&3.04&2&$3.04^{+0.27}_{-0.42}$&$2.84^{+0.22}_{-0.21}$&$1.62^{+2.25}_{-1.60}$&$3.14^{+0.21}_{-0.08}$&$3.13^{+0.33}_{-0.19}$&\nodata\\
SG-ALMA-71&53.056833&-27.798389&1.16\,$\pm$\,0.30&$11.85^{+0.08}_{-0.06}$&\nodata&\nodata&\nodata&\nodata&1.71&1&$2.03^{+0.25}_{-0.21}$&$1.72^{+0.19}_{-0.46}$&$0.28^{+5.89}_{-0.25}$&$1.71^{+0.02}_{-0.07}$&$1.73^{+0.11}_{-0.15}$&$1.79^{+0.01}_{-0.02}$\\
SG-ALMA-72&53.119917&-27.743111&1.11\,$\pm$\,0.29&$12.94^{+0.04}_{-0.05}$&\nodata&\nodata&\nodata&\nodata&1.92&2&$1.92^{+0.78}_{-0.29}$&$1.85^{+0.09}_{-0.10}$&$1.23^{+0.49}_{-0.50}$&$3.76^{+0.55}_{-0.29}$&$3.06^{+-0.00}_{-0.87}$&\nodata\\
SG-ALMA-73&53.142833&-27.874083&1.07\,$\pm$\,0.17&$12.38^{+0.15}_{-0.09}$&\nodata&\nodata&3.470&13&3.470&0&$3.04^{+0.22}_{-0.26}$&$2.70^{+0.24}_{-0.23}$&$1.73^{+5.75}_{-1.30}$&$2.20^{+0.03}_{-0.10}$&$3.12^{+0.18}_{-0.11}$&\nodata\\
SG-ALMA-74&53.093625&-27.826444&0.93\,$\pm$\,0.23&$11.72^{+0.01}_{-0.02}$&\nodata&\nodata&0.732&6&0.732&0&$0.76^{+0.06}_{-0.04}$&$1.41^{+0.04}_{-0.04}$&$0.82^{+2.87}_{-0.80}$&$0.77^{+0.01}_{-0.05}$&$0.75^{+0.03}_{-0.03}$&$0.71^{+0.02}_{-0.00}$\\
SG-ALMA-75&53.074833&-27.787111&0.84\,$\pm$\,0.14&$12.89^{+0.10}_{-0.06}$&\nodata&\nodata&\nodata&\nodata&2.02&2&$2.02^{+0.44}_{-0.42}$&$2.16^{+0.19}_{-0.17}$&$1.43^{+0.40}_{-0.40}$&\nodata&$3.65^{+0.81}_{-0.77}$&\nodata\\

\enddata
\tablecomments{
Columns: 
(1) Source name;
(2) Right ascension;
(3) Declination;
(4) ALMA 870~$\mu$m flux and error;
(5) Infrared luminosity measured from {\tt mbb} fits (errors are 68\% confidence interval); 
(6) Spec-z obtained or confirmed by ALMA;
(7) Flag denoting $z_{\rm spec}$ quality (same as Table~\ref{tab:line_detections});
(8) Literature spec-z;
(9) Reference for literature spec-z (see below);
(10) Best redshift;
(11) Type of redshift (0 = spec-z, 1 = \citetalias{straatman16} photo-z, 2 = {\tt MAGPHYS} photo-z);
(12) Photo-z (median and 68\% confidence interval) from {\tt MAGPHYS};
(13) Photo-z (median and 68\% confidence interval) from {\tt mbb};
(14) Photo-z (median and 68\% confidence interval) from {\tt MMPZ};
(15) Photo-z (median and 68\% confidence interval) from \citetalias{straatman16} ({\tt EAZY});
(16) Photo-z (median and 68\% confidence interval) from \citetalias{eisenstein26} ({\tt EAZY});
(17) Grism-z (median and 68\% confidence interval) from \citetalias{momcheva16}.
\tablerefs{The literature spec-z references in Column~9 are as follows:
(1) Redshift from our own MOSFIRE observations; 
(2) \citet{vanzella08};
(3) \citet{kurk13}; 
(4) \citet{kriek15}; 
(5) \citet{mclure18, pentericci18, garilli21}; 
(6) \citetalias{cowie18}; 
(7) \citet{franco18}; 
(8) \citet{gonzalez-lopez19}; 
(9) \citet{birkin24};
(10) \citet{barrufet25}; 
(11) \citet{lapasia25}; 
(12) \citet{deugenio25}; 
(13) \citet{curtis-lake25, scholtz25};
(14) Taken from the DJA NIRSpec catalog\footnote{https://dawn-cph.github.io/dja/spectroscopy/nirspec/} (v4; \citealt{heintz25}).
}
}
\end{deluxetable*}
\end{longrotatetable}

\appendix
\section{Cutouts for the ALMA sources}

{In Figure~\ref{fig:cutouts}, we show color cutouts of the 75 DSFGs in our sample. Unless noted on the panels, we use the JWST/NIRCam F444W, F277W, and F150W filters for the images. The source name, 870~$\mu$m flux density, and redshift (spec-z or photo-z) are also labeled on each panel.}

\begin{figure*}[!ht]
    \centering
    \includegraphics[width=\linewidth]{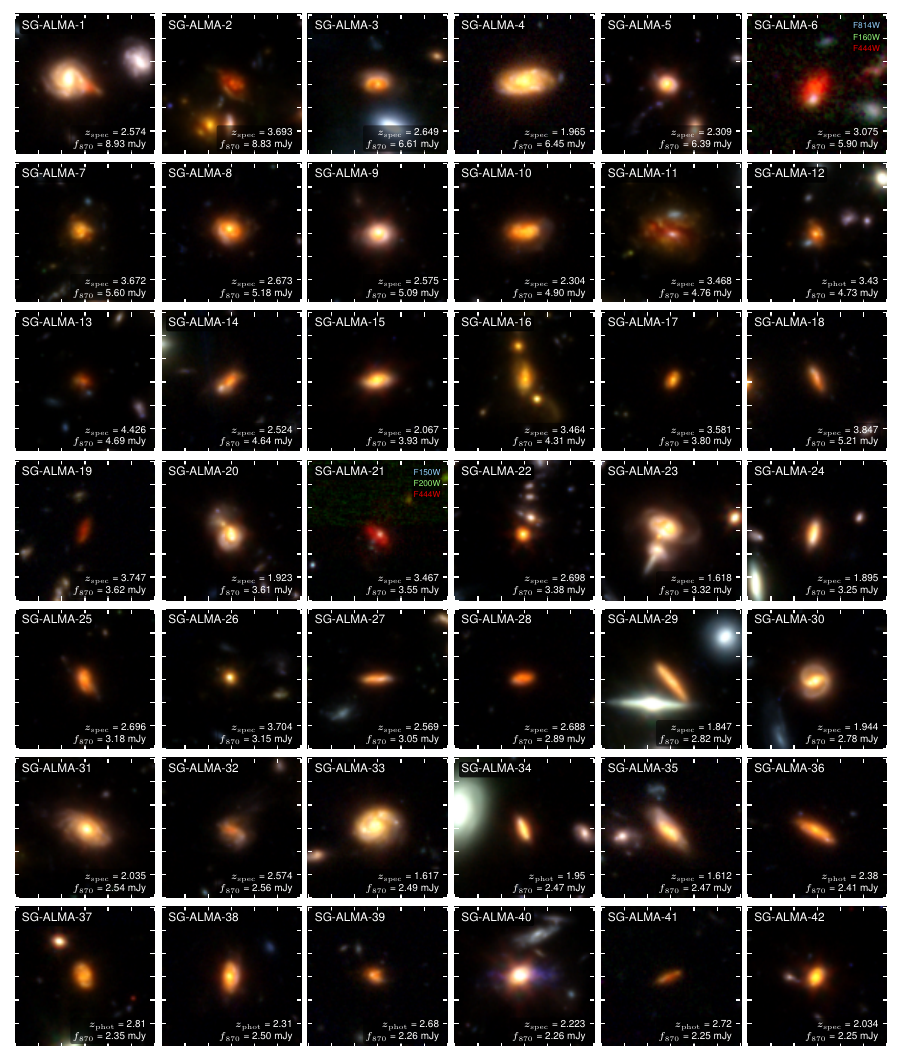}
    \caption{Color images of the ALMA sample. Unless otherwise noted, the filters are from JWST/NIRCam with R = F444W, G = F277W, and B = F150W. Where sources are not covered in these bands, we use other available HST and/or JWST bands and give the filter names on the panels. Cutouts are {$6''\times6''$}; north is up and east is to the left. We label the source name, 870~$\mu$m flux density, and redshift on each panel.
    }
    \label{fig:cutouts}
\end{figure*}
\begin{figure*}[!ht]
    \figurenum{10}
    \centering
    \includegraphics[width=\linewidth]{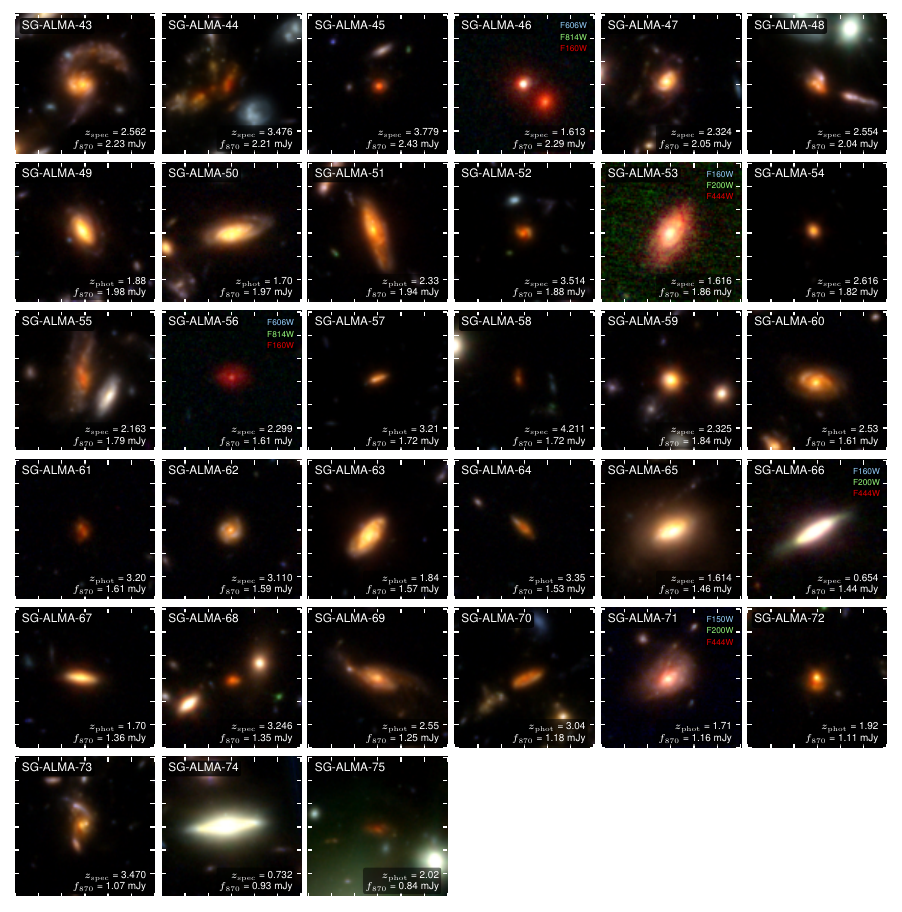}
    \caption{Continued.}
    \label{fig:cutouts2}
\end{figure*}

\end{document}